\documentclass[iop]{emulateapj-rtx4}
\usepackage{natbib,amsmath,lscape}
\usepackage{lscape}
\usepackage{graphicx}
\usepackage{natbib}
\usepackage{amssymb}
\usepackage{float}

\usepackage{makeidx}
\usepackage{amsmath}
\usepackage{rotating}
\usepackage{longtable}
\usepackage{subfigure}
\usepackage{subfloat}
\usepackage{morefloats}
\usepackage{captcont}
\nocite{*}

\newcommand{\mmsun}{M_\odot}

\newcommand{\etal}{et~al.\/}

\newcommand{\kms}{\mbox{km\thinspace s$^{-1}$}}

\newcommand{\nhi}{Log \mbox{$N_{\rm HI}$}}

\newcommand{\nlow}{\mbox{$N_{\rm Low}$}}
\def\ltk{\left [ \,}
\def\ltp{\left ( \,}

\def\rtk{\, \right  ] }
\def\rtp{\, \right  ) }

\def\sci#1{{\; \times \; 10^{#1}}}

\shortauthors{Werk \etal}
\shorttitle{Physical CGM: Mass and Density}

\begin{document} 
\slugcomment{v9.0, accepted to ApJ}

\title{The COS-Halos Survey: Physical Conditions and Baryonic Mass in the Low-Redshift Circumgalactic Medium}

\author{Jessica K.\ Werk\altaffilmark{1,2},
J. Xavier Prochaska\altaffilmark{1},
   Jason Tumlinson\altaffilmark{3}, 
   Molly S. Peeples\altaffilmark{3}, 
  Todd M. Tripp\altaffilmark{4}, 
 Andrew J. Fox\altaffilmark{3},
 Nicolas Lehner\altaffilmark{5}, 
 Christopher Thom \altaffilmark{3},
  John M. O'Meara\altaffilmark{6}, 
  Amanda Brady Ford\altaffilmark{7},
   Rongmon Bordoloi\altaffilmark{3},
   Neal Katz\altaffilmark{4}, 
   Nicolas Tejos\altaffilmark{1},
   Benjamin D. Oppenheimer\altaffilmark{8, 9},
   Romeel Dav{\'e}\altaffilmark{10}, and 
   David H. Weinberg\altaffilmark{11}
  }

\altaffiltext{1}{UCO/Lick Observatory; University of California, Santa Cruz, CA $jwerk@ucolick.org$}
\altaffiltext{2}{Hubble Fellow}
\altaffiltext{3}{Space Telescope Science Institute, 3700 San Martin Drive,  Baltimore, MD}
\altaffiltext{4}{Department of Astronomy, University of Massachusetts, Amherst, MA}
\altaffiltext{5}{Department of Physics and Astronomy, University of Notre Dame, South Bend, IN}
\altaffiltext{6}{Department of Chemistry and Physics, Saint Michael's College, Colchester, VT}
\altaffiltext{7}{Astronomy Department, University of Arizona, Tucson, AZ 85721, USA }
\altaffiltext{8}{Leiden Observatory, Leiden University, NL-2300 RA Leiden, Netherlands}
\altaffiltext{9}{CASA, Department of Astrophysical and Planetary Sciences, University of Colorado, Boulder, CO 80309}
\altaffiltext{10}{University of the Western Cape, Bellville, Cape Town 7535, South Africa}
\altaffiltext{11}{Department of Astronomy, The Ohio State University, Columbus, OH}
\begin{abstract}

We analyze the physical conditions of the cool, photoionized (T $\sim 10^4$K)
circumgalactic medium (CGM) using the COS-Halos suite of gas column density
measurements for 44 gaseous halos within 160\,kpc of $L \sim L^*$
galaxies at $z \sim 0.2$.
These data are well described by simple photoionization models, with
the gas highly ionized (n$_{\rm HII}$/n$_{\rm H} \gtrsim 99\%$) by
the extragalactic ultraviolet background (EUVB).  Scaling by estimates
for the virial radius, R$_{\rm vir}$, we show that
 the ionization state (tracked by the dimensionless ionization
 parameter, U) increases with distance from the host galaxy. The ionization parameters imply a decreasing volume density
 profile n$_{\rm H}$ = (10$^{-4.2 \pm 0.25}$)(R/R$_{\rm vir})^{-0.8\pm0.3}$.  Our derived gas volume densities
 are several orders of magnitude lower than predictions from standard two-phase
 models with a cool medium in pressure equilibrium with a hot,  
 coronal medium expected in virialized halos at this mass scale.  Applying the ionization corrections to the  \ion{H}{1} column
 densities, we estimate a lower limit to the cool
 gas mass M$_{\rm CGM}^{\rm cool} > 6.5 \times 10^{10}$ M$_{\odot}$
 for the volume within R $<$ R$_{\rm vir}$.  Allowing for an additional warm-hot, OVI-traced phase, the CGM accounts for {\emph{at least}} half of the baryons purported to be missing from dark matter halos at the 10$^{12}$  M$_{\odot}$ scale. 
   
\end{abstract}

\keywords{galaxies: halos -- galaxies:formation -- intergalactic medium --- quasars:absorption lines}

\section{Introduction}
\label{sec:intro}

Baryons account for 17\%
of the gravitating mass in the universe ($\Omega_b$ = 0.17 $\Omega_m$; Blumenthal et al. 1984; Dunkley et al. 2009\nocite{cdm84, wmap05}).  Yet, observational inventories reveal a shortage of baryons on both universal and galaxy-halo scales.  The first `missing baryon problem' is illustrated by counting up all the baryons
revealed by observations of stars, dust, and gas in galaxies and clusters ($\Omega_g$). The
total is  significantly less than the value  expected from the
widely-accepted Big Bang Nucleosynthesis model, weighing in at only
0.03 - 0.07$\Omega_b$ \citep{persic92, fhp98, bell03b}. Second, baryons are apparently missing from galaxies themselves in what is known as the galaxy halo
missing baryon problem \citep{mcgaugh07,bregman07, mcgaugh10}.  To
explain these baryon shortages one must invoke unseen or poorly-defined components: highly photoionized intergalactic hydrogen, known as the Ly$\alpha$ forest \citep{lynds71, sargent80, cen94}, the warm-hot intergalactic medium, or WHIM, \citep{co99,dhk+99} and the circumgalactic medium, or CGM \citep[e.g.][]{bergeron86, lbt+95}.   In cosmological hydrodynamical simulations, for instance, baryons are apportioned comparably between the Ly$\alpha$ forest (40\%), the CGM (25\%) and the WHIM (25\%, excluding the gas that is also CGM; Dav\'e et al. 2010\nocite{dave+10}). 
 
The present work concerns the halo missing baryon problem, which we
briefly summarize here.  Generally speaking, the condensed baryonic component of
galaxies, which dominates the energy output of the system, is
predicted to dynamically trace the underlying dark matter halo. 
Traditionally, baryon counting in this
regime has focused on a galaxy's stars,  cold ISM, and its
hot X-ray halo gas \citep{bell03b, klypin, baldry08, yang09,
  mcgaugh10, andersonbregman10, papastergis12, gupta12}. Compared to the cosmological
$\Omega_b/\Omega_m$ ratio, galaxies and their halos come up significantly short on baryons. For a Milky-Way luminosity galaxy, the
various estimates of the ratio in stellar mass to the dark matter mass
within the virial radius range from $M_*/M_{\rm DM} \approx 0.02-0.05$
\citep{behroozi10}; when we add the cold, neutral component from HI
surveys \citep{martin10}, this fraction
increases to only 0.07. Finally, when we add in the detected 
X-ray halo gas, the fraction is at most 0.08 \citep[but
see][]{gupta12, fang13}. Such a
deficiency is often expressed in terms of   $(M_{\rm stars,gas}/M_{\rm
  DM})/(\Omega_b/\Omega_m)$. In this representation, galaxy halos appear to
be missing approximately 60\% of their baryons, suggesting that they are structures nearly
devoid of baryons both in mass and spatial extent.

Models of the formation of galaxies like our Milky Way have long
predicted that the central galaxy contains only a modest fraction of
the available baryons (Klypin et al. 2011, and references therein\nocite{klypin}).  Galaxies are inefficient producers that have converted a small
portion of their available gas into stars.  In turn, theorists have
suggested a suite of physical processes to suppress star formation
and/or expel gas from the galaxies \citep{ds86, sprimack99,  od+10}.  
While evidence for outflowing gas from galaxies is common
\citep[e.g.][]{wcp+09,rubin13}, its impact on the efficiency of
galaxy formation is unclear.  Furthermore,
feedback processes are also required to explain the observed incidence 
of metal-line absorption along quasar sightlines
\citep[e.g.][]{dodorico91,od+11,booth} and to enrich the CGM of modern
galaxies \citep[e.g.][]{chen10,prochaska11, tumlinson11}.

Over the past twenty years it has become increasingly apparent that
galaxies also exhibit a diffuse baryonic component within the dark
matter halo that extends far from the inner regions to the virial
radius and beyond \citep{mwd+93, lbt+95, tripp+98,wakker09,prochaska11}.  This halo gas or CGM is
similar in concept to the intracluster medium revealed in X-ray
emission, but the CGM is observed via UV absorption lines and has much lower temperature and 
density \citep{werk13}.  As such, much of the CGM cannot be traced with X-ray
imaging nor any other radiative emission process:  it is simply too diffuse to
permit direct detection with any present-day telescope.

Our collaboration, COS-Halos, has been working to characterize this elusive multiphase medium \citep{tumlinson11, thom12, tumlinson13, werk12, werk13}. We have designed  and executed a large program
with the {\emph{Cosmic Origins Spectrograph}} (COS; Froning \& Green 2009\nocite{froning09}, Green et al. 2012\nocite{green12}) on the {\it Hubble Space Telescope (HST)} that observed halo gas of 44  galaxies, drawn from the imaging dataset of the Sloan Digital Sky Survey (SDSS),  whose angular offsets from quasar sightlines and photometric redshifts implied impact parameters (R $<$ 160 kpc) well inside their virial radii.  These data comprise a carefully-selected, statistically-sampled map of the physical state and metallicity of the CGM for L $\approx$ L$^*$ galaxies.

Of particular relevance to the halo missing baryon problem is the total baryonic mass contained in the multiphase CGM, as traced by absorption from hydrogen and metal lines in various ionization states (e.g. MgII, SiII, CII, SiIII, CIII, SiIV, OVI).\footnote{The commonly used temperature-based nomenclature for the CGM gas phases is different from that of the ISM. The circumgalactic gas in the temperature range 10$^4$ K $\le$ T $<$ 10$^5$ K is typically referred to as cool; the gas in the temperature range  10$^5$ K $\le$ T $<$ 10$^7$ K is called warm-hot; and gas above 10$^7$ K is termed hot, and would be observed via X-ray transitions. Each of these gas phases is highly ionized.  } Previous studies have attempted to estimate the total mass contribution of the CGM to a typical L$^{*}$ galaxy with varying degrees of success. Using absorber samples from HST and FUSE \citep{pss04, tripp+05, dsr+06, tripp08,tc08a, ds08, cm09}, and ground based follow-up spectroscopy to determine redshifts of galaxies along the lines-of-sight,  Prochaska et al. (2011) report a strong H I-traced CGM out to 300 kpc for all galaxy types. They estimate a  baryonic mass of 10$^{10.5\pm0.3}$ M$_{\odot}$ for an assumed constant total hydrogen column, N$_{\rm H}$ =  10$^{19}$ cm$^{-2}$. \cite{tumlinson11} determine the minimum mass of the highly-ionized CGM ($T$ $\approx$ 10$^{5 - 5.5}$ K) as traced by OVI absorption to be $>$ 10$^{9}$ M$_{\odot}$, based on the maximum possible value for the ionization fraction of OVI (f$_{\rm OVI}$ $<$ 0.2; but the fraction may be higher and the corresponding mass lower in some non-equilibrium scenarios; Vasiliev et al. 2013\nocite{vasiliev13}) and assuming the CGM extends to only 160 kpc. 

Based on HI measurements and a simple halo model that uses a power-law gas density profile exposed to a uniform ionizing background, \cite{thom12} estimate the total mass of the CGM could range from 10$^{9}$ - 10$^{11}$ M$_{\odot}$. \cite{zhu13} and \cite{lan14} use statistical techniques to assess the absorption from CaII and MgII in galaxy halos, and find an order of magnitude more cool gas in the CGM than in the interstellar medium of galaxies, implying a larger total gas mass in the CGM than in the ISM.  Stocke et al. (2013) model the ionization state and metallicity of T $\sim$ 10$^{4}$K CGM clouds using absorption line data from COS and STIS. They statistically associate late-type galaxies from SDSS imaging with the COS/STIS absorbers using virial radii estimated from photometry. Based on their assumed galaxy/absorber associations, they estimate that the low-ion CGM can account for  between 10\% and 15\% of the total baryonic budget of luminous spiral galaxies.  This estimate is a lower limit because of saturated HI absorption lines. 

Here, we refine these mass calculations by  modeling the photoionized gas of the CGM using a carefully selected sample of  L $\approx$ L$^*$ galaxies with precise, accurate redshift measurements from Keck and Magellan spectroscopy \citep{werk12} whose 10$^{4}$K  CGM is probed by {\emph{HST}}/COS spectroscopy.  Our sample covers and detects a large suite of ions  \citep{werk13, tumlinson13}. We rigorously determine the ionization state and metallicities for 33 of the COS-Halos sightlines that provide the best-determined measurements of HI and metal-line column densities.   With the constraints imposed by the data and models, we are able to provide a conservative mass estimate for an L $\approx$ L$^*$ galaxy's CGM, and show that the CGM is a dominant reservoir of baryons on galactic scales. 

 Section 2 summarizes the sample and data used in our analysis; in Section 3, we discuss the results of the photoionization modeling and tabulate all derived ionization parameters, metallicities, and total hydrogen columns of the individual lines of sight; in Section 4 we present our analysis of these results, including a mass estimate of the photoionized diffuse gas in the circumgalactic medium of L$\approx$L$^*$ galaxies; Section 5 presents a discussion of this result in the context of previous mass estimates, cosmological simulations, and simple hydrostatic solutions. We present a summary and conclusions in Section 6. We additionally provide an Appendix that details the photoionization modeling, explores additional sources of ionization, and discusses the results on a sightline-by-sightline basis.  To maintain consistency with previous COS-Halos results, throughout this work we assume the 5-year WMAP cosmology with $\Omega_{\Lambda}$ = 0.74, $\Omega_{m}$ = 0.26, and H$_{0}$ = 72 km s$^{-1}$ Mpc$^{-1}$ \citep{wmap05}. Distances and galaxy virial radii are given in proper coordinates. We use atomic data for absorption lines from Morton (2003), and the solar relative abundances of metals from \cite{asplund09}.
 
 \section{Sample, Observations, and Measurements}
 \label{sec:sample}
 	
 	We draw our sample from the COS-Halos survey of the CGM gas surrounding  L $\approx$ L* galaxies in the low-redshift Universe.  Using the {\emph{Hubble Space Telescope}}/ COS, COS-Halos observed 39 UV-bright quasars within an impact parameter R $< 160$\,kpc of 44 galaxies with 0.1L$^* <$ L $<$ 3 L$^*$  at z$\sim$0.2.   The primary motivation and the details of the survey design and execution are discussed in detail in Tumlinson et al. (2013). Previous papers describing the COS-Halos sample and data include the Werk et al. (2012) compilation of galaxy spectroscopy, the Tumlinson et al. (2011) study of O VI bimodality in galaxy halos, the Thom et al. (2012) study of H I in early type galaxies, the Werk et al. (2013) empirical description of the CGM as seen in metal absorption lines, and the Peeples et al. (2014) metal census. In this work we use the same 44-galaxy sample described extensively in Werk et al. (2013) and Tumlinson et al. (2013). 
	
            For every sightline, COS observations yielded a continuous spectrum spanning $\lambda \approx 1150-1800$\AA.  The exposure times were chosen to achieve a signal-to-noise (S/N) of $7-15$ per resolution element (FWHM~$\approx 15 \kms$) at $\lambda \approx 1300$\AA.   Keck/HIRES echelle spectra supplement the far-UV spectra from {\emph {HST}}/COS for all of the sightlines included in this work.   For galaxies at $z>0.1$, these data provide coverage  of the \ion{Mg}{2}~$\lambda\lambda$2796,2803 doublet,  an excellent diagnostic of cool ($T \le 10^4$\,K), metal-enriched gas. Both the COS and HIRES data have been described in the previous works mentioned above, and we do not repeat the details here. We use the column densities of the metal ion lines presented by Werk et al. (2013) which are derived using the apparent optical depth method \citep{savage91}. Non-detections are given as  2$\sigma$ upper limits. We use the HI column densities presented by Tumlinson et al. (2013), also based on the AODM calculations, except for damped systems which are based on Voigt profile fits. Of the 44 galaxies considered here, 40  (91\%) show HI absorption in the CGM out to 160 kpc with column densities of  15.0 cm$^{-2}$ $< $ Log N$_{\rm HI}$ $<$ 20.0 cm$^{-2}$. These values are presented in Table 1, along with  additional constraints on the upper and lower limits, as described more thoroughly in the Appendix. 

\begin{figure}[h!]
\begin{centering}
\hspace{-0.3in}
\includegraphics[height=1.1\linewidth,angle=90]{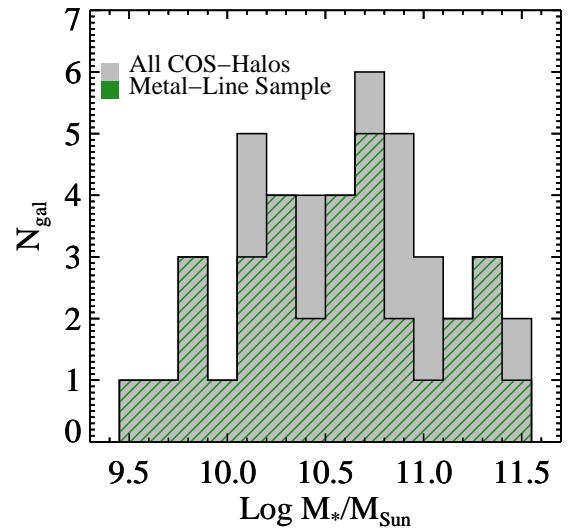}
\end{centering}
\caption{ A histogram of stellar masses for the full COS-Halos sample of 44 quasar/galaxy pairs (grey) and the 33 quasar/galaxy pairs that comprise our CLOUDY subsample (green lines), for which we are able to obtain solutions for the ionization parameter of the gas from CLOUDY photoionization modeling of the low ionization state metal absorption lines. The metal-line sub-sample spans the full range of stellar masses probed by COS-Halos, with a median value of 10$^{10.5}$ M$_{\odot}$, also the stellar mass of the Milky Way. 
}
\label{fig:masshist}
\end{figure}

  	Low and intermediate ionization state metal absorption lines (singly and doubly ionized species) are a common feature of the CGM for $L^*$ galaxies of all spectral types. Of the 44 sightlines in Werk et al. (2013) sample, 33 (75\%) show absorption from low/intermediate ionization state material that allow us to model the ionization state of the intervening gas. It is these 33 sightlines that we now analyze in greater detail to constrain the physical conditions of circumgalactic gas, and to provide a reliable baryonic mass estimate for the cool CGM of z$\sim$0 L$^*$ galaxies. Figure \ref{fig:masshist} shows the distribution of host galaxy stellar masses for the full COS-Halos sample, and for the sub-sample of sightlines with data that allow for an estimate of the gas ionization parameter. A  KS test indicates no statistically significant difference between the full COS-Halos sample and the CLOUDY-modeled sub sample in terms of stellar mass, or any other galaxy property.  
	
	Throughout this work we scale the projected distance from the sightline to the center of the host galaxy (impact parameter, R) to the virial radius of the galaxy, approximated here as R$_{\rm 200}$, the radius at which the halo mass density is 200 times the critical matter density of the universe.  At a given galaxy stellar mass determined by {\emph{kcorrect}} \citep{kcorrect} from the SDSS {\emph{ugriz}} photometry, we interpolate along the abundance matching relation of \cite{moster10} to find the halo mass (M$_{\rm halo}$). We then compute a virial radius with the relation: 
		\begin{equation}
			\rm R_{\rm 200}^{    3} = 3 \rm M_{\rm halo}/4\pi\Delta_{\rm vir}\rho_{\rm matter}
		\end{equation}
	where $\rho_{\rm matter}$ is the critical density at the spectroscopically determined galaxy redshift times $\Omega_{\rm m}$, and $\Delta_{\rm vir}$ = 200.  At the typical redshifts of the COS-Halos galaxies (z $\sim$ 0.2),  R$_{\rm 200}$ is slightly larger than the virial radius by a factor of $\sim$1.2.  Systematic errors in the stellar mass estimates and the scatter and uncertainty in the M$_{\rm halo}$ - M$_{*}$ relation gives an uncertainty in R$_{\rm 200}$ of approximately 50\%. Throughout this work we refer to this quantity R$_{\rm 200}$ as R$_{\rm vir}$, as is commonly done.

\section{Ionization State and Metallicity of the L$^{*}$  CGM} 
\label{sec:results}

	We model the CGM of 33 COS-Halos galaxies exhibiting absorption from low and intermediate ionization state metals (which we abbreviate here as low-ions and intermediate-ions, respectively) using version 13.03 of  CLOUDY \citep{ferlandetal98, ferland13}  with the Haardt \& Madau (2001) background radiation field from quasars and galaxies (HM 2001) as our ionization source for gas.  Specifically, we model the individual ionization states of multiple elements for each galaxy halo. For each absorber, we vary the ionization parameter (log U) and the metallicity ([X/H]; elements are assumed to have solar abundance ratios) to search for CLOUDY models that are consistent with the column densities of the HI and low and intermediate metal ions determined from the observations.  We thoroughly discuss the details of each sightline analysis, including systematic uncertainty, in the Appendix. The largest source of error arises from uncertainty in the extragalactic ultra-violet background (EUVB) and other potential sources of ionization (e.g. the galaxy itself; the cosmic ray background). We explore the effects of implementing different slopes and intensities of the background radiation field, including a comparison between \cite{hm01} and \cite{hm12}, and find the derived ionization parameters and metallicities remain consistent within a few tenths of a dex for a range of physically plausible radiation fields. The uncertainty in the slope of the EUVB therefore lends an additional systematic error of $\pm$0.3 dex to our derived gas parameters. In the Appendix, we additionally explore the effect of adding ionization from a central star-forming galaxy at various impact parameters from the line of sight. 
	
	Four key assumptions underlie our methodology: \begin{enumerate} \item { The low and intermediate ions  observed in the COS spectra (e.g. SiII, SiIII; CII, CIII) arise from a single gas phase with the same origin (i.e. are co-spatial).}  \item{ The CGM probed by this absorption is a cool medium, i.e. T $<$ 10$^{5}$ K, in which photoionization dominates.} \item{ The absorption from the low and intermediate ions  trace the majority of observed H I gas.} \item {The gas is in ionization equilibrium. }\end{enumerate} None of these assumptions is radical -- all four are now standard practice in analyses of quasar absorption-line spectra probing both the CGM and IGM (e.g. Prochaska et al. 2004\nocite{lehner13, prochaska04}). With respect to the first assumption,  \cite{werk13} has shown that  the kinematic component structure of the low-ion and intermediate-ion absorption profiles is similar. Generally, the intermediate ion absorption tends to be stronger than the low ion absorption in the CGM of L$^{*}$ galaxies, but there is no evidence that the low and intermediate ionic absorption arise from different gas phases based on their component structure alone. Second, our observations of low-ion transitions in the majority of the spectra considered here demands that the gas be cool (T $<$ 10$^{5}$ K).  \cite{tumlinson13} have further demonstrated this temperature constraint based on the HI line widths.  Collisional ionization of gas at T $ >$ 10$^{4.6}$K would yield negligible quantities of ions like MgII, CII, and SiII, even if at solar abundance and with a large total gas column (e.g. Gnat \& Sternberg 2007). Furthermore, the 9 galaxies that exhibit only intermediate ions (CIII, SiIII) are constrained to have T $<$ 2 $\times$ 10$^{5}$K to avoid extremely large gas surface densities. Put another way, this assumption of a cool medium is conservative with regard to a total mass estimate. Finally, these low and intermediate-ion metals trace the majority of observed HI gas, as evidenced by both their line-profiles and the observed trend between N$_{\rm HI}$ and metal column density  shown by \cite{werk13}. 

	To reduce systematics arising from different ionization processes, we have not attempted to model the OVI detected in these systems (Tumlinson et al. 2011). OVI is potentially produced by multiple ionization processes (i.e. photoionization and collisional ionization), and modeling the amount of gas from these two physically distinct origins is complex and beyond the scope of this paper. Recent simulations by Ford et al. (2013) predict that while absorption from low-ions in the CGM of  L $\approx$ L* galaxies arises from denser gas closer to galaxies, higher ions like OVI  trace hotter, more diffuse gas extending to larger radial distances. Similarly,  \cite{stinson12} find that their MaGICC halos display two distinct phases: a cool, T$\approx$10$^{4}$ K dense gas that follows the HI absorption, and a warm-hot T $>$10$^{5}$ K low density medium that creates the OVI absorption primarily through collisional ionization.  Observationally, \cite{fox13} have shown that OVI absorption exhibits a significantly higher velocity spread than CII, and conclude that low-ions and high-ions trace distinct gas phases in Lyman limit systems.  Even when the low-ions and high-ions are kinematically very similar, several recent studies have shown that the observed column densities of low and high ionization species cannot be reproduced with a single gas phase (e.g. Tripp et al. 2011; Meiring et al. 2013; Lehner et al. 2013\nocite{tripp+11, meiring13, lehner13}). 
	
	Thus, our mass estimate explicitly excludes gas in a highly-ionized OVI phase which may itself comprise a large component of the CGM \citep{tumlinson11}. As shown in the Appendix, our best models for each system tend to systematically  underestimate the column density of the OVI absorption (see Figures \ref{fig:cloudyfirst}$-$ \ref{fig:cloudylast}), which reinforces the conclusions reached in the studies above with a larger and more statistically uniform sample. That is, when we detect OVI in our COS observations, its typical ionic column density requires an ionization parameter between 1 and 2 dex higher than the solution for the lower ionization state lines.  Typically, OVI requires log U $>$ -1.5 if it is primarily photoionized. One caveat is that it may be possible to produce more OVI by considering ionizing photons from a nearby, central starburst galaxy, including X-ray binaries, as has been recently shown by \cite{vasiliev13}. The OVI fraction can reach 60\% (compared to 20\% from photoionization) over a wide temperature range when superimposing on HM 2001 a time-dependent radiation field of a nearby starburst galaxy with soft X-rays. We note that in this case, the ionization parameters required to simultaneously reproduce the low-ions are much higher than those we derive using a HM 2001 spectrum alone, which further serves to make our mass estimate of the $\sim$10$^{4}$ K CGM a lower limit. Finally, we also do not consider the time-dependent, non-equilibrium ionization models of Oppenheimer \& Schaye (2013) \nocite{os13} for OVI, where an AGN in the recent past photoionized metals with a significantly stronger field than that of HM 2001. They predict the amount of OVI can increase by at least 1 dex, if for example a COS-Halos galaxy was a Seyfert within the last 10 Myrs and has since turned off. For reference, none of the COS-Halos galaxy sample are defined as AGN according to the BPT diagram, and we expect this non-equilibrium fluctuating scenario is not relevant for these galaxies. 
	
	We consider SiIV to the extent that it can provide an upper bound on the ionization parameter of the gas\footnote{And, in one case, CIV, though for the majority of systems we do not have spectral coverage of the CIV lines at 1548 and 1550 \AA.}. For example, the photoionization models predict a column density of  SiIV depending on log U, and we require that the SiIV data must lie at or above this level. Typically, we can explain  the majority of the SiIV absorption with the photoionization modeling, and our models match the SiIV column densities (or limits) well. 

	In the 19 (of 33) cases for which the Lyman series absorption lines exhibit saturated profiles, we assume the lowest possible value of the HI column density. In 7 of these 19 cases,  we are forced to assume N$_{\rm HI}$ slightly higher than the AODM lower limit in order to determine a self-consistent solution with CLOUDY. We discuss these details in the Appendix. The best-fitting value of [X/H] is highly dependent on the assumed log N$_{\rm HI}$, and thus in cases hampered by HI lower limits we cannot reliably determine the gas metallicity. For this reason, we generally find upper limits to the metallicity. The ionization parameter, as discussed in the Appendix, is largely immune to  uncertainty in the HI column density.  This is because we can independently constrain log U from the CLOUDY models based on several different ionization states of the same metal line, and the detection of a number of different metal lines of various ionization states (e.g. CII, CIII, SiII, SiIII, SiIV, NII, NIII, MgII). Our choice to adopt the minimum N$_{\rm HI}$ consistent with the data leads to a conservatively low total N$_{\rm H}$ and minimizes the implied mass in some cases by as much as an order of magnitude. 
	
	
 	For each of the 33 galaxy absorbers included in this analysis, we detail the specific ions used in the solution for log U and [X/H] and the overall consistency, accuracy, and precision of this solution in the Appendix. In all cases, we provide a range of log U, gas volume density,  and [X/H]  allowed by the data, along with the corresponding allowed range of the total hydrogen column (photoionized + neutral)  along each line of sight. Galaxy properties are given in conjunction with constraints on the metallicity, total gas column density, and ionization state of their CGM in Table 1. 
		
	We have assessed the quality of the solution by visual inspection, parameterized by the quantity ``Q flag", which ranges from 1 to 5 (null solution to  well-constrained). This Q flag is based on a combination of an overall data-quality assessment, number and quality of detections of low and intermediate ions, the constraints on the HI column density, and the overall consistency of the solution for the multiple ions and metals. Sightlines that show no absorption from any metal ion are immediately assigned a Q flag of 1 (11 of 44 galaxies). All of these excluded galaxies lie in the optically thin regime.   We explore the extent to which excluding them may bias our results for the mass determination of the halos of L$\sim$ L* galaxies in Section \ref{sec:nondets}, and conclude it could have at most a modest impact.  A Q flag of 2 indicates that both log U and N$_{\rm HI}$ are poorly constrained by the data, owing to limited detections of metal ions (or many blends) and saturated HI (3  galaxies).  Generally, a Q flag of 3 signifies that one of log U or N$_{\rm HI}$ is moderately-constrained by the data and modeling, resulting in a range of  allowed total hydrogen column that spans approximately one dex (8 galaxies).  A typical galaxy that is rated with a Q flag of 4 shows saturated HI with a lower limit $>$ 10$^{16}$ cm$^{-2}$, and good-quality detections from various low and intermediate ions that constrain the solution from log U to better than a  dex (12 galaxies). Finally, we reserve our highest Q flag of 5 for solutions that are well-bounded  in metallicity and log U (on average, $\pm0.2$ dex) and consistent with a suite of column density measurements of several low and intermediate-ions (9 galaxies). Absorbers with Q flags greater than or equal to 2 are included in the analysis and the systematic uncertainties  are discussed in detail in the Appendix. 

\subsection{Ionization Trends with Gas Column Density and Impact Parameter}

\label{sec:ion}

\begin{figure}[h!]
\begin{centering}
\hspace{-0.3in}
\includegraphics[height=1.1\linewidth,angle=90]{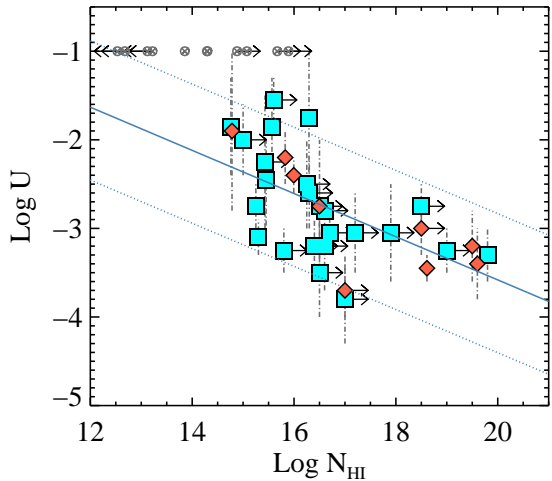}
\end{centering}
\caption{ The derived ionization parameter from the CLOUDY modeling versus HI column density for 44 COS-Halos absorbers within 160 kpc of an L$\sim$L$^{*}$ galaxy. The absorbers with star-forming host galaxies are shown as blue squares, while absorbers having host galaxies without any detectable ongoing star formation are shown as red diamonds. The 11 sightlines on which we are able to place no constraints owing to non-detections of all metal ions are shown as grey circled x's at an arbitrary Log U of -1.  Upper and lower limits on the measured HI column densities from the COS data are shown, respectively, by left-facing and right-facing arrows. The range of allowed log U values for each absorber are shown by vertical dash-dotted gray lines. We have fit a line in log-log space to these data points (shown in blue) using linear regression analysis for 1000 trials with the log U values randomly distributed along the range of allowed log U values. We reject the null hypothesis with a probability of 99.994\%, and therefore find a  correlation between Log U and Log N$_{\rm HI}$ with 4$\sigma$ significance. 
}
\label{fig:unhi}
\end{figure}

The dimensionless ionization parameter, U, is defined as the ionizing photon density divided by the total hydrogen number density (neutral + ionized). For our purposes we explore models with log U ranging between $-$1 and $-$5, with higher values corresponding to a higher ionized gas fraction and a lower gas density. We derive a mean log U of $-2.8$ for our sample of 33 absorbers, which corresponds to approximately 99\% of the hydrogen being ionized, on average. Our adopted values of log U range from $-3.8$ to $-1.6$, corresponding to a range in the neutral gas fraction between approximately  25\% and 0.01\%. Figure \ref{fig:unhi} shows log U versus  the HI column density for 44 COS-Halos absorbers within 160 kpc of an L$\sim$L$^{*}$ galaxy, with the 11 galaxies excluded from our analysis shown for reference as grey x's at an arbitrarily chosen value of log U = $-1$.  Upper and lower limits on the measured HI column densities from the COS data are shown, respectively, by left-facing and right-facing arrows. The colored data points (blue = star-forming; red = non-star-forming) for the 33 lines of sight with metal line data show a clear trend of decreasing ionization parameter with increasing HI column density.   \cite{lehner13} carried out a similar CLOUDY-based analysis for Lyman limit systems with log N$_{\rm HI}$ $>$ 16.0, and found a mean log U of -3.3$\pm$0.6. When we consider our absorption line systems at similar HI column densities, we also find a mean log U of -3.3. 

To assess the strength of the correlation between log U and HI column density, we perform a statistical analysis using the ASURV software package Rev 1.2 (LaValley, Isobe \& Feigelson 1992), which implements the methods presented in Feigelson \& Nelson (1985) and Isobe, Feigelson \& Nelson (1986).  We include censored data points (i.e. lower limits) in the HI column density. We implicitly assume that the limits are random with respect to the galaxies. Given the range of impact parameters sampled and that the quasars were selected without any knowledge of the absorption, this assumption should hold. We exclude the 11 galaxies with no HI and/or metal-line detections from this analysis, since those data points exhibit censoring in HI column density and log U, which would make any fit unconstrained. We reject the null hypothesis with a probability of 99.994\% by performing a Kendall Tau test of censored bivariate data in this parameter space. Thus,  log U and log N$_{\rm HI}$ are very likely anti-correlated. To derive the best linear fit to the observed correlation, we perform a  linear regression analysis for 1000 trials with the log U values randomly distributed along the range of allowed log U values for each absorber (shown by the gray lines in Figure \ref{fig:unhi}). The best-fit, which includes HI  column density lower limits, is shown as  a solid blue line, and is given by: \begin{equation}
\rm log~U = (-0.24\pm0.06)~ \rm log~N_{\rm HI} + (1.3\pm0.5)
\end{equation}The dotted lines show  1$\sigma$ confidence intervals of the fit, and the standard error of the regression is 0.38. 

 Interestingly, this anti-correlation between  log U and log N$_{\rm HI}$ is qualitatively consistent with predictions for photoionized clouds in (or not far from) local hydrostatic equilibrium. \cite{schaye01_lya} argues that over-dense absorbers, such as those we may observe in the CGM, have sizes of the order the local Jeans length, regardless of whether the cloud as a whole is in hydrostatic equilibrium. A relation between neutral hydrogen column density and the characteristic volume density naturally arises from this requirement (Equation 8 of Schaye 2001) such that as N$_{\rm HI}$ increases, n$_{\rm H}$ also increases (equivalently, log U decreases; see also Prochaska et al. 2004). We explore several hydrostatic solutions in Section 5.3.

\begin{figure}[h!]
\begin{centering}
\hspace{-0.3in}
\includegraphics[height=1.1\linewidth,angle=90]{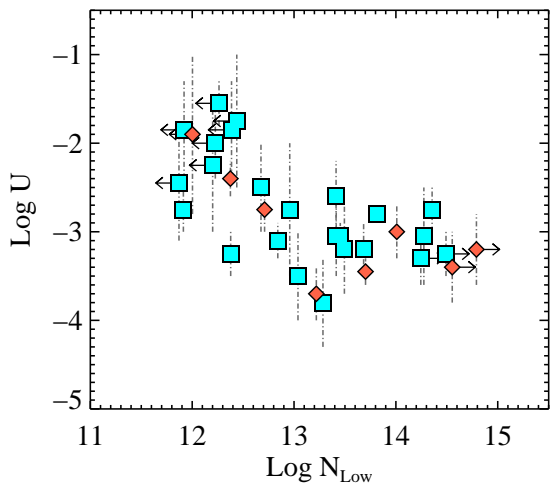}
\end{centering}
\caption{ The derived ionization parameter from the CLOUDY modeling versus low-ion column density for 33 COS-Halos absorbers, where \nlow\ is measured by either MgII or SiII, as described in Section 3.1. 
}
\label{fig:unlow}
\end{figure}

To compare log U to the low-ionization-state metal-line absorption, we derive the quantity \nlow~for each system, 
defined as follows \citep{werk13}:
(1) \nlow= N$_{\rm SiII}$, if the SiII measurement is a value or lower limit;
(2) \nlow= N$_{\rm MgII}$ if the SiII measurement is an upper limit (or there is none recorded) and a MgII measurement exists.
We choose SiII as the primary low ion because it has multiple
transitions in the far-UV bandpass with a range of oscillator
strengths yielding more reliable column density estimates.  The
\ion{Mg}{2} doublet, meanwhile, offers more sensitive upper limits.
Because we measure  N$_{\rm SiII}$ $\approx$ N$_{\rm MgII}$  in cases where
both are measured, we apply no offset when adopting one versus the
other.   In all, there are roughly half of the systems in each category. Figure \ref{fig:unlow} shows log U versus \nlow~for our sample, and bears a high degree of similarity to Figure \ref{fig:unhi} such that higher values of \nlow\ exhibit lower values of log U.  Werk et al. (2013) have shown that \nlow~is significantly coupled to \ion{H}{1} column density, with  low-ion detections essentially
requiring a non-negligible opacity at the \ion{H}{1} Lyman limit. Together, Figures \ref{fig:unhi} and \ref{fig:unlow} show that this observed trend follows from the ionization parameter of the
gas being higher at lower low-ion and HI column densities. We explore the physical significance of this correlation in Section \ref{sec:disc}.

\begin{figure}[h!]
\begin{centering}
\hspace{-0.3in}
\includegraphics[height=1.1\linewidth,angle=90]{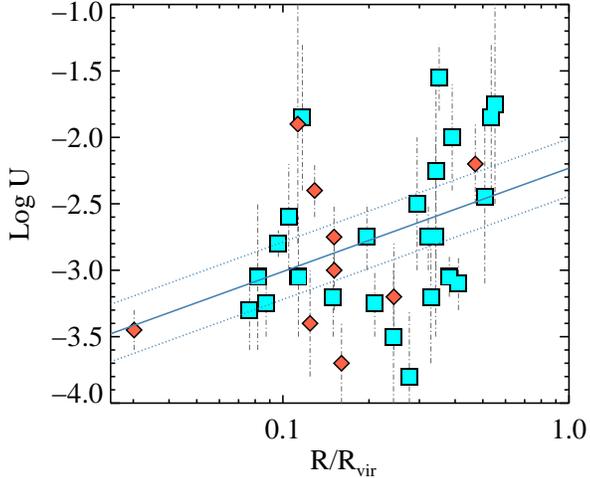}
\end{centering}
\caption{ Ionization parameter as log U  versus impact parameter in units of R/R$_{\rm vir}$ for 33 COS-Halos absorbers within 160 kpc of an L$\sim$L$^{*}$ galaxy.  The absorbers with star-forming host galaxies are shown as blue squares, while absorbers having host galaxies without any detectable ongoing star formation are shown as red diamonds. The ranges of allowed log U are shown by the vertical dash-dotted gray lines. We determine the mean log U by populating a random distribution of allowed log U 1000 times. The mean log U of the CGM of L$\sim$L$^{*}$ galaxies to 160 kpc is -2.8. We reject the null hypothesis at 97\%, indicating that log U and impact parameter are correlated. The best-fitting power law slope is U $\propto$ (R/R$_{\rm vir}$)$^{0.8\pm0.3}$. 
}
\label{fig:uvrho}
\end{figure}

 Figure  \ref{fig:uvrho} shows Log U vs.  R/R$_{\rm vir}$ for the same sample. Despite the large scatter,  Figure \ref{fig:uvrho} shows that ionization parameter and impact parameter are positively correlated. A Kendall-Tau test of 1000 trials with the log U values randomly distributed along the range of allowed values  rejects the null hypothesis with a probability of  97\%, indicating a positive correlation is present. Based on a linear regression analysis (as implemented in ASURV), we find that the best power-law fit to this correlation is: 
 \begin{equation}
  U = 0.006\pm0.003 (R/R_{\rm vir})^{0.8\pm0.3} \end{equation} with  a combined standard deviation of $\sim$0.5 (shown in blue, with 1$\sigma$ confidence intervals shown). Thus,  the CGM is more highly ionized farther from its host galaxy. We discuss the extent to which this trend is the result of a declining gas density gradient, in Section \ref{sec:density}.

\subsection{Limited Constraints on the Metallicity of the L$^{*}$ CGM}

As described throughout this work, and in detail in the Appendix, our constraints on the gas metallicity are generally poor owing to the line saturation of HI Ly$\alpha$ and other Lyman series lines in our COS data. Nonetheless, for 11 COS-Halos absorbers with well-constrained HI column densities, we are able to estimate the gas metallicity to better than $\pm$0.2 dex. Our values of [X/H] range from -1.5 to solar, and are determined for galaxies with either log N$_{\rm HI}$ $<$ 16.5 cm$^{-2}$ where Lyman series lines do not saturate in our COS data or log N$_{\rm HI}$ $>$ 18.5 where the presence of damping wings on Lyman series absorption lines allows for an estimate of the HI column density from Voigt profile fits \citep{tumlinson13}. The 11 systems that do not show absorption from metal ions do not offer any useful upper limits on the gas metallicity. We discuss the limited constraints implied by metal-line non-detections more fully in the next section.

\begin{figure}[h!]
\begin{centering}
\hspace{-0.3in}
\includegraphics[height=1.1\linewidth,angle=90]{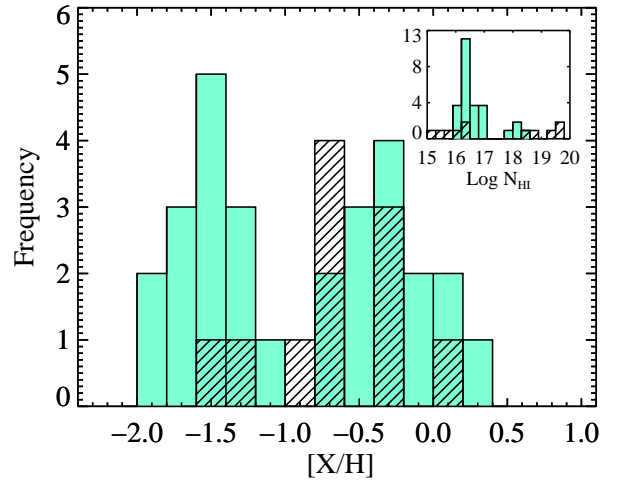}
\end{centering}
\caption{ A histogram of CLOUDY-derived metallicities for the Lyman Limit systems analyzed by Lehner et al. (2013; light green) and for the 11 COS-Halos absorbers that provide a constraint on the metallicity to better than 0.2 dex (diagonal black lines).  An inset in the upper right corner of this plot shows the HI column densities of the Lehner sample (light green) and the sample of COS-Halos absorbers included here (black diagonal). Most of this COS-Halos 11-absorber subsample extends to lower or higher HI column than the Lehner HI column densities and would not have been included in their analysis. While Lehner et al. find evidence of a bimodal distribution of CGM gas metallicity, we do not see evidence for bimodality here though we are limited by a small sample. 
}
\label{fig:goodmetals}
\end{figure}

 Figure \ref{fig:goodmetals} compares our gas metallicity for the CGM of L$^{*}$ galaxies within 160 kpc to the values of \cite{lehner13} who analyzed 28 Lyman Limit systems and found a bimodal metallicity distribution. We note that the Lehner et al. study examines a range of HI column density where constraints from COS-Halos are the weakest (16.2 $<$ log N$_{\rm HI}$ $<$ 18.5). A key difference between this study and that of \cite{lehner13} is that the COS-Halos target selection is based on galaxy properties while the Lehner et al. (2013) target selection is based on HI column density selection that allows them to sensitively probe both high and low metallicities.  The range of gas metallicities we find is similar to that of the Lyman Limit systems examined by \cite{lehner13}, but we do not have enough data to distinguish any bimodality in metallicity for COS-Halos galaxies. Moreover, we have not analyzed 11 cases in which no metals are apparent (Q = 1), and it is possible some of those cases would occupy the low metallicity branch.

With respect to a metallicity gradient with R,  we find no apparent trend between metallicity and  impact parameter for the 33 COS-Halos absorbers included in this analysis, albeit very large uncertainties and upper limits. We have performed a survival analysis in this parameter space using ASURV to test for any hint of a correlation. A Kendall-Tau test which includes censoring (upper limits) and 1000 trials with the [X/H] values lying along a randomly populated distribution defined by the range of allowed values reveals a z-value of 0.689 and  rejects the null hypothesis with a probability of only  41\%.  Because the gas metallicity is poorly constrained by our data, we see no evidence of a metallicity gradient with impact parameter.  The best fitting power-law slope is [X/H] $\propto$ (R/R$_{\rm vir}$)$^{0.2\pm0.4}$.  We explore whether this finding is consistent with theoretical predictions from a primarily wind-fed CGM and with naive expectations based on several possible origins of the CGM in Section \ref{sec:disc}.


\subsection{Constraints from Non-detections of Metal Ions}
\label{sec:nondets}

 For 7 of the 44 sightlines considered,  we have not detected absorption lines other than HI Ly$\alpha$ or Ly$\beta$. Four additional galaxy-halo sightlines are devoid of absorption from any ion, including HI Ly$\alpha$. In this section, we address whether these `undetected' systems are more likely to exhibit low metallicity and/or low N$_{\rm H}$ compared to the sample of 33 absorbers for which we can model the gas using CLOUDY.  To do this, we examine the typical upper limits to the column densities of several common metal ions in the parameter space of metallicity and N$_{\rm H}$.  We show that the 7 sightlines with HI absorption but no metal absorption can exhibit the full range of  physical conditions (ionization parameters, metallicities) for the gas that shows detected metal-ion absorption. 

\begin{figure}[h!]
\begin{centering}
\hspace{-0.3in}
\includegraphics[height=1.1\linewidth,angle=90]{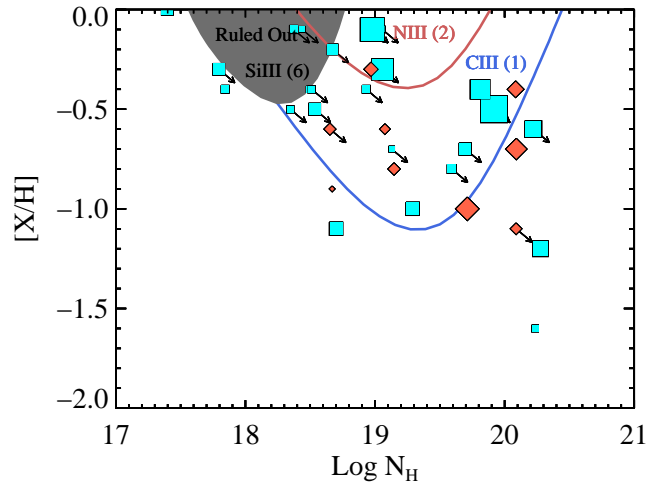}
\end{centering}

\caption{  Metallicity versus total hydrogen column density for the detected sightlines (blue squares = star-forming galaxies; red diamonds = non star-forming galaxies). The size of each point is inversely proportional to the uncertainty in each quantity. The filled grey region highlights the  region of parameter space that is ruled out for the 6 non-detections in SiIII. Additional constraints placed on total hydrogen column and gas metallicity by the 2$\sigma$ upper limits of other undetected intermediate ions (NIII, red and CIII, blue) are based on CLOUDY grid lines at log N$_{\rm HI}$ = 15.  The number of undetected systems  on which the 2$\sigma$ upper limit is based is given next to the ion name in each region. While CIII is the most constraining of the ions in assessing the state of the gas for this undetected sample, its upper limit is based on only one absorber with HI for which we did not detect CIII. The other 6 absorbers in this sample do not cover the line at 977 \AA. Four additional systems show no metal ion absorption and no detection of HI, which we cannot include on this figure. 
 }
\label{fig:nondets}
\end{figure}

In Figure \ref{fig:nondets}, we plot CLOUDY-derived gas metallicities as a function of total hydrogen column for the 33 systems showing HI and metal lines. The sizes of the data points are inversely proportional to the errors in their derived quantities, such that larger data points have well-constrained ionization parameters, HI column densities, and/or metallicity. Upper limits to metallicity and lower limits to the total hydrogen column density are shown as diagonal arrows to the bottom right, as the two quantities are degenerate. The filled grey region of this plot showcases the region of this parameter space that is {\emph{ruled out}} by the typical 2$\sigma$ upper limits  to the SiIII column density. These six limits have all been scaled to log N$_{\rm HI}$ $=$ 15.0 to show in this parameter space. SiIII is  the most commonly detected metal ion with the best coverage in the COS-Halos dataset, and in 6 of the 7 non-detections we can place reliable upper limits on its column density. By comparison, non-detections of the lower ionization state metal ions (e.g. MgII, SiII, and CII) are completely consistent with the full range of parameter space shown. Other intermediate ion non-detections, such as NIII and CIII, are only informative in one and two cases of non-detections. We show the constraints of those upper limits as solid colored lines of red and blue for CIII and NII, respectively. The regions not allowed by these two limiting cases would lie above the plotted curves. Next to the ion name we give the number of 2$\sigma$ upper limits (out of 7 possible) that were averaged.  

Figure \ref{fig:nondets} shows that the allowed metallicities and N$_{\rm H}$ values (i.e. unshaded area) for the `undetected' systems are largely consistent with the ranges exhibited by the 33 data points from the `detected' systems. Thus, there is no reason to assign them unusual N$_{\rm H}$.  In general, the 11 systems excluded from this analysis have low HI column densities, log N$_{\rm HI}$ $<$ 15.0. Based on the relation derived in Section \ref{sec:ion}, we might expect the optically thin gas along these sightlines to have a log U $\approx$ $-2$, and thus a total hydrogen column density of log N$_{\rm H}$ $\approx$ 10$^{19}$ cm$^{-2}$. At this  value of  N$_{\rm H}$, the upper limits to the column densities of intermediate ionization state lines for 6/7 of the `undetected' systems are consistent with the full range of metallicity considered, from 0.01 solar to solar. As shown, there is one case where the upper limit on CIII  can rule out a gas total hydrogen column and metallicity that most commonly describes our sample ([X/H] = -0.5; log N$_{\rm H}$ = 19.6 -- area above blue contour).  In other words,  this sightline requires the assumption of  a  low gas metallicity ([X/H] $<$ -1.0) if the total hydrogen column of the gas is near the median of the detected sample (log N$_{\rm H}$ $=$ 19.6 cm$^{-2}$).

Other than this one upper limit to N$_{\rm CIII}$, the metal ion non-detections can easily exhibit the same metallicity and total hydrogen column density as the gas we do detect. However, the inverse is also true. The 11 systems for which we detect no metal ion absorption can also exhibit lower gas metallicity and/or lower N$_{\rm H}$.  Essentially, we cannot draw any conclusions about the physical conditions of the gas (or lack thereof) for the 11 `undetected' systems. Hence, we do not have a compelling reason to believe we are biasing our results one way or the other by simply excluding them from our analysis.  
	
\section{Results}

The COS-Halos survey design and multi-wavelength, high-quality, spectroscopic dataset have allowed us to empirically characterize the CGM of L$^{*}$ galaxies with unprecedented accuracy \citep{werk13, thom12, tumlinson11,battisti12,  tumlinson13}. With the ionization modeling, we can now characterize the physical nature of the CGM without any additional model-based assumptions regarding, for example,  its origin or underlying density profile.  In this section, we describe the gas surface and volume density profiles implied by the ionization modeling and absorption-line data. We then use these density profiles to estimate the total amount of photoionized material in the CGM of L$^{*}$ galaxies. Given that 19 of 33 sightlines show saturated HI absorption lines, and the total hydrogen column scales roughly linearly with the neutral hydrogen column density, our derived CGM mass is a lower limit. However, this lower limit  is  more constraining than previous values as a result of our detailed CLOUDY analysis that establishes the ionization state of the gas. Previous estimates  using the COS-Halos sample (e.g. Tumlinson et al. 2011; Werk et al. 2013) have simply assumed the lowest possible ionized gas fraction in the presence of ionizing radiation from background quasars and galaxies at the characteristic temperatures of prominent ionization species. Other estimates using different samples have employed photoionization modeling (e.g. Lehner et al. 2013; Stocke et al. 2013), and require  different sets of assumptions. In the case of the \cite{lehner13} study, the sample was selected on HI absorption, and so they must assume a total volume. In the case of \cite{stocke13}, the sample is partially blind with respect to absorption as COS-Halos, and partially comprised of a `serendipitous' sample of absorbers for which the properties of the host galaxies are less homogeneous than the COS-Halos galaxies. 

\subsection{Surface and Volume Density Profiles}

\label{sec:density}

The total hydrogen column density, N$_{\rm H}$, is simply the sum of the neutral and ionized hydrogen column densities, where we determine the ionized hydrogen column density directly from the derived ionization parameter of the gas. Thus, N$_{\rm H}$ $=$ N$_{\rm HI}$ $+$ N$_{\rm HI}$/(1- $\chi$), where $\chi$ is the ionized gas fraction (mean $\approx$ 99\%). Table 1 lists both the range of allowed N$_{\rm H}$ and the adopted N$_{\rm H}$. The low value is based upon the HI column density from \cite{tumlinson13} (typically the AODM column density) and the lowest possible ionization parameter allowed by the photoionization modeling. The high N$_{\rm H}$ value is based on the preferred HI column density and the highest possible ionization parameter allowed by the photoionization modeling. This high value is still a lower limit because of HI saturation. In the Appendix, we discuss how we determine a `preferred' HI column density. In brief, it is equivalent to the measured AODM column density, with the exception of several lower limits which have been raised to make the gas metallicity consistent with being solar or below. The value for N$_{\rm H}$ we adopt is calculated using the preferred HI column density and the mean log U for the full range. 

Some values of N$_{\rm H}$ are determined to $\pm$0.2 dex, accounting for the systematic errors in the CLOUDY modeling, uncertainties in the derived column densities, and allowing for $\sim$0.1 dex uncertainty that arises if the elemental abundance ratios are non-solar. Uncertainty in the slope of the EUVB adds an additional $\pm$0.3 dex of systematic error. Shallower slopes than HM01 will yield higher ionization parameters, and steeper slopes result in smaller ionization parameters. We describe the details of the error accounting in the Appendix.  The mean uncertainty in  N$_{\rm H}$ is $\pm$0.5 dex. For the statistical tests and linear regression analysis described in this Section, we sample a random distribution   along the full `allowed' range of N$_{\rm H}$ 1000 times.  We `censor'  those sightlines for which HI is saturated, and thus include them as lower limits.

\begin{figure}[h!]
\begin{centering}
\hspace{-0.3in}
\includegraphics[height=1.1\linewidth,angle=90]{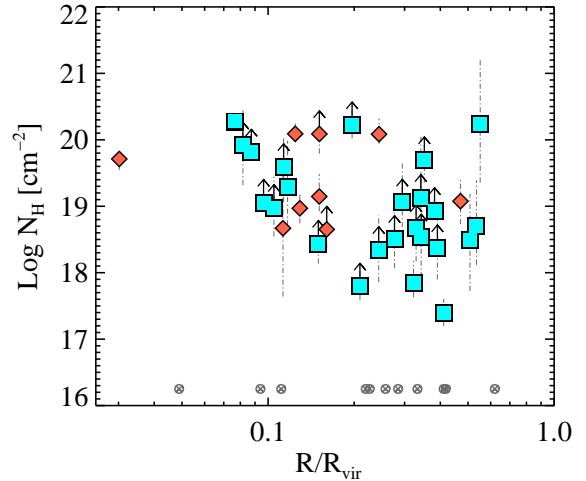}
\end{centering}
\caption{ The total hydrogen column  from the CLOUDY modeling versus impact parameter in units of R/R$_{\rm vir}$ for 44 COS-Halos absorbers within 160 kpc of an L$\sim$L$^{*}$ galaxy.  The absorbers with star-forming host galaxies are shown as blue squares, while absorbers having host galaxies without any detectable ongoing star formation are shown as red diamonds. The 11 sightlines on which we are unable to place constraints (owing to non-detections of all metal ions)  are shown as grey circled x's at an arbitrarily low log N$_{\rm H}$ of 17.2 for reference. We show the absorbers with lower limits to the  HI column density with up-facing arrows. The range of allowed total hydrogen columns at the adopted HI column densities are shown by the vertical dash-dotted gray lines. The mean log N$_{\rm H}$ of the CGM of L$\sim$L$^{*}$ galaxies calculated including censored data points is 19.6. 
}
\label{fig:nhvrho}
\end{figure}

\begin{figure}[h!]
\begin{centering}
\hspace{-0.3in}
\includegraphics[height=1.1\linewidth,angle=90]{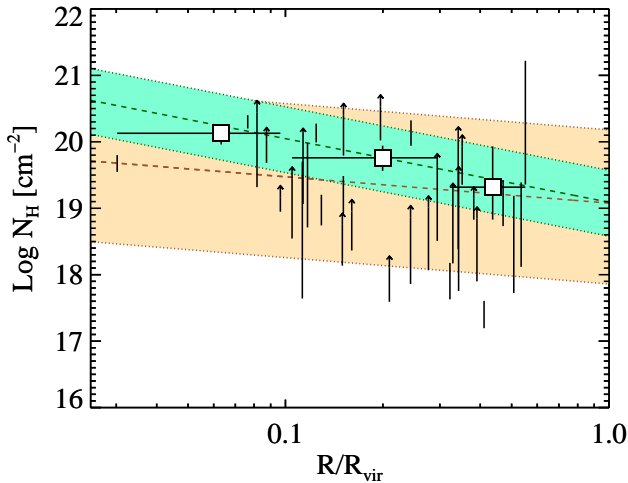}
\end{centering}
\caption{ The total hydrogen column  from the CLOUDY modeling versus impact parameter in units of R/R$_{\rm vir}$ for 33 COS-Halos absorbers within 160 kpc of an L$\sim$L$^{*}$ galaxy.  The range of  allowed log N$_{\rm H}$ values for each of the 33 absorbers is shown as a vertical black line. Absorbers with a lower limit on their HI column densities are indicated by up-facing arrows. We additionally divide the data in three bins of R/R$_{\rm vir}$, taking the mean log N$_{\rm H}$ in each bin. The error bars on log N$_{\rm H}$ come from the 1000 trials included in the statistical analysis. A Kendall-Tau test on these data rejects the null hypothesis with a confidence of  92\%.  The power-law fit from a linear regression analysis on the sample of 33 absorbers with metal line detections is  shown as a dotted green line, with the light green shaded area representing the 1$\sigma$ uncertainty in the fit parameters. One reason the binned points and the fit seem to lie above the distribution is that over half of the absorbers show saturated HI absorption lines. The lower limits have the impact of increasing the uncertainty of the best fit declining profile, decreasing the slope, and increasing the y-intercept. The beige shaded area on this figure shows the fit that results when the 11 non-detections are included, as described in the text. 
}
\label{fig:nhvrhobin}
\end{figure}

Figures \ref{fig:nhvrho} and  \ref{fig:nhvrhobin} show the total hydrogen column density as a function of impact parameter, and are effectively gas surface-density profiles. There is a 92\% chance that log N$_{\rm H}$ declines with impact parameter, and the implications of this trend are discussed more fully in Section \ref{sec:disc}. Figure \ref{fig:nhvrho} shows this result, with blue and red data points symbolizing star-forming and non-star-forming galaxies, respectively. Figure  \ref{fig:nhvrhobin} shows the same results, limited to the modeled systems,  with data divided into three bins of R/R$_{\rm vir}$, and the best power law fit with 1$\sigma$ errors shown in green. We perform a survival analysis on these data by populating a random distribution along a range of allowed hydrogen column density 1000 times, and including censored data points where the HI (and thus N$_{\rm H}$) is a lower limit. The best power law fit from a linear regression analysis is: \begin{equation} \label{eqn:profile} N_{\rm H, preferred} = 10^{19.1 \pm 0.5} (R/R_{\rm vir})^{-1.0\pm0.5}  \rm cm^{-2} \end{equation} We note that without the lower limits in the gas column densities, the null hypothesis is rejected with a confidence of 99\%. 

Additionally, in Figure \ref{fig:nhvrhobin} we show in the shaded beige area the best fit that now includes the 11 metal-line non-detections. We sample the full range of allowed total hydrogen column densities for the 7 systems showing HI but no metal lines. The range spans the HI column density on the lower side to the total possible N$_{\rm H}$ allowed by the 2$\sigma$ upper limits to the metal-ion column densities on the upper side. For a typical absorber in this sample, the full range of N$_{\rm H}$  is 10$^{15}$ $\sim$ 10$^{20.5}$. For the 4 systems that show no absorption at all, we include them at a N$_{\rm H}$ of 0. The best power-law fit, shown by the dashed brown line within the beige area that encompasses the 1$\sigma$ error to this fit is: 
\begin{equation} \label{eqn:lowprofile} N_{\rm H, low} = 10^{19.1 \pm 1.2} (R/R_{\rm vir})^{-0.4\pm1.3}  \rm cm^{-2} \end{equation} 

\begin{figure}[h!]
\begin{centering}
\hspace{-0.3in}
\includegraphics[height=1.1\linewidth,angle=90]{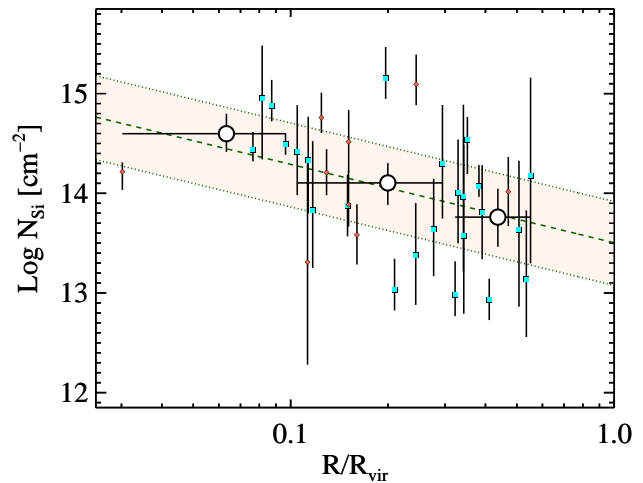}
\end{centering}
\caption{ The total silicon column density versus impact parameter, given as R/R$_{\rm vir}$.  The range of  allowed log N$_{\rm Si}$  directly corresponds to the range in log U allowed by the data, and is shown as a vertical black line for each absorber. Red diamonds indicate a  non-star-forming host galaxy, and blue squares indicate that the host galaxy is actively forming stars.  We divide the data in three bins of R/R$_{\rm vir}$, taking the mean log N$_{\rm Si}$ in each bin. The power-law fit from a linear regression analysis on the sample of 33 absorbers with metal line detections is  shown as a dashed green line, with the light beige shaded area representing the 1$\sigma$ uncertainty in the fit parameters. }
\label{fig:novrhobin}
\end{figure}

Finally, we determine a total low-ion silicon column density as a function of impact parameter in Figure \ref{fig:novrhobin}.  We calculate log N$_{\rm Si}$  for each CLOUDY-modeled sightline by applying the ionization corrections to the lowion metal lines we observe, and converting to silicon by assuming solar relative abundances. Since most lines of sight provide good estimates of low-ion metal column densities without saturation, this metal surface density determination is more robust than the total gas surface density determination. Metal surface density is more reliable than metallicity because it does not include uncertain HI column densities. The best power law fit from a linear regression analysis is: \begin{equation} \label{eqn:siprofile} N_{\rm Si} = 10^{13.5 \pm 0.3} (R/R_{\rm vir})^{-0.8\pm0.3}  \rm cm^{-2} \end{equation}  \cite{peeples13} explore the implications on the total metal mass of the cool CGM that this metal surface density implies. 

\begin{figure}[b!]
\begin{centering}
\hspace{-0.3in}
\includegraphics[height=1.1\linewidth,angle=90]{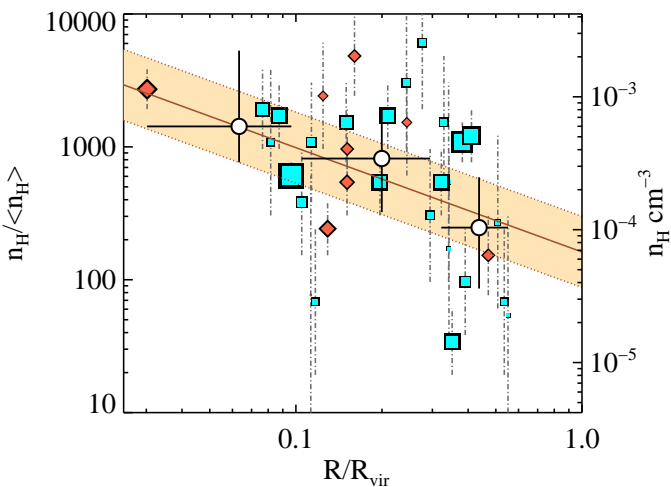}
\end{centering}
\caption{ The CLOUDY mean gas density as a function of impact parameter scaled to R/R$_{\rm vir}$. We find that gas density decreases with impact parameter at a 2.3$\sigma$ level.  The power law fit to the data from a linear regression analysis is shown as brown line, with the shaded beige area represented the 1$\sigma$ uncertainty of the fit parameters. The absorbers with star-forming host galaxies are shown as blue squares, while absorbers having host galaxies without any detectable ongoing star formation are shown as red diamonds. The range of density values for each absorber are shown by vertical dash-dotted gray lines, and are derived from the range in allowed log U from the CLOUDY modeling and COS data. A right-hand axis shows the corresponding values for density in cm$^{-3}$. We also show the data binned in three bins of R/R$_{\rm vir}$ in the white circles. 
}
\label{fig:pgrad}
\end{figure}

Along with ionization parameters and metallicities, the CLOUDY modeling of our COS data allows us to estimate the gas volume density along each sightline, where n$_{\rm H}$ = $\Phi$/ Uc. Here, $\Phi$ is the total flux of ionizing photons ($\sim$ 1.21$\times$10$^{4}$ cm$^{-2}$ s$^{-1}$), as defined by the  \cite{hm01} background radiation field from quasars and galaxies, and c is the speed of light.  This quantity n$_{\rm H}$ refers to the volume density of the gas that gives rise to the low-ionization-state metal lines and neutral hydrogen, and is thus not intended to represent the average volume or mass-weighted density of the CGM. The range of allowed n$_{\rm H}$ directly results from the range of allowed gas ionization parameters. In Figure \ref{fig:pgrad}, we show this gas number density, parameterized to the cosmic mean density of hydrogen at z$\sim$0.2 of 3.3$\times$10$^{-7}$ cm$^{-3}$, as a function of impact parameter scaled to the galaxy virial radius.  Reflecting  the increase in ionization parameter with R (Figure \ref{fig:uvrho}), there is an indication of declining gas volume density with distance from its host galaxy. A Kendall-Tau test rejects the null hypothesis at the 2$\sigma$ level, and the best power law fit is: \begin{equation}
n_{\rm H}/\langle n_{\rm H} \rangle = 2.2\pm0.25(R/R_{\rm vir})^{-0.8\pm0.33} \end{equation} 
\vspace{0.1in}

\subsection{Total Mass of the Photoionized CGM}
\label{sec:cgmmass}

We have shown that the photoionized gas in the CGM of L $\approx $L$^{*}$  galaxies out to 160 kpc is highly ionized, where the mean neutral gas fraction is $\sim$1\%. The total covering fraction of the gas is 90$\pm$4\% at N$_{\rm HI}$ $>$ 10$^{14}$ cm$^{-2}$ (Thom et al. 2012, Tumlinson et al. 2013, Werk et al. 2013). Here, we estimate the total contribution of the photoionized CGM to the baryonic budget of an L $\approx$ L$^{*}$ galaxy treating the COS-Halos sightlines as probes of a single `fiducial' galaxy halo. This method leverages the careful, unbiased sample selection of COS-Halos: galaxy-quasar pairs were selected purely on the basis of galaxy properties with no foreknowledge of absorption.  Effectively, we have generated the first statistical map of the CGM with 44 individual probes out to 160 kpc that allow us to calculate the mass by exploiting our knowledge of the gas surface density. This method requires no assumptions about volume filling factors and cloud sizes, and relies only on measured column densities of low-ionization state metal lines and HI, along with CLOUDY-derived ionization parameters based on these quantities. In this sense, it is the first unbiased estimate of the total baryonic budget of the photoionized L $\sim$ L$^{*}$ CGM, and is independent of models of halo gas density or dark-matter mass.  

We make two estimates for the total mass of the CGM based on the CLOUDY modeling: a strict lower limit, and a preferred lower limit. We do so in order to account for the large systematic errors associated with our analysis, discussed more fully in the Appendix. Each calculation is described below, and  relies  upon converting the total hydrogen column density distribution to a gas surface density distribution by mass. For an additional mass estimate, we calculate the mass of the CGM by estimating the individual cloud sizes (N$_{\rm H}$ / n$_{\rm H}$) and masses indicated by the absorption along each line of sight, and populating a CGM  with these clouds to 300 kpc as is observed \citep{prochaska11}.  

\subsubsection{Strict Lower Limit}

We base the strict lower limit to the baryonic mass of the photoionized, cool CGM on three very conservative assumptions regarding our data: (1) The AODM HI column density we measure from the COS spectra is the true HI column density, regardless of whether it is saturated in the data or whether adopting this value requires assuming a super-solar gas metallicity,   (2) The lowest ionization parameter (i.e. highest neutral gas fraction) allowed by the CLOUDY modeling is the true ionization parameter of the gas and  (3) As our observations include sightlines that only lie up to 160 kpc (0.55 R$_{\rm vir}$) in projection from the massive host galaxy, we assume the gaseous CGM abruptly `ends' beyond this value. We include the 11 non-detections in this estimate as described in Section \ref{sec:density}.  

The mean total hydrogen column assuming these minimal HI values and ionization parameters is log N$_{\rm H}$ = 19 cm$^{-2}$. The best power law fit  for a gas surface density profile based on these values, truncated to 160 kpc (0.55 R/R$_{\rm vir}$), is given by equation \ref{eqn:lowprofile}.  The corresponding gas surface density by mass is 1.4m$_{\rm p}$N$_{\rm H}$(r), abbreviated here as $\Sigma_{\rm gas}$(r), where the factor of 1.4 corrects for the presence of helium (the other metals make a negligible contribution to the baryonic mass). It then follows that the total mass is: 
	
	\begin{equation}\label{eqn:mlowions}
M_{\rm CGM}^{\rm cool}= \int 2\pi R ~\Sigma_{\rm gas}(R)\,{\rm dR},
\end{equation}

Integrating this equation from  0 to 0.55 R/R$_{\rm vir}$,  we find a  strict lower limit to the mass of the photoionized CGM of 2.1 $\times$ 10$^{10}$ M$_{\odot}$.  

\subsubsection{Preferred Lower Limit}

 Adopting the AODM lower limits to log N$_{\rm HI}$ in some cases requires a super-solar gas metallicity, which would be significantly larger than the metallicity of the typical host galaxy disk covered by COS-Halos (0.3 - 1.0 Z/Z$_{\odot}$; Werk et al. 2012). Furthermore, the ionization parameters allowed by the data extend significantly higher than the minimal values (described in detail in the Appendix). We now estimate the total mass of the cool CGM based on the preferred values for log N$_{\rm HI}$ and log U  we derive from the CLOUDY modeling and our absorption-line data. We calculate the mass in the same way as above, now excluding the 11 non-detections. In this case the best power-law fit for a gas surface density profile based on these values  is given by Equation \ref{eqn:profile}. 

 As COS-Halos specifically targeted the inner CGM of L$^{*}$ galaxies, our survey explicitly does not trace the full extent of the CGM. Data from studies that blindly probe IGM and CGM absorption and its connection to host galaxies find a gaseous extent of 300 kpc, independent of galaxy redshift and luminosity \citep{prochaska11, rudie12}. Furthermore,  some metal absorbers have been shown to lie very far from the nearest galaxy, with good completeness to L $>$ 0.04 L$^{*}$ \citep{tripp06, johnson13}. Accordingly, we integrate our gas surface density profile from R/R$_{\rm vir}$ = 0$-$ 1 (the mean value of  R$_{\rm vir}$ for COS-Halos is $\sim$ 300 kpc) to find the preferred lower limit of the mass of the photoionized CGM to be M$_{\rm CGM}^{\rm cool}$ =  6.5 $\times$ 10$^{10}$ M$_{\odot}$. This value is still a lower limit in the sense that the HI column densities used to derive this fit are lower limits. If, for example, we find that the true HI column densities of the saturated absorbers rise by a factor of 3, then the total mass of the photoionized CGM rises to M$_{\rm CGM}^{\rm cool}$ = 1.2 $\times$ 10$^{11}$ M$_{\odot}$. 

\subsubsection{Volume Filling Factor}
\label{sec:fv}
Here, we explore the effects of approximating individual cloud sizes
and deriving a volume filling factor of the CGM. Previous work to
constrain the total amount of gas in a photoionized, cool phase
of the CGM has relied upon cloud counting such as this and geometrical estimations of the
volume filling factor of the clouds in extended halos (e.g. Stocke et
al. 2013\nocite{stocke13}).

We estimate the mass in the photoionized phase of the CGM assuming that each line of sight probes one to a few gas clouds with
physical properties determined by the CLOUDY modeling. We assume a
cloud diameter (i.e. thickness along the line of sight), $\ell$,  of
N$_{\rm H}$/n$_{\rm H}$ for each cloud,  resulting in cloud sizes that range from 0.1 - 2000 kpc, with a median value of 10$^{+35}_{-10}$ kpc. There is  up to three orders of magnitude uncertainty in the quantity $\ell$ for the sightlines with the least-constrained CLOUDY solutions.  For reference, \cite{stocke13} find a typical cloud size of  $\sim$1 kpc, with a range of 0.1 - 10 kpc using the same methodology (hence, subject to the same systematics). 

In order to determine a volume filling factor, we must estimate a shadowing factor, S (see Stocke et al. 2013 for a full explanation of this factor),  which is approximately equal to an average number of discrete absorption components at separate velocities along each line of sight. Almost all of our sightlines show multiple absorbers within 600 km s$^{-1}$ of the galaxy systemic velocity (see Section \ref{sec:failure} for a discussion of how this impacts our mass estimate), with an average of 2.4 discrete components per line of sight when determined based on the low-ion metal-lines.  Because the COS spectral data have a velocity resolution of approximately 20 km s$^{-1}$, the number of discrete components, and thus the value of S should be considered a lower limit. In this calculation,  we also include a covering fraction, C, of 75\% based on 33 of 44 sightlines showing metal-line absorption with these properties,  and a total extent of the CGM, R$_{\rm CGM}$, of 160 kpc. We are performing this calculation solely for the inner portion of the CGM (R $<$ 0.5 R$_{\rm vir}$) where our data lie. We note that \cite{stocke13} find that this inner portion of the CGM contains approximately 75\% of the total mass. 

 
 In order to calculate the mass, we must assume that the cloud properties are representative of the full population of clouds in a given L$^{*}$ CGM. Furthermore, we include the full allowed range of cloud size for each absorber based on allowed ranges of N$_{\rm H}$ and n$_{\rm H}$. We find a volume filling factor that ranges between  1 and 100\% (v$_{\rm ff}$ = C$\times$S$\times\ell$/R$_{\rm CGM}$), where the largest allowed cloud sizes are actually larger than our assumed CGM! The median value of the volume filling factor, determined statistically with 1000 trials along the full range of allowed cloud sizes, and including lower limits, is 11$^{+15}_{-9}$\%. 
 
Assuming a geometry of  spherical clouds in a spherical CGM, the total number of clouds is N$_{\rm cl}$ = C$\times$S$\times$ R$_{\rm CGM}^{2}$/$\ell^{2}$ \citep{stocke13}. We find N$_{\rm cl}$ to range between 1 and 10$^{5}$, with the largest clouds being the fewest in number, and the smallest clouds representing the majority of the population by design. The median value of  N$_{\rm cl}$ is 440$\pm$400. Cloud masses,  M$_{\rm cl}$,  range from 100 - 10$^{11}$ M$_{\odot}$, with a  median value of 10$^{7.6}$ M$_{\odot}$. In this
case, the median values of N$_{\rm cl}$ $\times$ M$_{\rm cl}$ gives us a value of
3.2$\times$10$^{10}$ M$_{\odot}$ for the total amount of mass in the
photoionized CGM within 160 kpc. This mass estimate is remarkably consistent with the
calculation of Stocke et al. (2013) for their sample of super-L$^{*}$ galaxies, and our own surface-density-based estimate of the
lower limit to the CGM baryonic mass.  

However, the uncertainty in this estimate is very large (a factor of 10) and this calculation should be interpreted with caution. The large errors associated with this method of calculating the mass arise from the use of both N$_{\rm H}$ and n$_{\rm H}$ to calculate the cloud size. Many of our N$_{\rm H}$ values are lower limits, which translates to lower limits on cloud sizes (many of which are larger than 10 kpc). Such `clouds' would not fit into an L$^{*}$ halo, which makes it difficult to reconcile with the observation that there are at least 2.4 discrete components per line of sight.  Furthermore, the total hydrogen column and the total gas density are two quantities determined directly from the ionization parameter. At the lowest allowed volume density (i.e. highest gas ionization), the total hydrogen column is maximized, which results in a very large cloud size of hundreds to several thousands of kpc. The ranges in allowed cloud sizes for each line of sight span one full order of magnitude for the best-constrained absorbers, and three full orders of magnitude for the poorly-constrained absorbers.

 \section{Discussion}
 \label{sec:disc}
 
\subsection{Contribution of the Photoionized CGM to the Baryonic Budget of an L$^{*}$ Galaxy}

The median halo mass of our sample of 44 COS-Halos galaxies, based on abundance matching, is 1.6 $\times$10$^{12}$ M$_{\odot}$ (Moster et al. 2010\nocite{moster10}). Thus, the cosmological baryonic budget of  the typical COS-Halos galaxy is approximately 10$^{11.4}$ M$_{\odot}$ (17\% of the DM component), with the stellar disk contributing approximately 10$^{10.6}$ M$_{\odot}$, or 14\%. The gas in the ISM of these galaxies will vary from very little to nearly as much as the stellar contribution \citep{mcgaugh10, martin10}. We find that the contribution from the photoionized, bound CGM to the total baryonic budget of an L$\sim$L$^{*}$ galaxy is at least 25\% -- at least as much as the total contribution from the stars and gas in the disk of the galaxy. 
 
\subsection{Contribution from Additional, Unaccounted Gas Phases of the CGM}

Studies of warm-hot gas phases of the CGM indicate that 10$^{5}$ $-$10$^{7}$K gas likely comprises an additional, significant contribution to the baryonic content of galaxy halos (e.g. Tumlinson et al. 2011; Tripp et al. 2011; Anderson et al. 2012; Fox et al. 2013; Meiring et al. 2013\nocite{meiring13}). For example, our best CLOUDY models that fit the low ionization state metal absorption lines systemically underestimate the column density of OVI measured from our COS data (mean log N$_{\rm OVI}$ = 14.5 cm$^{-2}$), as described in Section \ref{sec:results} and shown in the figures of the Appendix. As a result, our best models require  that the majority of the observed OVI lie in a separate, more highly ionized gas phase which is consistent with the findings of previous studies. The mass estimates of this more highly ionized phase traced by OVI absorption are complicated by the lack of any additional metal line transitions near the ionization potential of OVI in the COS-Halos data. \cite{tumlinson11} estimate a lower limit to be $>$ 10$^{9}$ M$_{\odot}$, based on the maximum possible value for the ionization fraction of OVI (f$_{OVI}$ $<$ 0.2) and solar metallicty. \cite{peeples13} have refined this estimate by considering grids of  simple 1-D halo models that account for the observed OVI,  where the temperature ranges from 10$^{4}$ to 10$^{6}$ K, and the gas surface density profile drops as (R/300 kpc)$^{\alpha}$ with alpha values of  of $-1$ and $-2$. The typical value for the mass of the OVI-traced CGM is several times higher than the lower limit, and lies near  $\sim$10$^{10}$ M$_{\odot}$.  Thus the OVI gas phase contributes at least 5\% of the total baryons in the fiducial COS-Halos galaxy halo. For reference, if  the gas is  instead assumed to be 0.1 Z$_{\odot}$, this mass contribution rises accordingly by a factor of 10. 

The contribution of a diffuse, X-ray component to the CGM has been subject to debate, with estimates ranging from 10$^{9}$ $-$ 10$^{11}$ M$_{\odot}$ \citep{andersonbregman10, gupta12, anderson13}.   Stacked {\emph{ROSAT}} images indicate a mass of gas of $\sim$10$^{9}$ M$_{\odot}$ of 5 million degree gas within 50 kpc of an L$^{*}$ galaxy \citep{anderson13}. Extrapolating this value to 300 kpc, which may or may not be justified,  is complicated by the unknown slope of the hot halo gas profile, but will increase the total mass by a factor between 6 and 14. In this estimate, the mass of the X-ray traced CGM is approximately equal to that of the OVI-traced CGM phase.

 In contrast, \cite{gupta12} argue that the mass content of Milky Way gas at 2 $\times$ 10$^{6}$ K out to 160 kpc is greater than 10$^{11}$ M$_{\odot}$ based on OVII column densities from XMM Newton data of eight bright AGN and an average emission measure of the soft X-ray background (see also Fang et al. 2012). This result implicitly assumes that the OVII absorption lines and the background emission arise from the same gas phase at the same temperature and metallicity, an assumption which has generated some criticism \citep{wang12, fang13}. Taken at face value, this estimate would mean that the X-ray traced CGM to 300 kpc could comprise at least 50\% of the total baryonic budget of the Milky Way.  

Finally, recent observations indicate an extensive cool, dusty component of the CGM\nocite{menard10, roussel10, peek13}, possibly fed by a `slow flow' of dust that has coupled with the radiation from massive stars \citep{zahid13}. Based on a statistical analysis  of reddening using SDSS quasars behind galaxies, \cite{menard10} estimate a total CGM dust mass of $\sim$5 $\times$ 10$^{7}$ M$_{\odot}$. While this does not contribute substantially to the total baryonic budget of massive galaxies, we note that this mass implies a large fraction of the total metals in the CGM are in a solid, cool phase \citep{peeples13}. 

Summing the total contributions of all the distinct phases of the CGM, we estimate that the diffuse gas in galaxy halos accounts for at least 35\% of the total baryon budget for nearby, L$\sim$ L$^*$ galaxies. This quantity makes up more than half of the baryons purported to be missing ($\sim$60\%). Accounting for saturation in the HI column densities used in the cool CGM calculation may raise this contribution by an additional 20\%. Thus, the baryonic fraction of L$^{*}$ galaxy halos may be consistent with the cosmological baryon fraction.  
\nocite{mcgaugh12} and \nocite{mcgaugh10}\nocite{behroozi10}

\begin{figure*}[t!]
\begin{centering}
\hspace{0.35in}
\includegraphics[height=0.60\linewidth,angle=0]{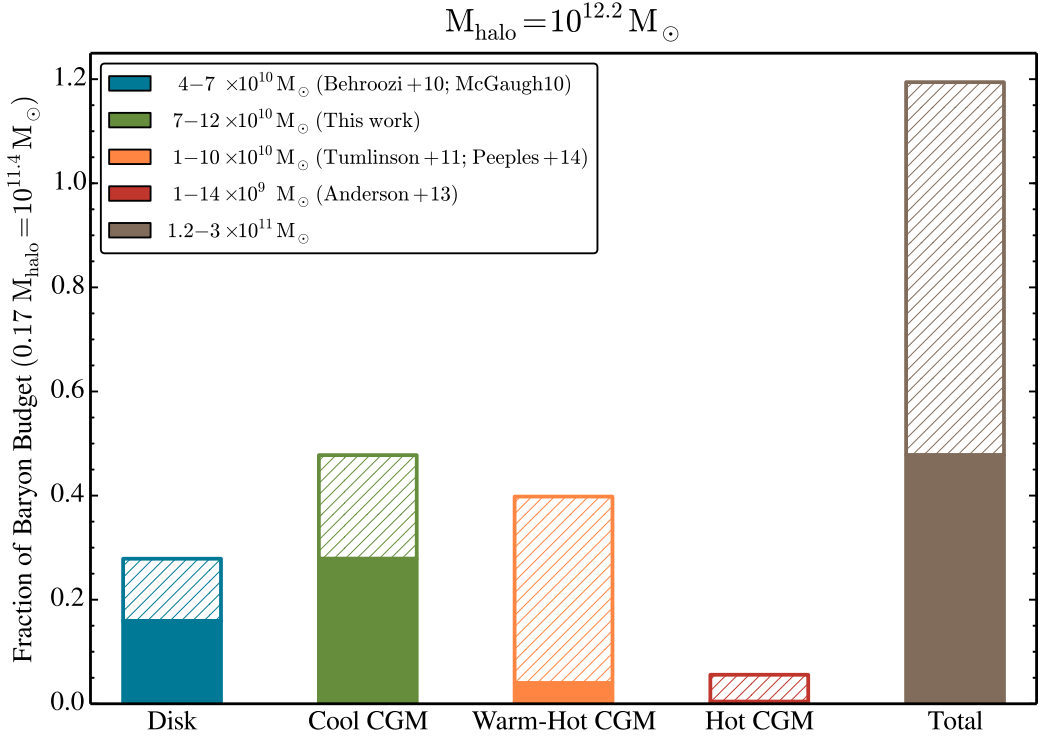}
\end{centering}
\caption{ The baryonic budget (0.17M$_{\rm halo}$) of the fiducial COS-Halos galaxy, at L $\approx$ L$^{*}$, represented as a bar chart showing the most massive baryonic components of the galaxy. The solid filled bars are lower limits to the fraction each component will contribute,  while the hashed area above the solid bars shows potential additional contributions allowed by the data. The stars and gas in the disk of the galaxy (green) make up between 14 and 24\% of the total budget, with the stellar contribution determined from abundance matching (e.g. Behroozi et al. 2010) and the gas contribution (hashed region)  estimated from HI surveys (Martin et al. 2010) and the baryonic Tully-Fisher relation (McGaugh et al. 2010, 2012). The cool CGM contribution of baryons ranges between 25 - 45\%. We have taken the preferred lower limit (solid blue, Section 4.2.2 ), and bounded  it on the top end by adding a factor of 3 to the HI column densities of the sightlines that show saturation.  The warm-hot CGM, traced by OVI, is poorly constrained, with a contribution of at least of 5\% (solid orange) from very conservative assumptions regarding the ionization fraction of OVI \citep{peeples13, tumlinson13} and assuming solar abundance, ranging up to 37\% which allows for gas metallicities of down to 0.1 Z$_{\odot}$. We take the contribution from hot gas at T $>$ 10$^{7}$ K from \cite{anderson13}, which ranges from 2\% to 6\% (red bar), depending on the distance to which the hot halo extends. The sum of these components is given by the black bar, illustrating that galaxies have anywhere between 45\% and 100\% of their baryons relative to the cosmological fraction. }
\label{fig:baryonbudget}
\end{figure*}

\subsection{Hydrostatic Solutions}

We now consider inferences that may be drawn on the nature of the CGM
in the context of simple, hydrostatic solutions.  Before proceeding, we will review the main characteristics of the CGM revealed by our
COS-Halos program and previous works.  First, the cool CGM is nearly
ubiquitous.  The covering fraction $f_C$ of \ion{H}{1} gas exceeds
90$\%$ for sightlines intersecting 0.55 R$_{\rm vir}$ \citep{tumlinson13},
and the incidence of lower ionization state metals is comparable
\citep{werk13}.  This material, therefore, is pervasive within the
dark matter halos of $L^*$ galaxies, but we allow that the volume
filling factor $f_V$ may be small (see Section \ref{sec:fv}).
Second, as emphasized in $\S$~\ref{sec:density}, the surface density of
cool gas is large N$_{\rm H}^{\rm cool}$ $>$ $10^{19}$ cm$^{-2}$ at
essentially all impact parameters  R $<$ R$_{\rm Vir}$.  
Third, the cool CGM is photoionized with $T \sim 10^4$K.
This is indicated by ratios like Si$^{++}$/Si$^{+}$ \citep{werk13} and
the detailed photoionization models presented here \citep[see
also][]{stocke13}.  
Fourth, the CGM of star-forming galaxies also exhibits a highly
ionized phase traced by \ion{O}{6} \citep{tumlinson11}.
As detailed in our Appendix, this material cannot be reproduced by
the photoionization models derived to match the lower ionization
states of metals observed.  This \ion{O}{6} gas presumably traces
another `phase' of the CGM.
Lastly, we assume the galaxies under study exist within dark matter
halos having $T_{\rm Vir} \gtrsim 10^6$K based on the stellar
mass estimates (see $\S$~\ref{sec:sample}).

 \subsubsection{Single-phase Solutions}

The simplest model to consider is a single-phase CGM 
in hydrostatic equilibrium with the dark matter halo potential.  For
the latter, we assume an NFW profile 

\begin{equation}
\rho_{\rm DM}(r) = \frac{\rho_S r_S^3}{r (r+r_S)^2}
\label{eqn:nfw}
\end{equation}
defined by a scale radius $r_S$ set by the 
concentration parameter $C_V \equiv r_V/r_S = 13$ and with $\rho_S$
set by the halo mass, which we take as $M_{\rm DM} = 10^{12} \mmsun$.

For the baryons, we assume the CGM is an optically thin medium, with
metallicity $1/10$ solar irradiated by the extragalactic UV background \citep{hm01}.
We allow for a total gas mass $M_g$ as large 
as $\Omega_b M_{\rm DM} / \Omega_m$ but also consider smaller masses
parameterized by $f_g$.

The astrophysical solutions for a baryonic plasma embedded within an
NFW potential have been considered many times previously
\citep[e.g.][]{makino98,suto98,capelo10}, primarly in the context
of hot gas around massive elliptical galaxies or the intracluster
medium.  Our scenario differs in that our fiducial halo is somewhat
less massive and, more importantly, we consider a plasma with
substantially lower gas temperature. 

To simplify the analysis, we make two standard approximations:
(i) the baryons do not contribute to the gravitational potential.
In the extreme case, $M_g$ may represent as much as
$\Omega_b/\Omega_m \approx 0.2$ of the total mass which is a modest
contribution; 
(ii) the gas follows a polytropic law $P \sim \rho^\Gamma$.
For an optically thin and photoionized gas, the temperature is
relatively insensitive to the gas density.  Examining the output of
the photoionization models presented in $\S$~\ref{sec:results} for
the EUVB background and a 1/10 solar metallicity, we find that
$T \propto n_{\rm H}^{1/5}$ for $\log n_{\rm H} \approx -5$ to $-1$.
This gives a polytropic index of $\Gamma = 0.8$.

We consider first the isothermal case ($\Gamma = 1$) with
$T_0=2\sci{4}$K.  Following the formalism of \cite{capelo10}, the
density profile is given by

\begin{equation}
\rho(r) = \rho_0 \, \exp \ltk -\Delta_{\rm NFW} \, \ltp 1 -
\frac{\ln(1+r/r_S)}{r/r_S} \rtp \rtk
\end{equation}
with $\Delta_{\rm NFW} = -\phi_0 \rho_0/P_0 = -\phi_0 \mu m_p/kT_0$
and the central gravitational potential $\phi_0 \approx 10^5 \, {\rm
  cm^2 \, s^{-1}}$ for our fiducial halo.
Therefore, we have $\Delta_{\rm NFW} \approx 1000 (T/10^4 \, {\rm
  K})^{-1}$ and the density falls off very steeply with radius
resulting in a negligible value in the outer halo ($r>r_s$).
We conclude that an isothermal, cool CGM cannot reproduce the
observations.

For the polytropic case, we have 
\begin{equation}
\rho(r) = \rho_0 \, \ltk 1 - \frac{\Gamma-1}{\Gamma} \Delta_{\rm NFW} \, \ltp 1 -
\frac{\ln(1+r/r_S)}{r/r_S} \rtp \rtk^{1/\Gamma-1}
\end{equation}
This leads to a slightly shallower density profile for $\Gamma = 0.8$
but still one where $\rho(r_S) \approx \rho_0/10^{10}$.  
The outer halo is very nearly a vacuum.

We conclude that a single-phase CGM with $T \approx 10^4$K in
hydrostatic equilibrium with an NFW potential cannot reproduce the
observations.  We are motivated, therefore, to consider more complex
(and realistic) scenarios.

%

 \subsubsection{Two-phase Models}

Guided by the observations for a wide
range of ionization states in halo gas (e.g. \ion{Si}{2}, \ion{Mg}{2},
\ion{Si}{3}, \ion{C}{4}, \ion{O}{6}), it is likely that the medium has multiple phases.  In their seminal paper,
\cite{mm96} introduced a two-phase model composed of cool clouds ($T
\sim 10^4$K) in pressure equilibrium with a more diffuse, hot halo
gas ($T \sim 10^6$K).  They presented solutions for the density
profile of the hot phase, placed constraints on the masses of the
cool clouds, and tracked the dynamics (i.e.\ infall kinematics) of the
cool clumps.  In turn, they demonstrated that this two-phase model
could reproduce some of the basic observables of halo gas at $z<1$.

Other authors have since developed hydrostatic solutions in the
context of high velocity clouds of the Milky Way \citep{sternberg02}.
These treat the EUVB radiation field to model the \nhi\ profile of the
clouds and also examine higher ionization states of the gas.
In all of these models, the cool phase is assumed to have `condensed'
out of the hot phase, via hydrostatic instabilities.
While there is theoretical support for this assumption
\citep{field65}, other analyses have argued that galactic halos are
generally stable to such condensations (Binney et al. 2009; but see McCourt et al. 2012 \nocite{binney09,mccourt12}).

In the following, we adopt the formalism of Maller \& Bullock (2004; hereafter MB04\nocite{mb04}) who expanded upon the \cite{mm96} treatment.  Our goal is
to examine whether such clumpy, two-phase scenarios reasonably
reproduce the gas volume densities and surface densities estimated from
our dataset.

MB04 assumed that the hot gas originally follows the NFW dark matter
profile (Equation~\ref{eqn:nfw}) with an inner core \citep{frenk99}.
Within a characteristic cooling radius $r_C$, a fraction of the mass
takes the form of cool, pressure-supported clouds and the hot gas
evolves adiabatically to a new density profile:

\begin{equation}
\rho_h(x) = \rho_c \left\{ 1 + \frac{3.7}{x} \ln(1+x) + 
  \frac{3.7}{C_C} \ln(1+C_C) \right\}^{3/2}
\label{eqn:hotgas}
\end{equation}
with $x = r/r_S$ and $C_C = r_C/r_S$.  Given this density profile for
the hot gas (and a related expression for the temperature which has a
small variation), we estimate the cool gas density as:

\begin{equation}
\rho_{\rm cool}(r) = \rho_h(r) \frac{T_h(r)}{T_{\rm cool}(r)}
\end{equation}
and we adopt $T_{\rm cool}$ = 2 $\times$ 10$^{4}$K in what follows.
All of the remaining variables relate to the assumed properties of the
halo.  For our fiducial model, we take $M_{\rm halo}$ = 10$^{12}$ M$_{\odot}$, $r_V = 199$\,kpc, $T_{\rm halo}$ = 1.3 $\times$ 10$^{6}$K, 
$C_V = 13$, and a halo gas metallicity
$Z_{\rm halo} = 0.1 Z_\odot$.
Also, we perform the calculation at $z=0$ but note the results are
similar for any $z \ll 1$.

To compare against the measurements along the quasar sightlines which
intersect halos at fixed impact parameters $R$, we must project the
gas density profile.  Because our analysis is weighted by column
density (e.g.\ we sum all of the gas along each sightline), we 
calculate a density-weighted value, which closely approximates projection effects: 

\begin{equation}
\rho(R) = \frac{\int \rho^2(r) ds}{\int \rho(r) ds}
\end{equation}

\begin{figure}[h!]
\begin{centering}
\hspace{-0.3in}
\includegraphics[height=1.1\linewidth,angle=90]{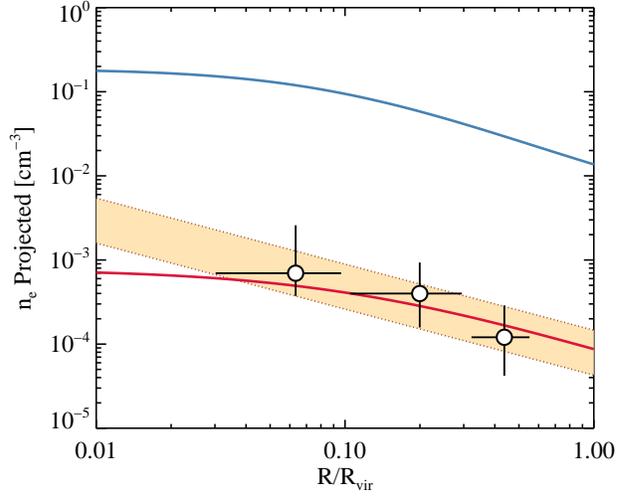}
\end{centering}
\caption{ Projected electron density profiles for gas in a 10$^{12}$ M$_{\odot}$ halo following the formalism presented by MB04. The hot gas profile is given by the solid red curve given by Equation 12, while the cool gas profile follows the solid blue curve given by Equation 13.  For comparison, we show the binned values and best fit from Figure \ref{fig:pgrad} that describe our data, with hydrogen densities converted to electron densities using the ionized gas fractions of hydrogen and helium (mean correction factor of 1.16 to n$_{\rm H}$). The T $\sim$ 10$^{4}$K gas we observe appears to follow the hot halo gas electron density profile rather than the cool gas electron density profile.  }
\label{fig:halo_model}
\end{figure}

The model is compared against the measurements in
Figure~\ref{fig:halo_model}, with each of our derived n$_{\rm H}$ values converted to an electron density based on the ionized gas fractions of hydrogen and helium.
The best fit to our observations, represented by the shaded light brown area and binned data points, shows electron densities approximately two orders of magnitude below the predicted cool gas electron density profile of MB04. Counter to expectation,  the data follow the hot gas electron density profile very well. Figure~\ref{fig:halo_model} shows that the inferred n$_{\rm H}$ values of the CGM lie approximately two orders of magnitude below predictions for a standard, simple two-phase scenario.  Thus, in these simplified one- and two-phase cases, we have ruled out static solutions for the cool CGM. 

\subsubsection{The Failure of the Two-Phase Solution}
\label{sec:failure}

We now consider the robustness of this result, in particular, aspects of our analysis that impact the gas density estimates. One consideration in addressing the density discrepancy noted above is that an increase in the total integrated ionizing flux ($\Phi_{\rm tot}$) from the background radiation field would correspondingly increase our gas electron density measurements. This relation derives directly from the definition of U, the dimensionless ionization parameter, in which n$_{\rm H}$ $=$ $\Phi_{\rm tot}$ / (U $\times$ c).  In the Appendix we discuss the dependence of our results on the slope and magnitude of the radiation field. We have considered both a UV background from quasars and galaxies \citep{hm01}, and the effect of adding the ionizing flux from a star-forming galaxy at a given impact parameter from our sightlines. We note that the ionization parameters are not impacted by the addition of ionizing radiation from a host galaxy,  but $\Phi_{\rm tot}$ increases with a total magnitude dependent on impact parameter ($\propto$ 1/R$^{2}$), star formation rate ($\propto$ SFR), and the escape fraction of ionizing photons ( $\propto$ f$_{\rm esc}$).  

An increase of two orders in magnitude in the gas densities  would result if, for example: (1)  The galaxy SFR exceeds 50 M$_{\odot}$ yr$^{-1}$ for sightlines with  R$<$ 75 kpc considering an escape fraction of 5\%, or (2) The SFR exceeds 20 M$_{\odot}$ yr$^{-1}$ for sightlines at all impact parameters with f$_{\rm esc}$ $>$ 10\%. For reference, the average SFR for the COS-Halos star-forming galaxies is approximately 1 M$_{\odot}$ yr $^{-1}$ and the average impact parameter is 72 kpc. At our average values, assuming an f$_{\rm esc}$ of 1\% (e.g. Inoue et al. 2006\nocite{fesc}), $\Phi_{\rm tot}$ increases by merely a factor of $\sim$ 3.  In addition, we note that a change in the slope of the ionizing radiation field would impact our derived ionization parameters. However, physically plausible slopes above  1 Ryd affect the ionization parameters at the level of $\pm$0.3 dex, and thus cannot be responsible for the two-order-of-magnitude discrepancy. 

Another source of uncertainty in the gas densities results from modeling all of the absorption along the line of sight within $\pm$600 km/s of the galaxy systemic velocity as a single phase. In reality, most of the UV-absorption data exhibit multiple component structure within this velocity range with an average of 2.4 components per absorption system seen in the lowions (see Section \ref{sec:fv} and Werk et al. 2013). The complications of modeling each component separately are many, including: saturation in HI, making it difficult to determine log N$_{\rm HI}$ for each component and blending between components, complicating line fitting and hence column density measurements.

 Here we comment on the effect on our derived gas densities of lumping the characteristics of many little ``clouds" into one measurement. First, we consider the trend shown in Figure \ref{fig:unhi} such that lower column density gas tends to exhibit higher ionization parameters.  The properties of the absorption we derive are largely dominated by the properties of the maximal component, which is the component with the highest HI column density, and, presumably, the lowest ionization parameter. Any lower density components along the line of sight typically do not drive the solutions we derive from our Cloudy modeling. Thus, if a bias exists in our results, it is a bias toward lower ionization parameters and higher gas volume densities, the opposite trend we would need in order to reconcile our observations with the MB04 two-phase solution. 

In light of these considerations, we  conclude that a two-phase model composed of cool clouds ($T\sim 10^4$ K) in pressure equilibrium with virialized hot halo
gas ($T \sim 10^6$ K) is a poor description of the origin of the 10$^{4}$ K CGM at z$\sim$0. In a linear analysis of an MB04 scenario,  \cite{binney09} notes that unless the entropy profile of the hot ambient halo gas is very flat, any cool cloud that might condense out of it would be disrupted before they could cool by the buoyancy of thermally unstable gas. However, non-linear perturbations, such as those that may result from filamentary cold flows do allow for cool clouds to form at densities $>$ 10$^{-3}$ cm$^{-3}$ \citep{mccourt12, joung12}. While this density is closer to matching those indicated by our observations, it still exceeds them by an order of magnitude. Thus, the gas densities we derive for our absorption line systems are very difficult to reconcile with this simple two-phase picture, and we must consider a different physical description of the gas.

 \subsection{Constraints on the Origin and Nature of the Cool CGM}
 
Here, we explore whether the CGM gas masses and densities we measure are consistent with the gas having originated in a wind from the central host galaxy.   Several recent cosmological simulations that examine the CGM incorporate various feedback prescriptions, yet all of them indicate that most of  the observed absorption in the halos of galaxies is due to gas that is, or was at some point, in a galaxy-scale wind \citep{shen12,stinson12, joung12, cen13, ford13, hummels13}. Furthermore, \cite{cen13} predict that over half of the gas within 150 kpc is in the cool 10$^{4}$K phase, consistent with our observations. The idea that winds eject a substantial amount of material into galaxy haloes is echoed at least in part by the recent finding that galaxies themselves seem to be missing over half of the metals their stars have produced over the course of their lives \citep{zahid12, peeples13}. Toward galaxies themselves, the imprinted signatures of inflows and/or outflows on absorption line profiles have been studied over the last decade for modest samples of individual galaxies (e.g. Heckman et al. 2000; Martin et al. 2005; Rubin et al. 2011\nocite{hls+00, martin05, rubin11}), in co-added spectra of many galaxies (e.g. Weiner et al. 2009, Bordoloi et al. 2013 \nocite{wcp+09, bordoloi13}), and most recently, for  a few hundred galaxies with HST imaging \citep{martin12, rubin13}.  These studies have shown that the Doppler blueshift of MgII of FeII absorption (indicative of outflows) in a galaxy spectrum is stronger with higher star formation rates, is more common for galaxies that are oriented face-on, and that such outflows have the capacity to transport a significant amount of mass into the CGM and are likely to be highly collimated.  

Based on absorption line profiles of MgII and FeII  for 105 individual star-forming galaxies at 0.3 $<$ $z$ $<$ 0.7,  \cite{rubin13} estimate a mass outflow rate of  at least 1 M$_{\odot}$ yr$^{-1}$. Assuming the outflow rate remains constant from z $\sim$ 1 to 0, and that none of the material is re-accreted, this rate implies over 10$^{10}$ M$_{\odot}$ of cool, photoionized material in the CGM of present day galaxies, consistent with our estimates for the mass of the cool CGM. Additionally, the implied total metal mass of the CGM is roughly equivalent to the metal deficiencies determined analytically for L$^*$ galaxies \citep{zahid12, peeples13}. Thus, the simple, seemingly self-consistent picture that emerges is one in which most of the cool gas observed in the halos of galaxies originates from the galaxy itself, building up over time to create a massive reservoir of 10$^{4}$K halo gas.  

In light of the gas densities indicated by the photoionization modeling, however, this simple picture unravels. Gas densities between 10$^{-4}$ and 10$^{-3}$ cm$^{-3}$ such as we determine are not only inconsistent with the MB04 two-phase model, but are also at least an order of magnitude lower than most cosmological simulations seem to require for the cool clouds to survive on cosmological timescales.   Regardless of the feedback prescription and origin, simulations of the CGM at all redshifts typically predict that the cool, 10$^{4}$K gas is at least 10$^{-2}$ cm$^{-3}$ \citep{stinson12, shen12, cen12, hummels13}. Effectively, the problem amounts to a lack of pressure support. At densities below 10$^{-2}$ $-$10$^{-3}$ cm$^{-3}$, cool clouds do not survive longer than several Myr \citep{joung12}. Thus, it appears that at the densities we measure, the gas would not survive on Gyr timescales, slowly assembling into a massive reservoir of cool circumgalactic gas. However, recent work by \cite{ford13} indicates that low ionization state metal lines primarily arise from  so-called recycled outflows -- gas that has been ejected, re-accreted, and ejected again from the central galaxy. The typical density of recycled material in this simulation is $>$ 10$^{-2}$ cm$^{-3}$ within 30 kpc, but consistent with 10$^{-3}$ -  10$^{-4}$ beyond these innermost regions. They claim that because the gas is not in hydrostatic equilibrium, it is falling back down onto the galaxy. In this picture, much of the low-ionization state material was enriched at early times and ejected to distances far from the galaxy (into the IGM), and is now falling back down. Our results are not inconsistent with this picture. 

Owing to the details of the sample selection, COS-Halos galaxies are typically fairly isolated compared to the general population at z$\sim$0.2 \citep{tumlinson13, werk12}. As a final note, we point out that group environments significantly complicate any interpretation on the origin of the CGM since absorption profiles are typically kinematically very complex \citep{araciletal06, trippint08, burchett13, stocke14}. Galaxy environment and interactions surely play some role in the observed properties of the CGM \citep{chg+10, yoon13}, though a comprehensive study of the role of environment has not yet emerged. Finally, the absence of OVI and yet the presence of cool gas in the halos of non-star-forming galaxies \citep{tumlinson11} remains a puzzle, though may be due to different origins for cool and warm-hot halo gas \citep{ford13}. 
  
\section{Summary and Conclusions}

 We have assessed the physical conditions and mass of highly ionized, cool (T $\approx$ 10$^4$ K) CGM gas observed within 160 kpc of low-redshift, L $\approx$ L$^*$ galaxies drawn from the COS-Halos survey. The column densities of HI and low-ionization state metal absorption lines require a characteristic total hydrogen column density of  N$_{\rm H}$ $>$ 10$^{19}$ cm$^{-2}$ in the CGM of these galaxies (\S~4.1; Figure 7). We have leveraged our unique dataset of 44 COS spectra of quasars selected to be within 160 kpc of the nearest L$^*$ galaxy to construct the first maps of the physical state of the CGM at low redshift. Our key findings are:  
 
 \begin{enumerate}
 \item There is a 4$\sigma$ anti-correlation between ionization parameter and HI column density (\S~3.1; Figure 2). The low-ionization state metal line column densities also follow this trend (Figure 3). This result is qualitatively consistent with photoionized clouds in hydrostatic equilibrium where higher column density clouds have a greater total gas volume density. 
 \item We find a 2$\sigma$ correlation between ionization parameter and the projected distance from the galaxy (Figure 4), which is driven by a declining gas volume density with impact parameter (Figure 10). Gas is more highly ionized further from the host galaxy because the gas is lower density at large radii, and thus less shielded from the EUVB. 
 \item We construct gas surface density profiles of hydrogen (Figure 8) and metals (Figure 9), and find they decline out to 160 kpc (0.55 R/R$_{\rm vir}$) with  power-law slopes of  $-1.0\pm$0.5 and $-0.8\pm$0.3, respectively (\S~4.1). These 2$\sigma$ correlations are derived from a survival analysis including censoring in the HI column densities (lower limits). 
 \item We provide a strict lower limit to the total mass of material in the CGM of low-redshift L$^{*}$ galaxies (\S~4.2.1). This limit does not allow for line saturation and truncates at 160 kpc.   There is at least 2 $\times$ 10$^{10}$ M$_{\odot}$ of cool material in the CGM of these galaxies in the most conservative limit. 
 \item We provide a more realistic lower limit to the mass of low-ionization-state material in the halos of L $\approx$ L$^{*}$ galaxies that allows for line saturation in HI (lower limits) and extends to 300 kpc: M$_{\rm CGM}^{\rm cool}$ $>$ 6.5 $\times$ 10$^{10}$ M$_{\odot}$ (\S~4.2.2).  We emphasize that this mass estimate is a lower limit because of saturation in the HI absorption lines for over half of our sample. This mass  of material suggests that over 25\% of the baryon budget of an L $\approx$ L$^{*}$ halo is accounted for by cool, photoionized gas in the CGM.  When we sum the conservatively-estimated contributions from observed hotter, more highly ionized gas phases (OVI, X-ray) we conclude that galaxies may not be baryon-depleted at all relative to the cosmological baryon fraction (Figure 11).  
 \item Finally, we analyze our derived gas volume densities in the context of simple hydrostatic one- and two-phase models (\S~5.3). Each of these models predicts higher gas volume densities by at least a two orders of magnitude. We conclude that the gas we observe is not in hydrostatic equilibrium with a hot gas phase at the virial temperature of the galaxy halo (Figure 12). There may be other means of supporting this gas  (e.g. turbulence, magnetic fields), or else the very large amount of gas we observe has no support at all and is very short-lived in its observed state, such as might occur if it is cycling to and from galaxies on timescales that are very short compared to the dynamical times of dark matter halos. 
   \end{enumerate}

 \section{Acknowledgements}
   
  We thank the referee, Gary Ferland, for very constructive comments and suggestions that improved this manuscript. Support for this work was provided by NASA through program GO11598, and through Hubble Fellowship grant \# 51332 from the Space Telescope Science Institute, which is operated by the Association of Universities for Research in Astronomy, Inc., under NASA contract NAS 5-26555.  Optical data used in this study were obtained at the W.M. Keck Observatory, which is operated as a scientific partnership among the California Institute of Technology, the University of California and the National Aeronautics and Space Administration. The Observatory was made possible by the generous financial support of the W.M. Keck Foundation. The authors wish to recognize and acknowledge the very significant cultural role and
reverence that the summit of Mauna Kea has always had within the indigenous Hawaiian community. We are most fortunate to have the opportunity to conduct observations from this mountain. JKW would like to especially thank Bill Mathews, Jabran Zahid,  Mike Anderson, Josh Peek, Gwen Rudie, Mary Putman, James Bullock, John Stocke, Mike Shull, and Joel Bregman for very useful discussions related to this work and comments on early drafts. MSP acknowledges support from the Southern California Center for Galaxy Evolution, a multi-campus research program funded by the University of California Office of Research. Neal Katz acknowledges support from NASA ATP grant NNX10AJ95G. 

    {{\it Facilities:} \facility{HST: COS} \facility{Keck: LRIS} \facility{Magellan: Mage} }

    \bibliographystyle{apj}
\bibliography{cgmmass_all}
\clearpage

\begin{deluxetable}{lcccccccccccc}
\tablewidth{0pc}
\tablecaption{Derived Physical Parameters\label{tab:cloudytab}}
\tabletypesize{\footnotesize}
\tablehead{\colhead{SDSS Field} & \colhead{Galaxy ID}& \colhead{z} & \colhead{R} & \colhead{R/R$_{\rm vir}$} &\colhead{log M$_{*}$}  & \colhead{Q} &\colhead{N$_{\rm HI}$} &\colhead{N$_{\rm HI}$} &
\colhead{Log U} &\colhead{[X/H]} & \colhead{Log N$_{H}$} & \colhead{N$_{\rm H}$}\\
 &  &  & (kpc) & &  & flag & Allowed & Adopted  &  &  &  & Adopted }
\startdata
J0226+0015&268\_22&  0.23& 78& 0.22& 10.8& 1& 14.2$\pm$ 0.03&\nodata & \nodata  & \nodata & \nodata  & \nodata   \\
J0401-0540&67\_24&  0.22& 83& 0.35& 10.1& 5&[ 15.6, 16.5]& 15.6&[ -1.8 ,  -1.3] &[ -1.2 ,  -0.4] &[ 19.4 ,  19.9]& 19.7\\
J0803+4332&306\_20&  0.25& 77& 0.11& 11.3& 2& 14.8$\pm$ 0.04& 14.8&[ -2.8 ,  -1.0] &[ -1.7 ,   0.0] &[ 17.6 ,  19.7]& 18.7\\
J0910+1014&34\_46&  0.14&112& 0.34& 10.6& 3&[ 15.8, 18.5]& 16.5&[ -3.5 ,  -2.0] &[ -1.3 ,  -0.2] &[ 18.4 ,  20.0]& 19.1\\
J0910+1014&242\_34&  0.26&135& 0.16& 11.4& 4&[ 16.5, 18.2]& 17.0&[ -4.0 ,  -3.4] &[ -1.8 ,   0.0] &[ 18.4 ,  19.0]& 18.7\\
J0914+2823&41\_27&  0.24&101& 0.51&  9.8& 3& 15.5$\pm$ 0.03& 15.5&[ -3.1 ,  -1.8] &[ -1.4 ,  -0.5] &[ 17.7 ,  19.2]& 18.5\\
J0925+4004&196\_22&  0.25& 83& 0.12& 11.3& 3& 19.5$\pm$ 0.15& 19.6&[ -3.8 ,  -3.0] & -0.7$\pm$0.2&[ 20.0 ,  20.3]& 20.1\\
J0928+6025&110\_35&  0.15& 91& 0.24& 10.8& 4& 19.4$\pm$ 0.15& 19.5&[ -3.6 ,  -2.8] & -0.4$\pm$0.2&[ 19.9 ,  20.3]& 20.1\\
J0935+0204&15\_28&  0.26&110& 0.26& 11.0& 1&$<$12.68&\nodata & \nodata  & \nodata & \nodata  & \nodata   \\
J0943+0531&106\_34&  0.23&121& 0.34& 10.8& 2&[ 15.4,20.0]& 15.4&[ -3.0 ,  -1.5] &[ -1.0 ,   0.0] &[ 17.8 ,  19.5]& 18.5\\
J0943+0531&216\_61&  0.14&149& 0.33& 11.0& 1&[ 14.9, 17.0]&\nodata & \nodata  & \nodata & \nodata  & \nodata   \\
J0943+0531&227\_19&  0.35& 92& 0.55&  9.6& 3& 16.3$\pm$ 0.03& 16.3&[ -2.5 ,  -1.0] & -1.6$\pm$0.2&[ 19.4 ,  21.2]& 20.2\\
J0950+4831&177\_27&  0.21& 91& 0.15& 11.2& 4&[ 17.5, 18.5]& 18.5&[ -3.3 ,  -2.7] &[ -1.5 ,   0.0] &[ 19.8 ,  20.4]& 20.1\\
J1009+0713&204\_17&  0.23& 60& 0.29&  9.8& 3&[ 16.0, 18.2]& 16.2&[ -3.0 ,  -2.0] &[ -2.2 ,   0.0] &[ 18.5 ,  19.6]& 19.1\\
J1009+0713&170\_9&  0.36& 44& 0.20& 10.2& 4&[ 18.0, 18.9]& 18.5&[ -3.0 ,  -2.5] & -0.6$\pm$0.2&[ 20.0 ,  20.5]& 20.2\\
J1016+4706&274\_6&  0.25& 23& 0.10& 10.2& 5&[ 16.6, 18.5]& 16.6&[ -2.9 ,  -2.7] & $<$  -0.1&[ 18.9 ,  19.2]& 19.1\\
J1016+4706&359\_16&  0.17& 44& 0.15& 10.5& 5&[ 16.4, 18.2]& 16.4&[ -3.5 ,  -2.9] & $<$  -0.1&[ 18.1 ,  18.7]& 18.4\\
J1112+3539&236\_14&  0.25& 53& 0.21& 10.3& 5&[ 15.8, 17.5]& 15.8&[ -3.5 ,  -3.0] & $<$  -0.3&[ 17.6 ,  18.1]& 17.8\\
J1133+0327&110\_5&  0.24& 18& 0.03& 11.2& 5& 18.6$\pm$ 0.06& 18.6&[ -3.6 ,  -3.3] & -1.0$\pm$0.2&[ 19.5 ,  19.8]& 19.7\\
J1133+0327&164\_21&  0.15& 55& 0.23& 10.1& 1&[ 15.8, 18.0]&\nodata & \nodata  & \nodata & \nodata  & \nodata   \\
J1157-0022&230\_7&  0.16& 19& 0.05& 10.8& 1&[ 15.6, 17.6]&\nodata & \nodata  & \nodata & \nodata  & \nodata   \\
J1220+3853&225\_38&  0.27&154& 0.47& 10.7& 5& 15.8$\pm$ 0.05& 15.8&[ -2.5 ,  -1.9] & -0.6$\pm$0.2&[ 18.7 ,  19.4]& 19.1\\
J1233+4758&94\_38&  0.22&132& 0.38& 10.8& 5&[ 16.7, 18.3]& 16.7&[ -3.2 ,  -2.9] & $<$  -0.4&[ 18.8 ,  19.1]& 18.9\\
J1233-0031&168\_7&  0.32& 32& 0.12& 10.5& 3& 15.6$\pm$ 0.02& 15.6&[ -2.4 ,  -1.3] &[ -1.3 ,   0.0] &[ 18.7 ,  20.0]& 19.3\\
J1241+5721&199\_6&  0.21& 20& 0.08& 10.2& 3&[ 16.9, 18.5]& 17.9&[ -3.6 ,  -2.5] &[ -1.2 ,   0.0] &[ 19.3 ,  20.4]& 19.9\\
J1241+5721&208\_27&  0.22& 93& 0.41& 10.0& 4& 15.3$\pm$ 0.06& 15.3&[ -3.3 ,  -2.9] &  0.0$\pm$0.2&[ 17.2 ,  17.6]& 17.4\\
J1245+3356&236\_36&  0.19&112& 0.54&  9.8& 3& 14.8$\pm$ 0.04& 14.8&[ -2.4 ,  -1.3] &[ -1.8 ,  -1.1] &[ 18.1 ,  19.4]& 18.7\\
J1322+4645&349\_11&  0.21& 37& 0.10& 10.8& 4&[ 16.3, 18.3]& 16.3&[ -3.0 ,  -2.2] &[ -1.9 ,  -0.3] &[ 18.5 ,  19.4]& 19.0\\
J1330+2813&289\_28&  0.19& 87& 0.33& 10.3& 4&[ 16.5, 18.5]& 16.6&[ -3.7 ,  -2.7] & $<$  -0.2&[ 18.2 ,  19.2]& 18.7\\
J1342-0053&157\_10&  0.23& 35& 0.09& 10.9& 5&[ 18.3, 19.5]& 19.0&[ -3.5 ,  -3.0] &[ -0.5 ,   0.0] &[ 19.7 ,  20.0]& 19.8\\
J1342-0053&77\_10&  0.20& 32& 0.11& 10.5& 1&$<$12.43&\nodata & \nodata  & \nodata & \nodata  & \nodata   \\
J1419+4207&132\_30&  0.18& 88& 0.28& 10.6& 2&[ 15.4, 18.1]& 17.0&[ -4.3 ,  -3.3] &[ -1.9 ,   0.0] &[ 18.1 ,  19.0]& 18.5\\
J1435+3604&126\_21&  0.26& 83& 0.32& 10.4& 4& 15.3$\pm$ 0.06& 15.3&[ -3.0 ,  -2.5] & -0.4$\pm$0.2&[ 17.6 ,  18.2]& 17.8\\
J1435+3604&68\_12&  0.20& 39& 0.08& 11.1& 5& 19.8$\pm$ 0.10& 19.8&[ -3.6 ,  -3.0] & -1.2$\pm$0.2&[ 20.2 ,  20.4]& 20.3\\
J1437+5045&317\_38&  0.25&143& 0.62& 10.1& 1& 14.5$\pm$ 0.12&\nodata & \nodata  & \nodata & \nodata  & \nodata   \\
J1445+3428&232\_33&  0.22&113& 0.42& 10.4& 1& 15.1$\pm$ 0.06&\nodata & \nodata  & \nodata & \nodata  & \nodata   \\
J1514+3619&287\_14&  0.21& 47& 0.24&  9.7& 4&[ 16.5, 18.4]& 16.5&[ -4.0 ,  -3.0] & $<$  -0.5&[ 17.9 ,  18.9]& 18.3\\
J1550+4001&197\_23&  0.31&102& 0.15& 11.3& 5& 16.5$\pm$ 0.03& 16.5&[ -3.0 ,  -2.5] & -0.8$\pm$0.2&[ 18.9 ,  19.5]& 19.1\\
J1550+4001&97\_33&  0.32&150& 0.41& 10.9& 1& 13.9$\pm$ 0.09&\nodata & \nodata  & \nodata & \nodata  & \nodata   \\
J1555+3628&88\_11&  0.19& 34& 0.11& 10.5& 4&[ 16.8, 18.3]& 17.2&[ -3.5 ,  -2.6] &[ -1.6 ,   0.0] &[ 19.1 ,  20.0]& 19.6\\
J1617+0638&253\_39&  0.15&101& 0.09& 11.5& 1&$<$13.12&\nodata & \nodata  & \nodata & \nodata  & \nodata   \\
J1619+3342&113\_40&  0.14& 97& 0.39& 10.1& 4&[ 15.0, 17.5]& 15.0&[ -2.4 ,  -1.6] & $<$  -0.1&[ 17.9 ,  18.8]& 18.4\\
J2257+1340&270\_40&  0.18&116& 0.28& 10.9& 1&$<$12.53&\nodata & \nodata  & \nodata & \nodata  & \nodata   \\
J2345-0059&356\_12&  0.25& 46& 0.13& 10.8& 4& 16.0$\pm$ 0.04& 16.0&[ -2.6 ,  -2.2] & -0.3$\pm$0.2&[ 18.7 ,  19.2]& 19.0\\
\enddata
\tablecomments{{}  (1) SDSS Field Identifier 
 and (2) Galaxy Identifier, given as PA ($^{\circ}$)
 and  angular separation ($"$) from the QSO, respectively
(3) Spectroscopic Redshift (Werk et al. 2012) 
(4) Projected separation in kpc, calculated in the galaxy restframe.
(5) Projected separation scaled to the virial radius of the galaxy
(6)  Log stellar mass from {\emph{kcorrect}} (Blanton et al. 2007) 
(7) Quality Flag; 1 - 5 (low to high) where absorbers having Q $>$ 2 are included in the analysis.
(8) Range of Log N$_{\rm HI}$ allowed by COS data
(9) Adopted HI column density.   (10) Range of Log U allowed by data
 (11) Range (or estimate) of [X/H] 
  (12) Range of total hydrogen column density (log cm$^{-2}$): low value is based on AODM HI column density and lowest Log U; high value is from adopted HI column density and highest allowed Log U  (13) Log of the Adopted Total Hydrogen Column, calculated using the adopted HI column density and the mean Log U for each sightline. } 
\end{deluxetable} 


\appendix
\section{Details of the Photoionization Modeling}

We model the ionization state of circumgalactic gas using the CLOUDY spectral synthesis code (version c13; Ferland et al. 2013), in which the gas is assumed to be a uniform slab in thermal and ionization equilibrium. We use the background radiation field from quasars and galaxies as our ionization source for gas at galactocentric distances between 10 - 160 kpc, implementing the Haardt-Madau UV background (Haardt \& Madau 2001; HM 2001) at z$=$0.2 in our CLOUDY inputs. We examine the outputs of a photoionization model grid  to find the set of models that are consistent with the constraints set by the ionic column densities determined from the observations.  Our models vary gas metallicity, log N$_{\rm X}$/N$_{\rm H}$  - log (X/H)$_{\odot}$ between 0.001 and the solar value (Asplund et al. 2009), and the ionization parameter, log U =  log n$_{\gamma}$/n$_{H}$ = ionizing photon density / total hydrogen number density (neutral + ionized) between $-$1 and $-$5. All of the measurements of and limits on the low-ionization metal lines that comprise this analysis (primarily SiII, SiIII, CII, CIII, NII, NIII, OI, and MgII) are detailed, tabulated, and provided in Werk et al. (2013). 

\subsection{The Addition of Ionizing Radiation and Cosmic Rays from a Central Star-Forming Galaxy}

\begin{figure}[h!]
\begin{centering}
\hspace{0.8in}
\includegraphics[height=0.75\linewidth,angle=90]{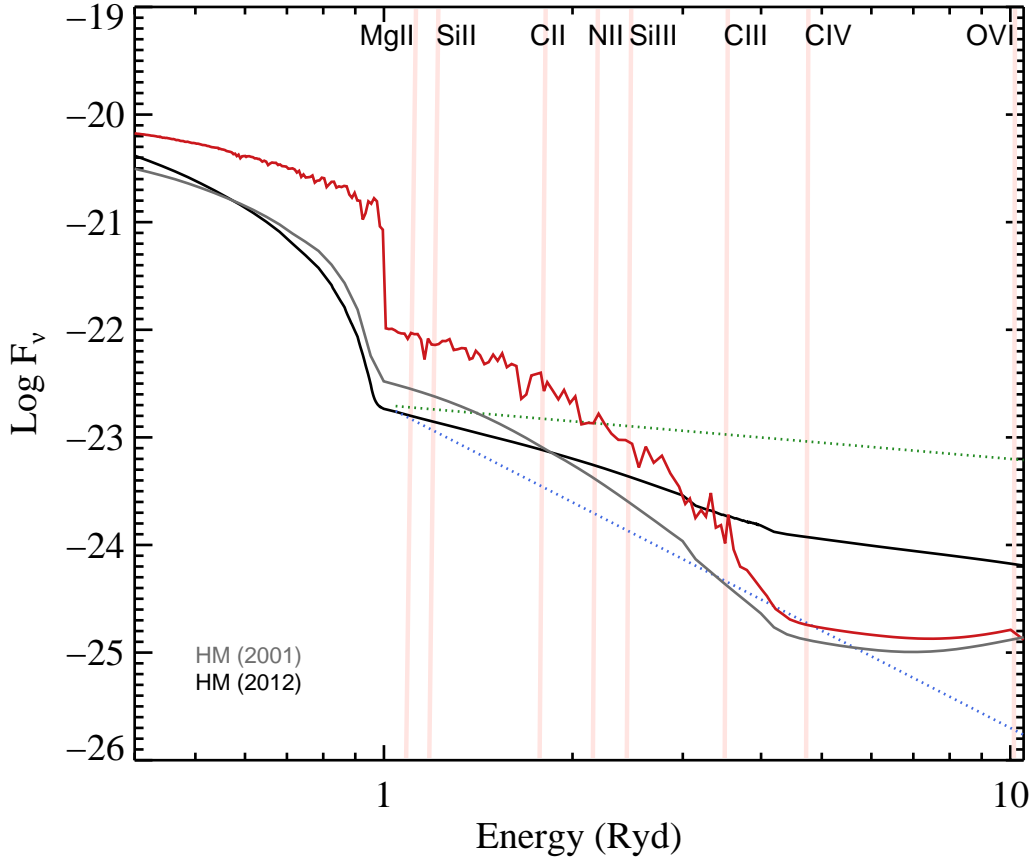}
\vspace{0.3in}
\end{centering}
\caption{ Several spectral energy distributions for the ionizing background radiation field: The Haardt-Madau (2001; HM 2001) UV ionizing background radiation field from quasars and galaxies (dark gray), the \cite{hm12} UV ionizing background radiation field from quasars and galaxies (black line),  and a galaxy SED from Starburst99 added to the HM 2001 background (red). We chose the galaxy to have properties of a typical COS-Halos target galaxy, with a SFR of 1 M$_{\odot}$ yr$^{-1}$, 72 kpc from the gas cloud, and with an escape fraction of ionizing photons of 0.1. We show the ionization potential energies of several common metal ions used in our analysis as vertical rose-colored lines. For the analysis described in the Appendix, we apply the HM 2001 ionizing background only.  We investigated the effects of including escaping ionizing radiation from a nearby galaxy, as well as varying the slope of the spectrum above 1 Ryd. We analyzed both a shallow slope, with a power law exponent of -0.5 (green dotted line),  and a steep slope with a power law exponent of -3.0 (blue dotted line). For reference, the HM 12 spectrum (black line) has a power law index of -1.57 above 1 Ryd. Ultimately, we determine that uncertainty in the EUVB radiation field contributes systematic error in our analysis of $\pm$ 0.3 dex. }
\label{fig:cldyspec}
\end{figure}

We examined a similar model grid that incorporates the addition of ionizing photons from the central galaxy to the Haardt-Madau UV background as our input CLOUDY spectrum. We show the spectra of these sources of ionization in Figure \ref{fig:cldyspec} for reference. Although the radiation from the starburst99  galaxy SED (Leitherer et al. 1999; d = 72 kpc; SFR = 1 M$_{\odot}$ yr$^{-1}$; both median values for the COS-Halos galaxy sample; Werk et al. 2012\nocite{lsg99}) dominates the HM 2001 UV background, the slopes of the SEDs between 1 and 4 Rydberg are very similar.  The extent to which the galaxy SED dominates HM 2001 depends on the escape fraction of ionizing photons (assumed to be 5\%), the distance from the galaxy, and the SFR of the galaxy.  Since the influence of the galaxy radiation field scales as SFR / d$^{2}$, we explored a wide range of parameter space for this additional ionization source. The total contribution to the ionizing radiation field increases substantially with lower impact parameter and higher SFR, dominating the extragalactic UV background radiation field below $\sim$50 kpc for modest SFRs (SFR $<$ 1 M$_{\odot}$ yr$^{-1}$). The overall slopes of the HM 2001 and S99 SED spectra remain approximately equivalent over the 1 - 4 Rydberg range.  The only effect of including ionizing photons from the central galaxy on our results is to push the derived ionization parameter  at most $\sim$0.5 dex higher at column densities below $\sim$10$^{16.5}$ cm$^{-2}$, which modestly increases the estimate of the total amount of gas (neutral + ionized) in the CGM.  

Within a galaxy virial radius, cosmic ray heating could be a significant supplemental heating and ionization source to photoionization \citep{wiener13}.  Additionally, cosmic ray feedback theory has shown that cosmic rays may impart a significant amount of  momentum to the ISM in a direction away from the galaxy, potentially driving a large-scale galactic wind \citep{socrates08}.  At some low value of the gas volume density, heating due to the cosmic ray background (CRB; H$_{\rm CRB}$ $\propto$ n$_{\rm H}$) becomes more important than photoelectric heating due to the EUVB  (H$_{\rm EUVB}$ $\propto$ n$_{\rm H}^{2}$). Thus, there would be a corresponding minimum gas density below which the CGM gas succumbs to a CRB-driven thermal runaway to extremely high, nearly relativistic temperatures. We may assess the contribution of the CRB to the heating of CGM gas using the built-in CRB in C13, which is based on observations of H$^{+}_{3}$ in the diffuse ISM of the Milky Way \citep{indriolo07}.  The local CRB can constitute as much as 85\% of the total heating for a gas volume density of 10$^{-3.5}$ cm$^{-3}$ and 50\% of the total heating for gas volume density of 10$^{-2.5}$ cm$^{-3}$, but these numbers are highly uncertain given the large uncertainty in the local CRB. Nonetheless, these contributions imply that at the low gas densities we derive, we may be very close to a CRB-driven thermal runaway. Thus, if the CRB in the CGM of L$^*$ galaxies is similar to the local background, then heating due to cosmic rays could have a significant impact on the results we present here. 

\begin{figure}[h!]
\hspace{-1.1cm}
\subfigure[ Measured Vs. Preferred HI Column Densities]{\includegraphics[width=10.3cm]{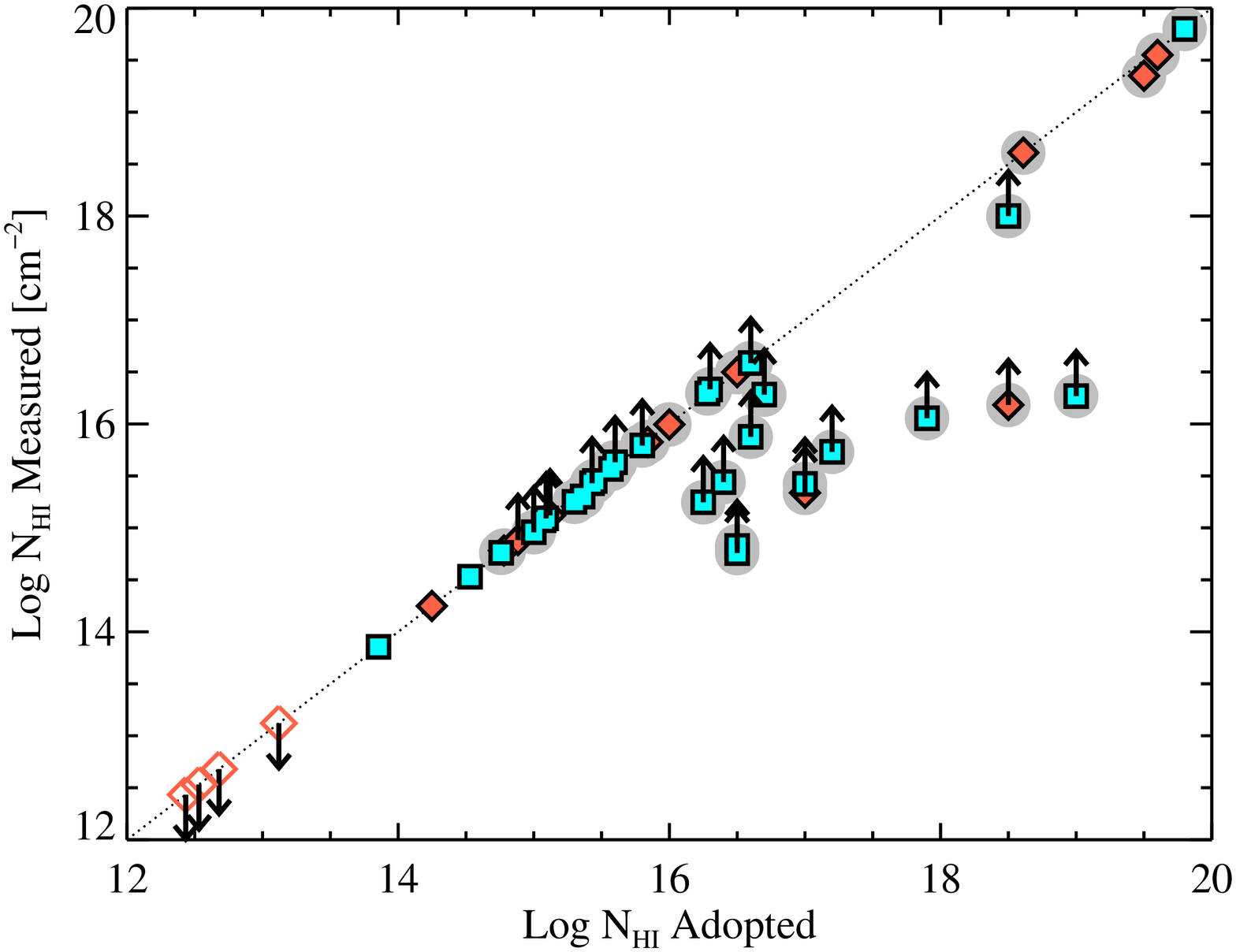} }
\hspace{-1.0cm}
\subfigure[ COS-Halos HI Column Densities Vs. Impact Parameter ]{\includegraphics[width=10.3cm]{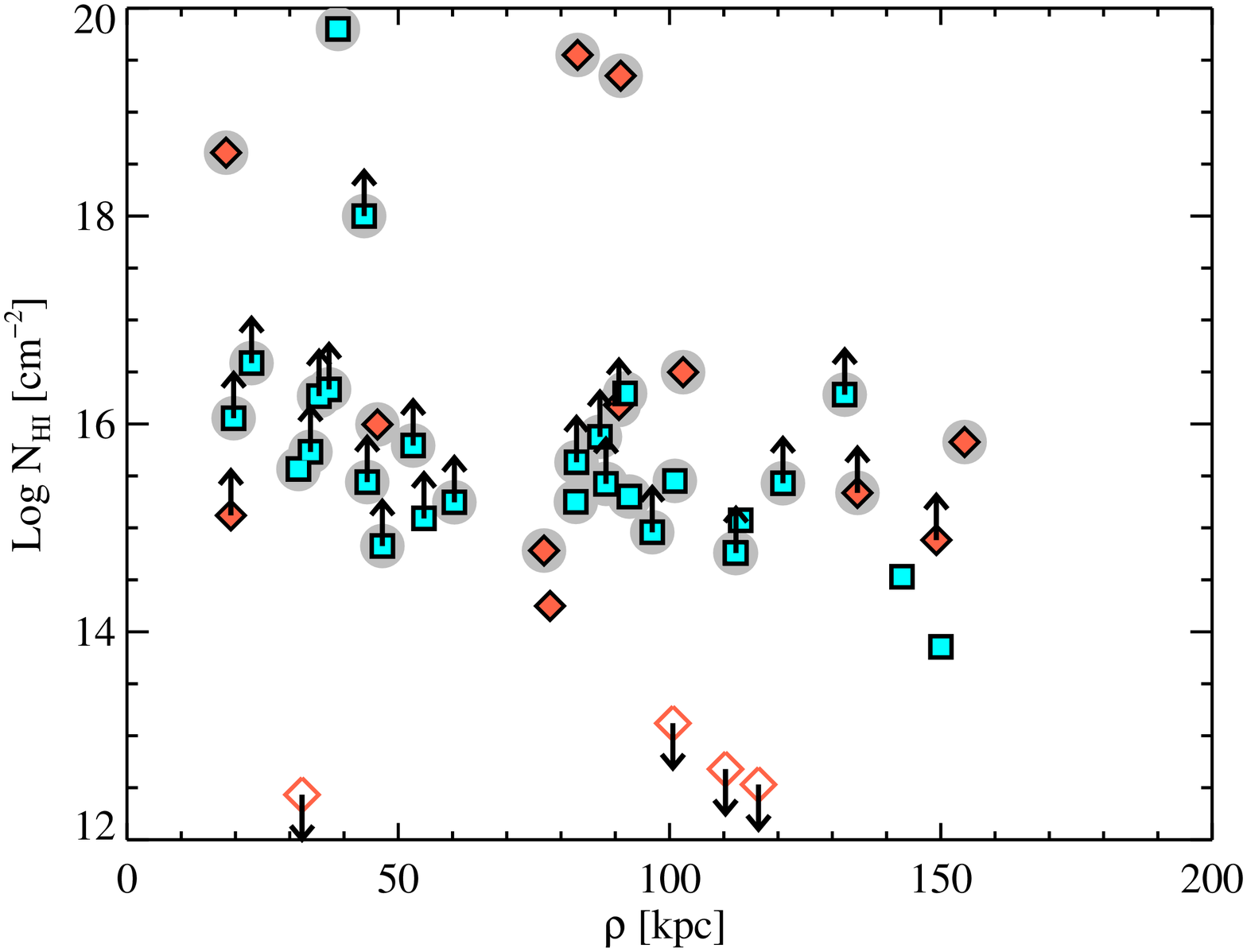} }
\caption{Two Figures showing (a) the HI values adopted in this analysis, and (b) the distribution of the HI column densities with the projected distance of the quasar sightline from the galaxy (R). We show the 33 sightlines included in our photoionization modeling analysis highlighted with light gray circles. As a practical matter, when log N$_{\rm HI}$ falls below 10$^{15}$ cm$^{-2}$, we are unable to assess the ionization state of the gas because there are no detected transitions of metal ions (Werk et al. 2013). Since most of the HI column densities from the COS-Halos data are lower limits owing to line saturation (see Tumlinson et al. 2013), we must sometimes adopt  an additional nominal (``preferred") value of the HI column density to perform the CLOUDY analysis. We detail our prescription for choosing a ``preferred" HI value in both Appendices, on a sightline-by-sightline basis. In general, we assumed as low a column density as possible that would be consistent with the observed absorption line strengths and physical limitations on the gas metallicity and ionization state (0.001 $<$ Z/Z$_{\odot}$ $<$ 1, and  -5 $< $ log U $<$ -1).  Ultimately, the sightlines we incorporate in our analysis span the full range of COS-Halos impact parameters, from  10 - 160 kpc.  }
\label{fig:hifigs}
\end{figure}

\subsection{Systematic Uncertainty Arising from Uncertainty in the Slope of the EUVB}

To carry out our analysis, we used the HM 2001 EUVB spectrum from galaxies and quasars for ease of comparison with previous results. Yet,  \cite{hm12} have updated their 2001 synthesis models with the addition of several new components. We show the 2012 updated HM EVUB (HM 2012) as a black line in Figure \ref{fig:cldyspec}. There are significant differences between HM 2012 and HM 2001. Most notably, HM 2001 exhibits a lower UV flux above $\sim$1.5 Ryd,  smaller spectral breaks at 1 and 4 Ryd, and a flatter soft X-ray spectrum. These differences arise primarily because of reduced HI and HeII Lyman Continuum absorption from a `sawtooth modulation' by the Lyman series of HI and HeII  that becomes more and more pronounced with increasing redshift. At low redshift, the differences between HM 2001 and HM 2012 are less pronounced than at high redshift. 

The most important difference for our purposes is that the spectral slope of HM 2001 at z$\sim$0.2 declines more steeply between 2 and 4 Ryd than that of HM 2012 (shown in Figure \ref{fig:cldyspec}). Below 2 Ryd, the slopes of HM 2001 and HM 2012 are approximately the same. Above 4 Ryd, HM 2001 is flatter than HM 2012. Thus, compared to HM 2012, HM 2001 is underproducing ions like SiIII and CIII  relative to the lower ionization potential ions (MgII and SiII are just above 1 Ryd in Figure \ref{fig:cldyspec}). Thus, gas ionization parameters need to be higher for HM 2001 to produce the observed SiIII  and CIII column densities. Repeating our analysis with HM 2012 generally has the effect of systematically lowering our gas ionization parameters by between 0.1 and 0.4 dex. Additionally, with our prior that gas not be super-solar, the new `preferred HI' must be raised by 0.2 - 0.4 dex for the cases in which this prior comes into play. If the HI is known, the preferred metallicity must be raised by 0.2 - 0.4 dex to be consistent with ionic column densities of higher ionization state ions. 

The change in the range of log U varies on a sightline by sightline basis and depends on which ions are used for the solution. For instance, sightlines that rely on SiII/SiIII for a determination of log U do not change significantly when re-analyzed with HM 2012 because SiIII has a an ionization potential of just over 2 Ryd, which is very close to where the slopes of HM 2001 and HM 2012 start to diverge. On the other hand, sightlines that rely on CII/CIII to estimate log U exhibit a greater change in their determined log U values because the ionization potential of CIII falls at an energy where the slopes of HM 2012 and HM 2001 are most different. 

In Figure \ref{fig:cldyspec}, we show with blue and green dotted lines a range of physically plausible slopes of the EUVB above 1 Ryd. We have varied the slope of the EUVB above 1 Ryd with  power law exponents between $-0.5$ (green)  and $-3.0$ (blue) to assess an overall systematic error in our methodology.  In general, a shallower slope of the EUVB tends to decrease the best-fitting ionization parameter to the measured ionic column densities of adjacent low-ions. However, the best fitting HI column density (for cases in which the AODM lower limits would give rise to a super-solar gas metallicity) also rises equivalently for many sightlines. The average systematic error in log U that arises from uncertainty in the slope of the EUVB is $\pm$0.3 dex, on average. The effect on the baryonic mass estimate is less pronounced. For example, a the total baryonic mass estimates of the cool CGM are a factor of 2.4 lower when we do the analysis with the shallowest slope considered (green dotted line in Figure \ref{fig:cldyspec}, power law exponent of $-0.5$) than the steepest slope considered (blue dotted line in Figure \ref{fig:cldyspec}, power law exponent of $-3.0$). Using HM 2012 lowers our baryonic mass estimates made with HM 2001 by a factor of 1.25. We also examined  models with non-equilibrium cooling and collisional ionization equilibrium from Gnat \& Sternberg (2007),  neither of which offer solutions consistent with our low-ion absorption line data.

\subsection{Additional Sources of Uncertainty}

\begin{figure}[h!]
\vspace{-1.5in}
\hspace{0.1in}
\begin{centering}
\includegraphics[height=0.90\linewidth,angle=90]{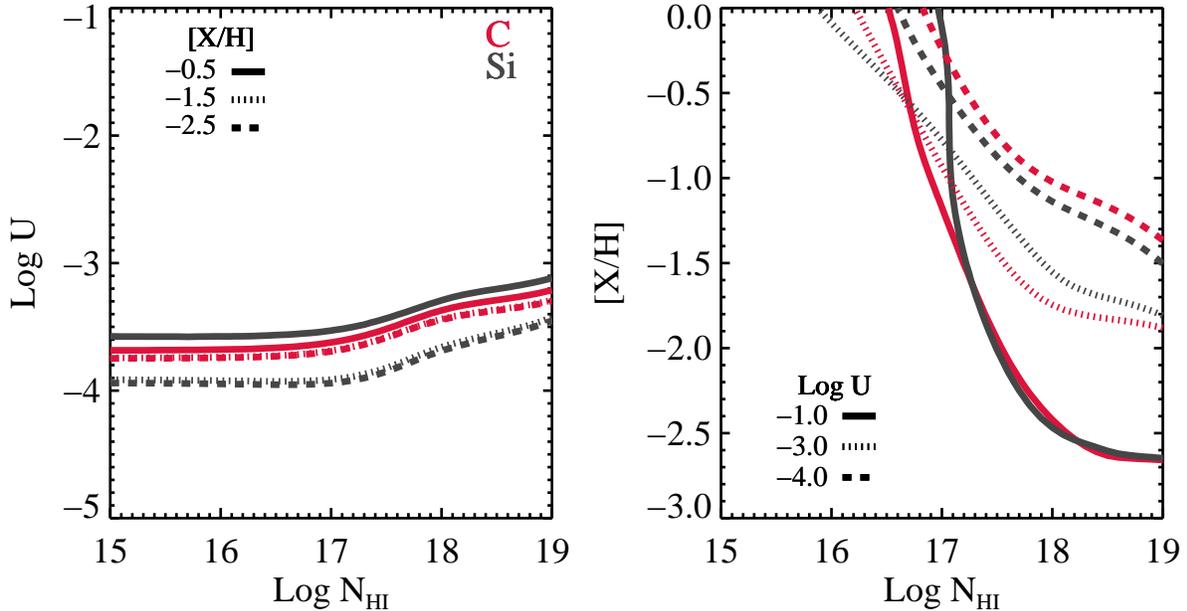}
\end{centering}
\caption{ log U vs. log  N$_{\rm HI}$ (left) and  [X/H] vs. log  N$_{\rm HI}$ (right) for CLOUDY grids at the typical column densities and ionic ratios for CII, CIII and SiII and SiIII.  On the left, we have shown several values of  [X/H], for reference, though the metallicity scarcely seems to affect the independence of log U with log N$_{\rm HI}$. log U is determined almost entirely from the ionic ratios of different ionization states of metal absorption lines. On the right, we show several values of log U in this parameter space which highlights that [X/H] becomes more dependent on log N$_{\rm HI}$ with increasing ionization parameter. Regardless of the ionization parameter, [X/H] is strongly dependent on the HI column density. }
\label{fig:meduz}
\end{figure}

A similar analysis incorporating CLOUDY modeling of absorption line data was carried out by Lehner et al. (2013) for 28 Lyman Limit systems at $z < 1$, with a focus on the metallicity distribution of the gas of HI-selected sightlines. Our analysis is distinct in several important ways.  First, our sample selection is based on the  COS-Halos survey which is galaxy-selected (Werk et al. 2012; Tumlinson et al. 2013), and  probes  circumgalactic gas of L $\sim$ L* galaxies at distances of 10 - 160 kpc from the host galaxies. As a result, we probe a larger range of HI column densities (Figure \ref{fig:hifigs}). Second, many of our HI column density measurements are limited by line saturation, and we  do not always obtain spectral coverage of the full Lyman series. The main side-effect of the uncertainty in the HI column densities is a concomitant uncertainty in metallicity measurements. Thus, we limit the bulk of our analysis to a discussion of the ionization parameter of the gas, which is mostly unaffected by the uncertainty in the HI column density. Figure \ref{fig:meduz} highlights that while the solution for the ionization parameter of the gas is the same across several orders of magnitude of HI column density, the solution for the metallicity is over 1.5 dex lower for the higher HI-column density gas. Given the uncertainty in the HI measurements, the metallicity of the gas is highly uncertain.  The ionization parameter, however, is mostly  independent of the choice of HI and metallicity. 



In this analysis, we have constrained the CLOUDY photoionization modeling with three key priors: 1. we assume solar relative abundances of the metals considered (C, N, O, Mg, Si, Fe); 2. we  do not consider gas metallicities above solar; 3.  we assume a log U less than $-$1, corresponding to a total gas density (n$_{\rm h}$) $>$ $\sim$10$^{-5.5}$ cm$^{-3}$.  Regarding the first prior, we acknowledge the possibility of non-solar abundance ratios by up to a factor of 2 (i.e. from dust depletion), which could in practice influence the derived metal abundances by the same factor. The ionization parameter of the gas is largely unaffected by departures from solar relative abundance ratios for two reasons. The first is that we frequently obtain column densities for two or more ions of the same element. We obtain the same ionization parameters (within the uncertainty given) when we consider only same-element pairs as we do when we consider all available ions. The second is that the CLOUDY model curves of ionic column density with log U are very steep, such that small changes in relative abundances have a minimal effect on the derived log U.  Regarding the second prior, we also acknowledge the possibility of super-solar circumgalactic gas metallicities, for example, in material recently ejected by a galaxy wind, or material stripped from a nearby interacting galaxy.  In general,  the COS-Halos sample was selected against galaxies with any indication of a recent merger event (Tumlinson et al. 2013), though at redshifts $>$ 0.1, it can be difficult to obtain the detailed morphological measurements necessary to rule them out completely. We note that none of our systems analyzed require super-solar abundances, and many of them do not allow it given the measured ionic ratios. This constraint affects approximately 1/5 of our estimates on the ionization parameter, providing a lower bound to our estimate of log U (and thus log N$_{\rm H}$) in these cases. In the following section, we note the specific cases for which this constraint comes into play.  Regarding the third prior, the upper bound on log U of $-$1 rarely affects our analysis as most of the ionic ratios we consider require log U $<$ -2. As a rule, gas with a larger (less negative) value of log U has a greater ionized gas fraction.  A greater ionized gas fraction, in turn, implies a greater total mass of gas contained in the CGM of  L $\sim$ L* galaxies. Thus, this limitation, if it has any impact on our analysis,  has the effect of making our final mass estimates more conservative.

In cases of saturated HI absorption lines, we attempt to place additional constraints on the HI column density limits from the COS-Halos survey in several ways, described in detail in the sightline-by-sightline analysis, and summarized broadly here. In general, we assume the lowest value of the HI column density allowed by the COS data (Thom et al. 2012). To obtain an additional upper limit on the HI column density, we determine whether the HI absorption line profile shows the presence of visible damping wings, the production of which generally occurs at column densities greater than 10$^{18.5}$ cm$^{-2}$.  In the absence of damping wings, we place this additional upper limit  on HI column densities. Second, we incorporate an allowed range of Doppler $b$ values, parameterizing the width of the absorption components to be between 1 and 70 km/s and model the Voigt profiles of the observed HI absorption lines within the allowed range of values. This limitation is most valuable in placing additional lower limits on the HI column density, where a single saturated component cannot have a $b$ value greater than 70 km/s. This analysis is consistent with the fits to the HI absorption lines presented in Tumlinson et al. (2013), which often give column densities above the simple AODM method determinations. Though we do not limit ourselves to the results of the Voigt profile fitting, we are  guided by them. Finally, on a case by case basis, we examine the column densities of the low ionization state metal absorption lines to see if their strengths are consistent with our model grid ranges for the adopted HI value.  In general, we choose a value for the adopted  HI to be as low as it can be while being consistent with the line profile fits presented by Tumlinson et al. (2013) and the requirements outlined above. We show these  adopted values as they compare to the COS-Halos AODM column densities in Figure \ref{fig:hifigs}a. Overall, these conservative choices mean our overall mass estimate for the CGM of L* galaxies will also be conservative.

\section{Notes on Individual Lines of Sight}

	 The figures shown below for each system display a cross section of the CLOUDY grid at the preferred \nhi~ and [X/H] determined by examining the absorption present in the COS data for each sightline and comparing it against our CLOUDY grid of output models in several different parameter spaces. These figures show the column density of the ions in the grid as a function of the most relevant quantity to our CGM mass determination, log U.  On the top x-axis, we show the gas volume density, n$_{\rm H}$ = $\Phi$/ Uc. Here, $\Phi$ is the total flux of ionizing photons ($\sim$ 1.21$\times$10$^{4}$ cm$^{-2}$ s$^{-1}$), as defined by the  \cite{hm01} EUVB,  and c is the speed of light.  We hold both the HI column density and the metallicity constant at their preferred values (described above), given in the lower right of the figure.  We incorporate our measured values for specific lower ionization state metal absorption lines into these figures by outlining in bold the allowed column density from the COS data over the grid column density curves. In cases where we have a good column density constraint on an absorption line, we place in bold the value plus or minus the error on the measurement. Upper and lower limits are indicated in bold over the full allowed range of column density.  On all of these plots, we have placed a bright yellow stripe of the range of log U values that fit all the available data given the preferred \nhi~ and [X/H]. In some cases, the stripe may appear to be larger than the intersection of the low ion column density measurements constrain it to be since we add additional uncertainty depending on how well we can constrain \nhi~ and [X/H]. Here we assess each line of sight and its solution in the CLOUDY model grid, ultimately rating it with a quality flag between 1 and 5 (5 being the best, described in Section 1 of the main text body). We show OVI model lines and column density measurements for reference, but do not require that the selected ionization parameter account for the total observed column density.

	 We describe which metal ions are most useful in constraining the ionization parameter and metallicity, and how we arrived at our adopted HI  and [X/H] values, and ultimately how we constrain the ionization parameter of the circumgalactic gas. Higher values of log U imply higher ionization corrections to the HI, and higher CGM mass. For our conservative mass estimate, we therefore adopt the lowest allowed log U, the range of which is given in Table A1. When \nhi~ is not well constrained by the COS-Halos data, we chose a solution for log U at the lowest \nhi~ possible. The effect of this conservative choice often requires adopting the highest allowable value of  the gas metallicity.

\begin{figure}[h!]
\vspace{0.15in}
\begin{centering}
\includegraphics[width=0.95\linewidth]{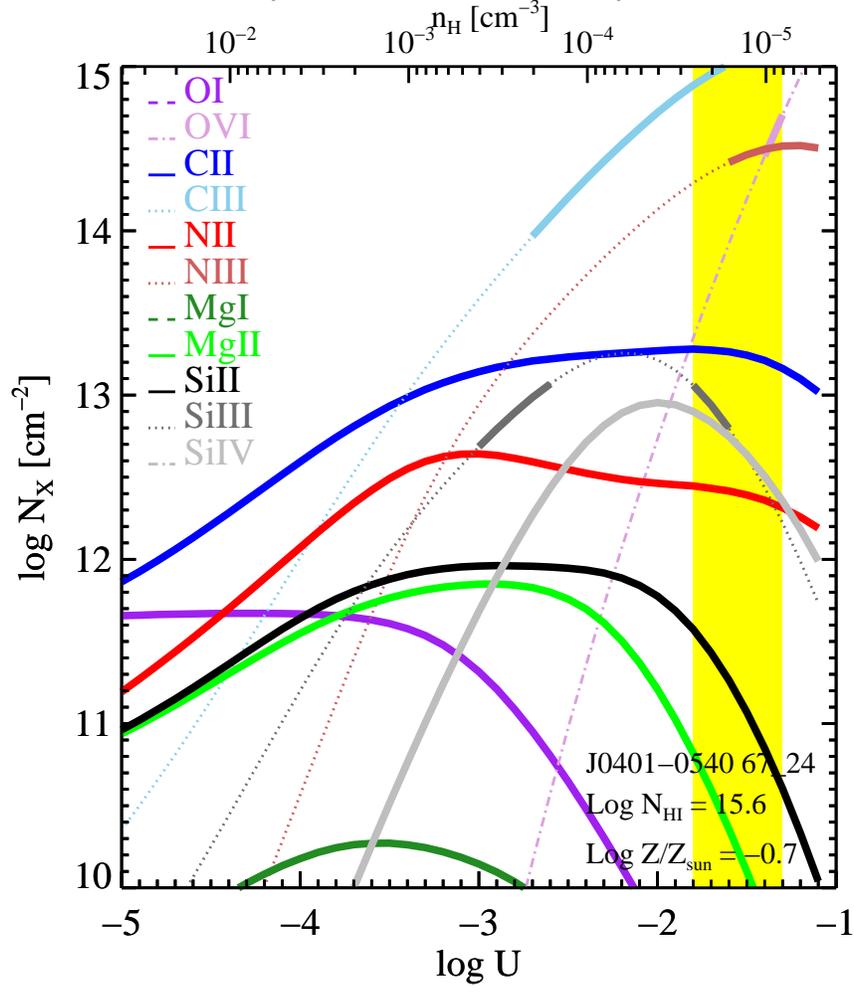}
\end{centering}
\caption{J0401-0540 67\_24: The COS spectral data cover the Lyman series down to Ly$\epsilon$ ($\lambda$ 937 \AA), placing a lower limit on \nhi~of 15.6 since all lines are saturated, though not to a large degree. The absence of damping wings in this system lead us to adopt an  \nhi~ of 15.6. A combination of good measurements of the  SiIII and NIII column densities, along with the lower limit on CIII absorption from the saturated line at 977 \AA~ allow us to constrain the solution for log U  to be between -1.8 and -1.3.  At the preferred HI column density of 15.6, the solution for [X/H] is -0.7, constrained tightly by the small overlap region between NIII and SiIII column densities in the cloudy parameter space of log U vs. [X/H]. If we were to adopt a higher value of \nhi~, the solution for metallicity could drop substantially, but the solution for log U would remain near -1.5 (see figure \ref{fig:meduz}). We characterize this solution for log U as being of high quality, and thus rate it a 5. 
}
\label{fig:cloudyfirst}
\end{figure}

\begin{figure}[ht]
\vspace{0.15in}

\begin{centering}
\includegraphics[width=0.95\linewidth]{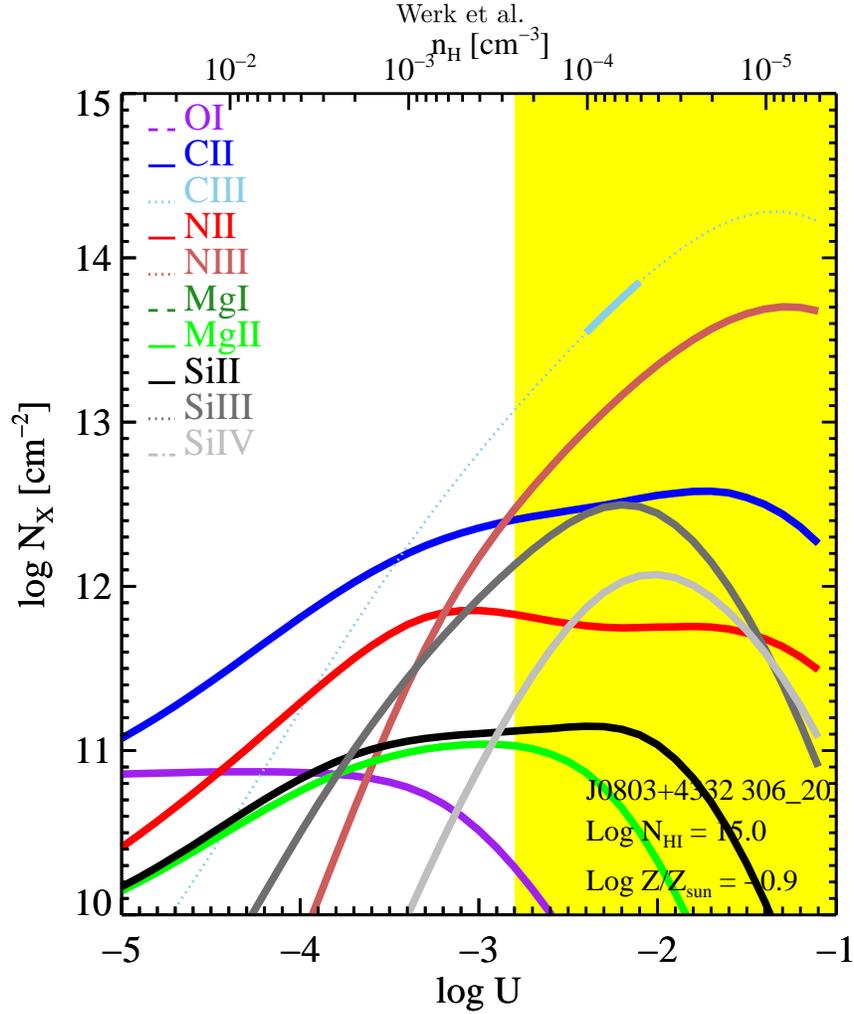}
\end{centering}
\caption{J0803+4332 306\_20: 
 The COS data cover both Lyman $\alpha$ (saturated) and Lyman $\beta$ (good detection), allowing us to constrain \nhi~ to be 14.8 (in the optically thin regime, and rescaled to 15.0 in this figure). The only low ionization state metal line we detect is CIII, which along with the upper limit on CII,  can only constrain  log U to be greater than -3.5. The additional constraint on log U comes from the prior that at the adopted HI column density, the gas is unlikely to be super-solar, and thus the range of allowed log U is additionally constrained to lie above -2.8. The shape of the carbon absorption line traces that of the HI and lies over the same velocity range. We rate this system as only mediocre, giving it a quality flag of 2 since the solution is based upon a single detection of CIII combined with a non detection of  SiIII, and does not tightly constrain log U. 
}
\label{fig:cloudyb}
\end{figure}

\begin{figure}[ht]
\vspace{0.15in}

\begin{centering}
\includegraphics[width=0.95\linewidth]{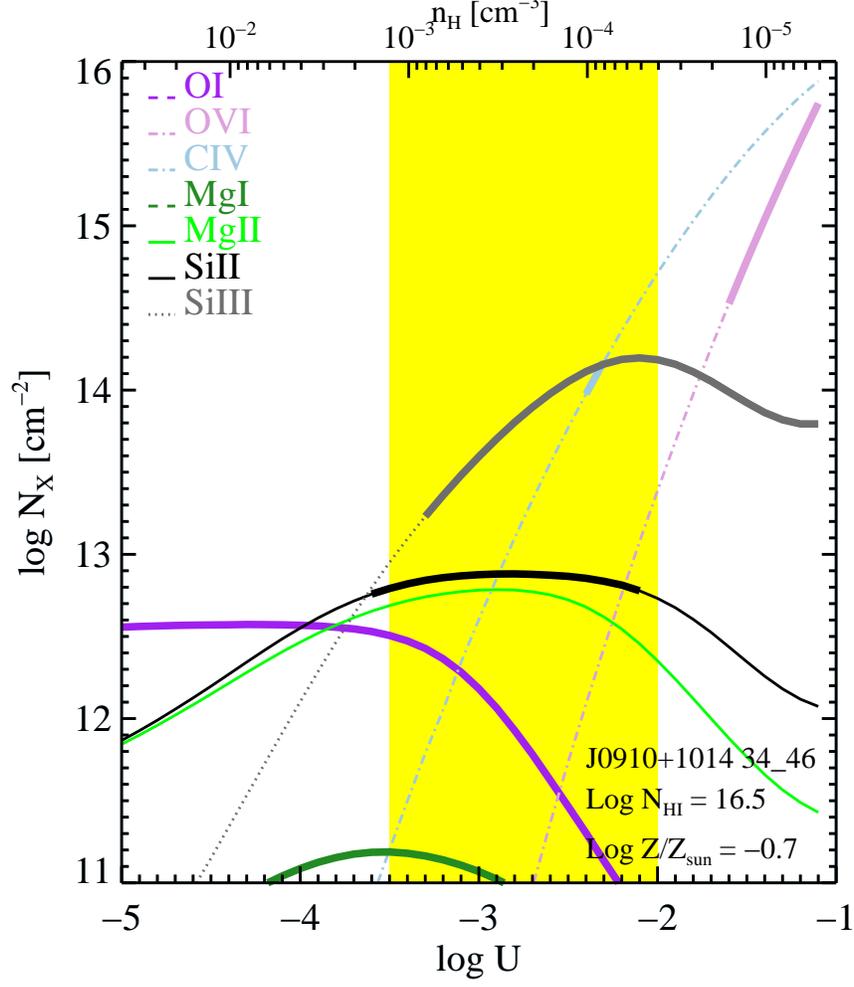}
\end{centering}
\caption{J0910+1014 34\_46: The COS Data cover only Ly$\alpha$ and Ly$\beta$, both of which are badly saturated.  The AODM method gives a lower limit to \nhi~ of 15.0, though the Voigt profile fits indicate the column density is closer to 16. The absorption exhibits no damping, and so we adopt 18.5 as our upper limit to \nhi. The lower limit of MgII column density requires that the \nhi~ be greater than 16.5 cm$^{-2}$ in order for the gas not to be super solar. However, the solution at all allowed values of \nhi~ is inconsistent between SiII and MgII, with SiII preferring lower metallicity at the same value of \nhi~ as MgII over the full range of allowed HI column density. We show the lower metallicity solution here, so we can compare the ionic ratio of the same element (SiII/SiIII). The solution for log U  based on MgII/SiIII and MgII/CIV  lies consistently between -3.5 and -2.5, which we adopt here as our range. We note that MgII/SiIII exhibits the same ionization parameter range only for higher adopted model metallicities. CIV is also consistent with this solution at the upper end of the log U range.  We rate this system a 3 due to the inconsistency between MgII and SiII. }
\label{fig:cloudyc}
\end{figure}

\begin{figure}[ht]
\vspace{0.15in}

\begin{centering}
\includegraphics[width=0.86\linewidth]{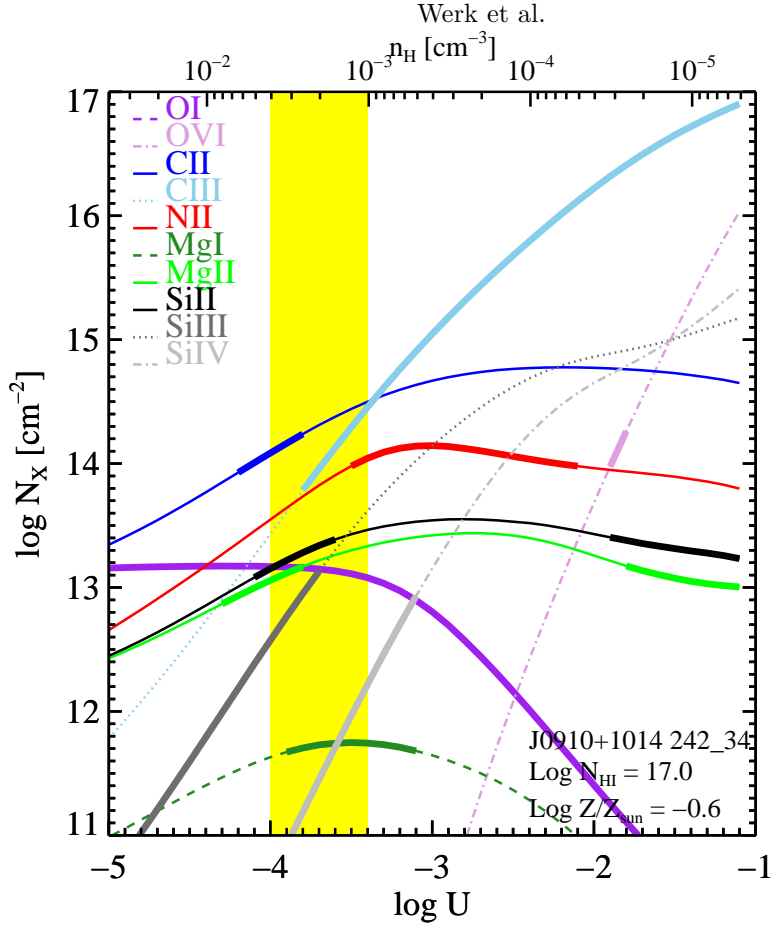}
\end{centering}
\caption{J0910+1014 242\_34: The COS Data cover nearly the full Lyman series in this system, though the signal to noise ratio drops precipitously below Ly$\gamma$, which is saturated. The AODM method gives a lower limit to \nhi~ of 15.3, though the Voigt profile fits indicate the column density is closer to 16.5. Here, we are justified in adopting a higher value of \nhi, since the measured column density of NII requires a \nhi~ of 16.6 under the assumption that the gas is not super-solar in its metallicity and the solution for the full set of lower ionization state metal lines here does not exhibit self-consistency below an \nhi~ of 17.  The HI absorption does not exhibit damping wings, and so we can place an additional upper limit on \nhi~ of 18.5.  The HI absorption is well fit by four components, with the component at a velocity offset from the host galaxy of $\sim$-180 km/s being the strongest. The saturated CIII absorption exhibits the same shape as the HI over the full velocity range. However,  the SiII, MgII, and NII are detected only in the strongest component at -180 km/s. The solution for log U lies between -4 and -3.4, based on MgII, SiII, SiIII, CII, CIII, and NII detections. The NII detection is slightly inconsistent with the other low ions in this adopted model, which is why the adopted range of log U is large here.  [X/H] is set to be -0.7 at the adopted \nhi, though at higher values of \nhi~ [X/H] drops accordingly.   Over the range of \nhi~ we consider here, 17 - 18.5, the solution for log U remains constant. We rate the solution for this system with a 4, which is good, but complicated by both the HI saturation, and the moderate offset of NII from the solution given by the rest of the low ions. 
}
\label{fig:cloudyc}
\end{figure}

\begin{figure}[ht]
\vspace{0.15in}

\begin{centering}
\includegraphics[width=0.95\linewidth]{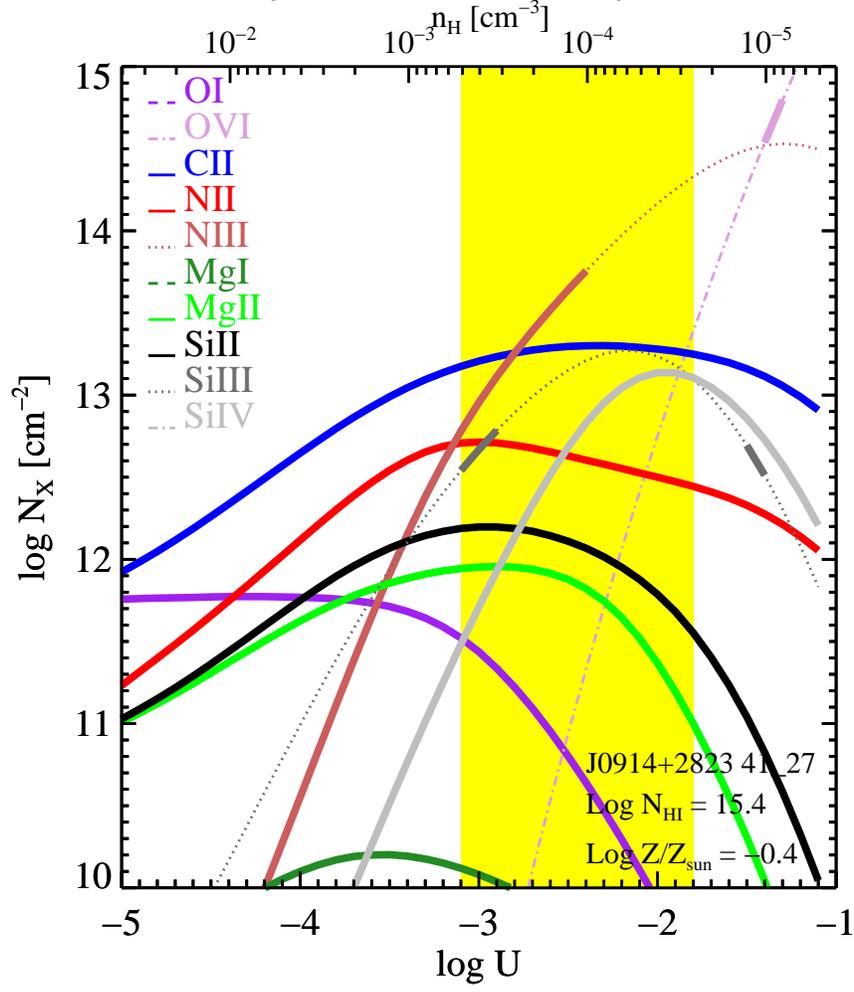}
\end{centering}
\caption{J0914+2823 41\_27:  \nhi~ is well determined  to be 15.45 from the COS data, with a detection of the Lyman series down to 930 \AA. A detection of absorption from SiIII, along with a non-detections of NIII constrain log U to be below -1.8 over the full range of metallicity considered.  Additionally, the non-detection of MgII requires [X/H] $<$ -0.4 at this \nhi~ in order for this limit to be consistent with the SiIII and NIII. At this [X/H], the lowest possible value of log U, based on SiIII is then $-$3.1, which is the lower bound we adopt here.  Because the constraints on log U are not particularly tight here, we rate this solution as mediocre, and give it a quality flag of 3. 
}
\label{fig:cloudye}
\end{figure}

\begin{figure}[ht]
\vspace{0.15in}

\begin{centering}
\includegraphics[width=0.95\linewidth]{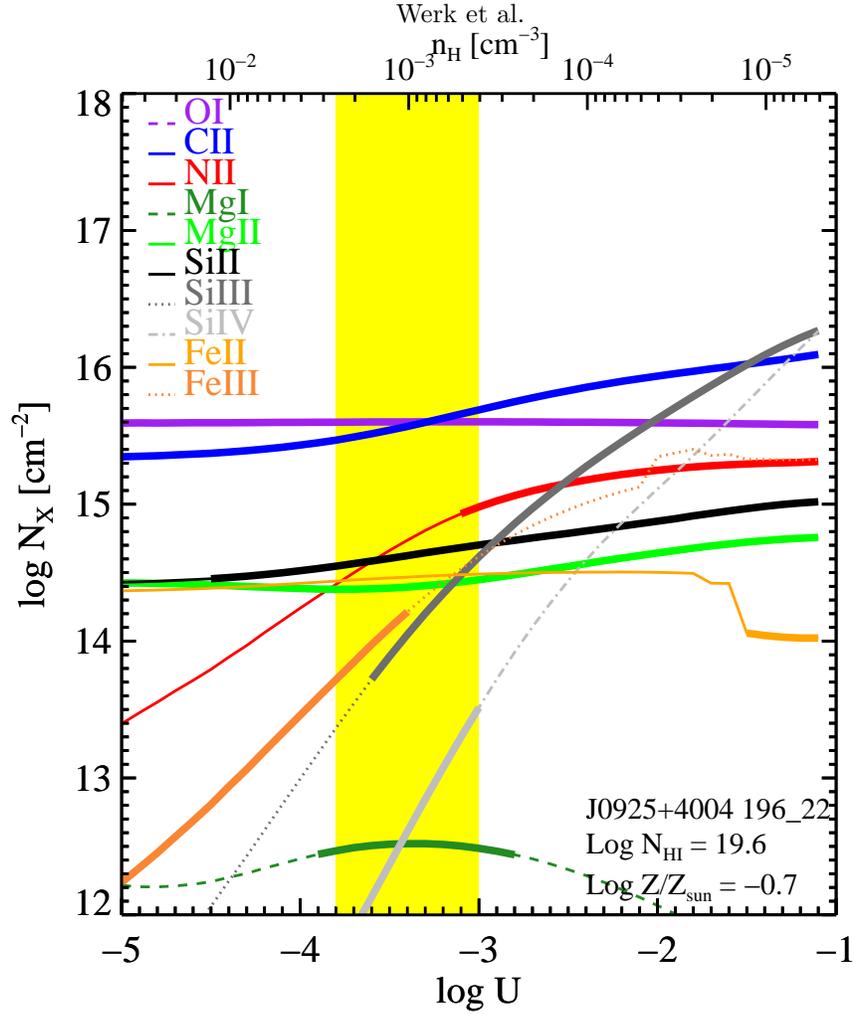}
\end{centering}
\caption{J0925+4004 196\_22: The HI Ly$\alpha$ absorption of this system has well-defined damping wings, allowing a measurement of  \nhi~ to be 19.6 from the Voigt profile fitting. Many of the low ionization metal species are also saturated, offering lower limits to their column densities. The one metal ion we can measure well is FeII  $\lambda$1144\AA, with log N of 14.1. Because OI is saturated, with a lower limit on its column density of 15.7, [X/H] is constrained to be greater than  -0.7 at this HI column density. The lower limit on MgI is also consistent with this metallicity. However, at [X/H] of -0.7 and higher, the one FeII detection is inconsistent with the lower limits on the other lower ionization state metal lines. We consider that this inconsistency may be due to the depletion of iron onto dust grains. Ignoring the FeII measurement constrains the log U to lie between -3.8 and -3.2, the upper bound set by SiIV, and the lower bound set by both NII (at higher metallicity) and SiIII. Due to the necessity of invoking depletion of iron for the solution to be self-consistent, we rate this system to be mediocre, with a quality flag of 3. 
}
\label{fig:cloudyd}
\end{figure}

\begin{figure}[ht]
\vspace{0.15in}

\begin{centering}
\includegraphics[width=0.95\linewidth]{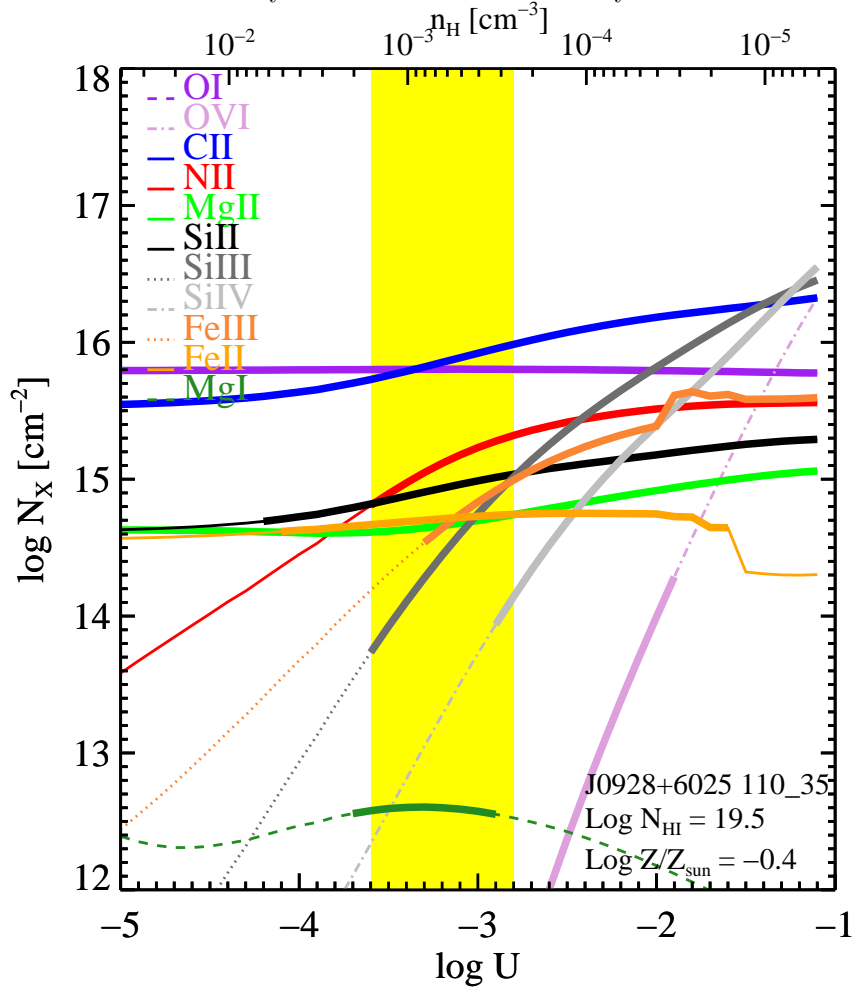}
\end{centering}
\caption{J0928+6025 110\_35: Like the previous system, the HI Ly$\alpha$ absorption here has well-defined damping wings, allowing a measurement of  \nhi~ to be 19.5 from the Voigt profile fitting. Nearly all of the low ionization state metal lines are saturated in the data, except for MgI at 2852 \AA, with a log N of 12.75. At these columns, with \nhi~ of 19.5, the solution for [X/H] is between -0.5 and solar, with a preference at the lower end due to better consistency with the MgI absorption. At higher metallicities, MgI tends to move toward lower log U, whereas the lower limits for the other low-ions push the solution to higher log U.  We  then set the lower bound on log U based on the SiIII and NII  lower limits and the MgI measurement,  and the upper bound on log U based on the MgI line allowing a maximum log U of $-$3 at this \nhi.  In past versions of CLOUDY, single-phase ionization models tended to under-predict the MgI strength relative to the MgII strength, and a separate lower temperature/higher density phase was proposed to be associated with MgI (e.g. Churchill et al. 2003). However, newer models of CLOUDY (as the one used here) are now able to recover MgI and MgII in a single phase as a result of improved rate coefficients for charged transfer reactions (Klingdon et al. 1996; Narayanan 2008). We rate this system with a quality flag of 4, since the solution for log U is well-bounded, but based on a limited set of ions owing to saturation. 
}
\label{fig:cloudye}
\end{figure}

\begin{figure}[ht]
\vspace{0.15in}

\begin{centering}
\includegraphics[width=0.95\linewidth]{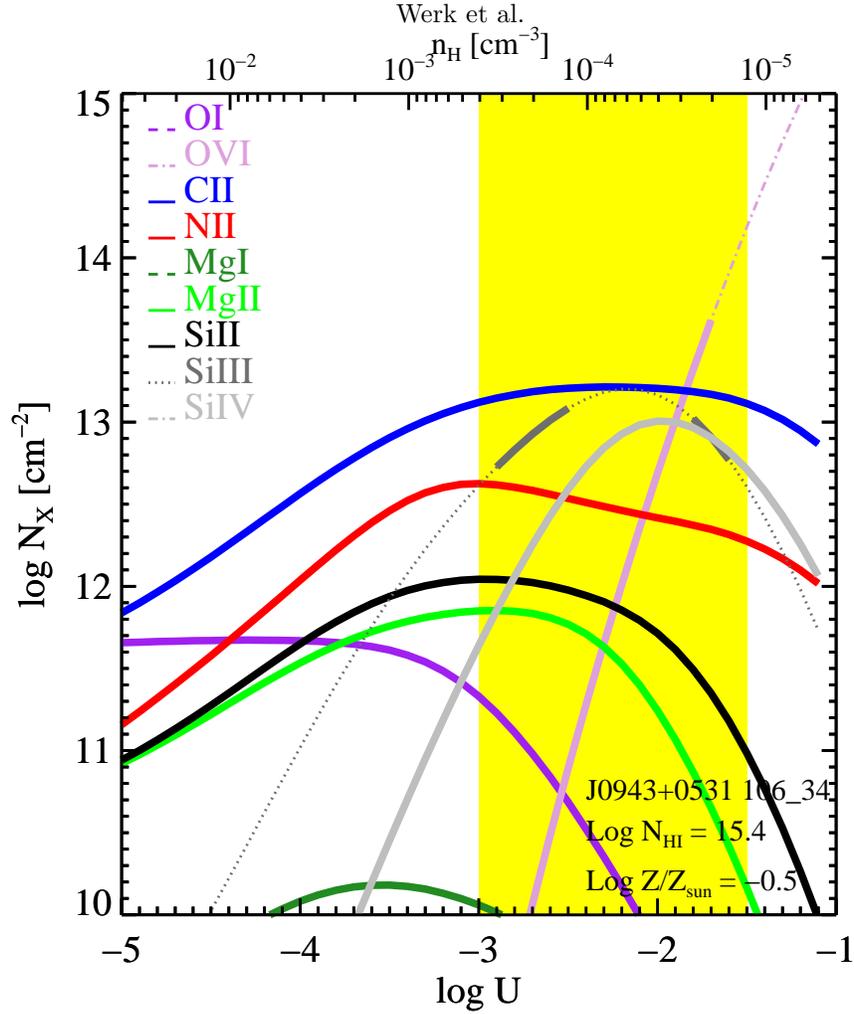}
\end{centering}
\caption{J0943+0531 106\_34:  The HI of this system is covered down to Ly$\delta$ in the COS data, which is still saturated to a small degree. The Voigt profile fits to the Lyman series show two components, with the strongest absorption feature at 150 km/s. The fits are consistent with the lower limit from the AODM method, giving a \nhi~ of 15.4, which we adopt here. The only low ion detection is SiIII, with its absorption profile matching that of the strongest HI absorption component at 150 km/s. From the SiIII detection, combined with the non-detection of several other species (CIII is blended with the Milky Way NI triplet, so we do not use it here, and NIII is wiped out by Milky Way Ly$\alpha$), including a non-detection of MgII, constrain [X/H] to be between -1.0 and -0.1, with a log U between -3 and -1.5. Given the poor constraint on log U, we rate this system with a quality flag of 2. 
}
\label{fig:cloudyf}
\end{figure}

\begin{figure}[ht]
\vspace{0.15in}

\begin{centering}
\includegraphics[width=0.95\linewidth]{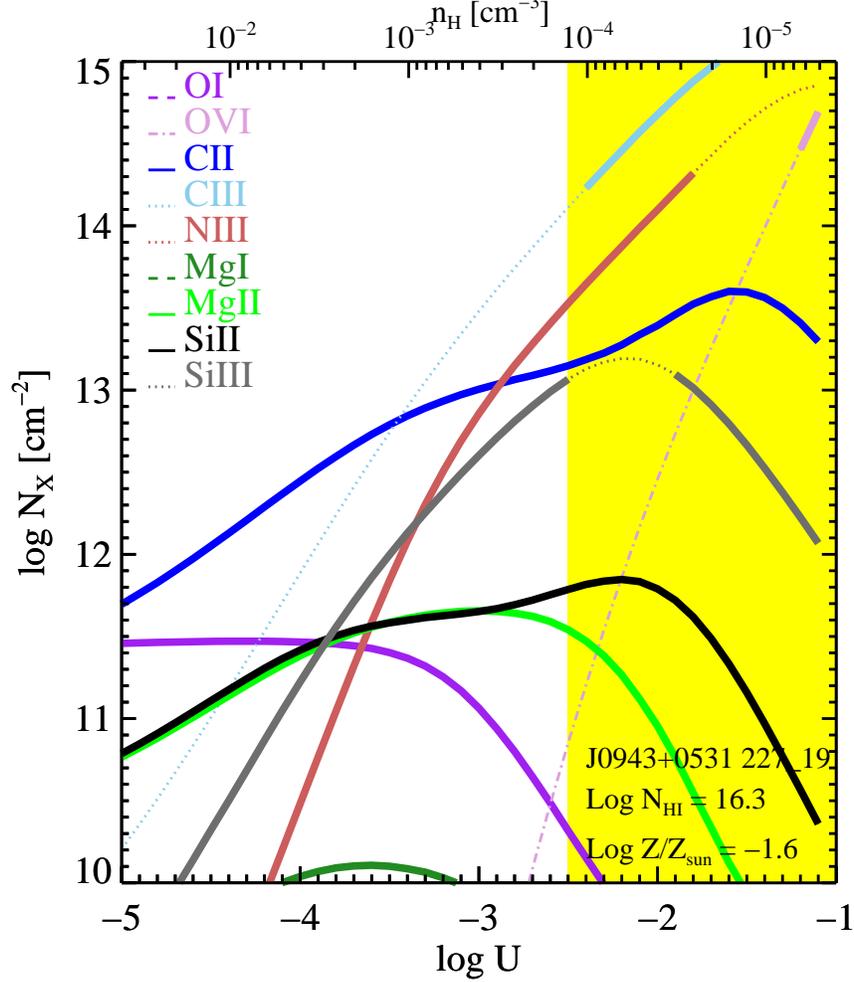}
\end{centering}
\caption{J0943+0531 227\_19:  We detect the HI Lyman series absorption lines down to 926 \AA, allowing a measurement of \nhi~ to be 16.3. The HI absorption for this system is fit by many components, ranging in centroid velocities from 65 km/s to 800 km/s. The strongest component of the absorption is at 350 km/s, which is also where we see the strongest absorption in CIII. Unfortunately, due to the high redshift $z\sim0.35$) of this system compared to the majority of our targets (median $z\sim0.2$), many of the Silicon lines fall in a part of the COS detector where the S/N is not very high, which means our analysis is slightly hampered by upper limits. The CII of this system at 1036 \AA~ is blended with a detection OVI at 1037 \AA, and so we incorporate it here as an upper limit. As it turns out, the range of non-detections combined with the CIII lower limit work to constrain log U to be between -2.5 and -1. Setting the lower bound on U is primarily the  CIII lower limit combined with the SiIII upper limit.  The [X/H] of this system is constrained to be lower than -1.6, mostly due to the SiIII upper limit which becomes inconsistent with the CIII and NIII solution at [X/H] above this value. We show the solution at this value of metallicity to show where the solution begins to diverge, as at lower [X/H], everything becomes considerably more self-consistent at log U $>$ -2.5. Given the range of allowed log U values, along with the consideration that this range is based on only upper and lower limits on low ion absorption, we rate this system with a quality flag of 3. }
\label{fig:cloudyg}
\end{figure}

\begin{figure}[ht]
\vspace{0.15in}

\begin{centering}
\includegraphics[width=0.95\linewidth]{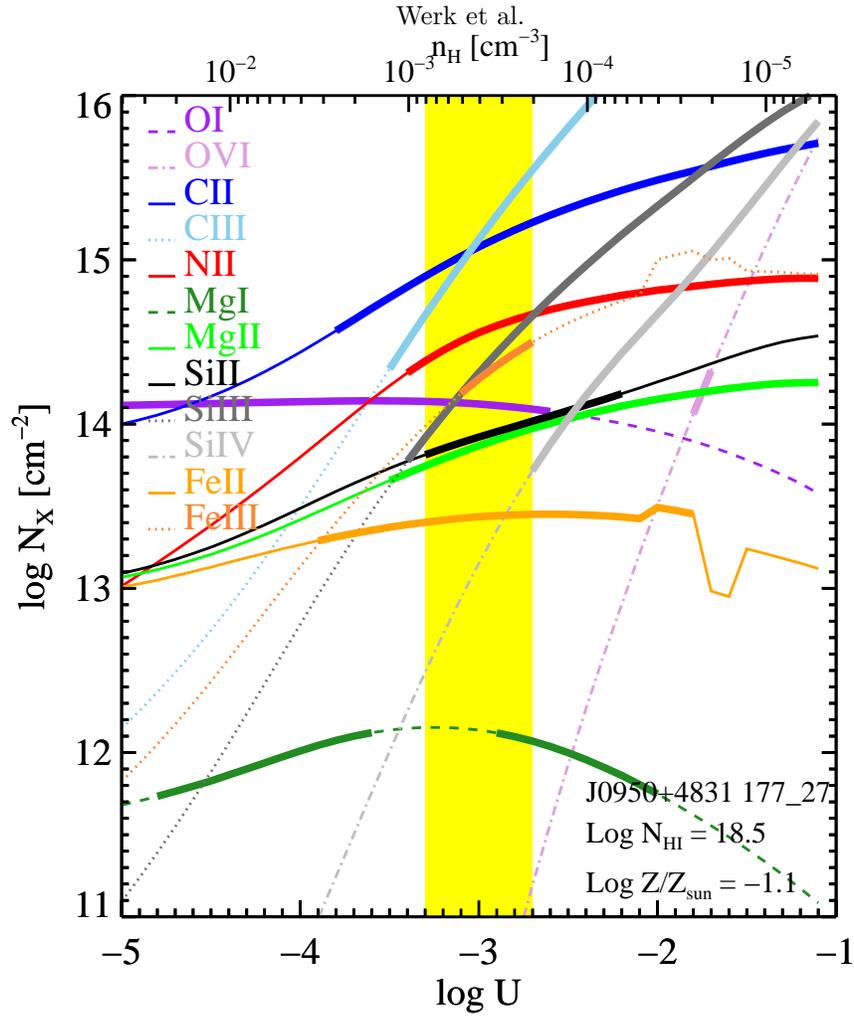}
\end{centering}
\caption{J0950+4831 177\_27: The HI of this system is detected  and saturated in the Lyman series down to 937 \AA~ in the COS data, with an AODM lower limit on the \nhi~ of 16.2. The absorption is best fit by two components, with a total \nhi~ of 16.9. The strongest component has a rather large Doppler $b$ value in this fit of 60 km/s, and thus the column of 16.9 should be seen as a strict lower limit in this case. The system may have a slight amount of damping in its strongest Ly$\alpha$ component. Furthermore, good detections of OI  and MgI absorption lines are consistent with each other at a \nhi~ of 18.5, and a [X/H] of $\sim-1$. Solutions at lower \nhi~ and higher [X/H] diverge considerably, and thus we consider 18.5 to be the lowest value of \nhi~ at which the CLOUDY models fit the data.  Many  low ion detections, including the ratio of FeII/FeIII, help to better constrain the metallicity and ionization parameter to be near a tenth solar, and between -3.3 and -2.7, respectively.  Although the HI column density measurement is made uncertain by saturation in the COS data, the level of consistency of the low ion detections with a solution at  \nhi~ of 18.5, and a [X/H] of $\sim-1$ is very convincing, and we rate this solution to be good, with a quality flag of 4. 
}
\label{fig:cloudyh}
\end{figure}

\begin{figure}[ht]
\vspace{0.15in}

\begin{centering}
\includegraphics[width=0.95\linewidth]{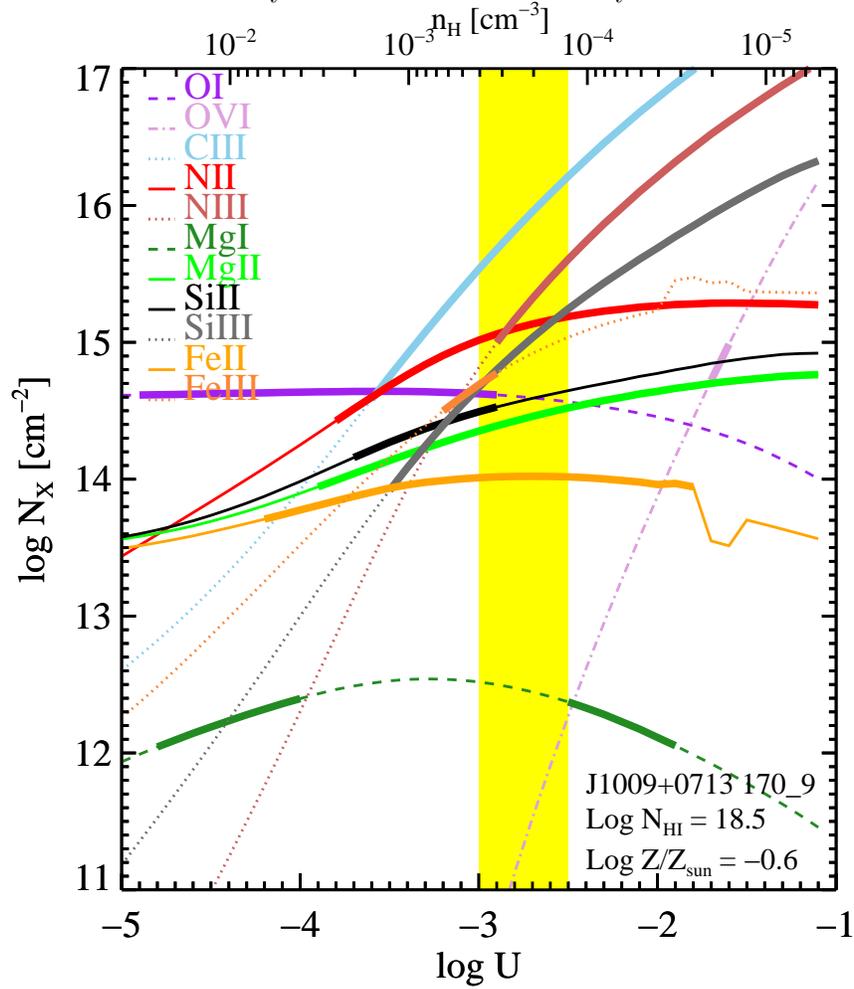}
\end{centering}
\caption{J1009+0713 170\_9:  The lower limit on \nhi~ is determined to be 18.0 from the saturated COS data, with the Lyman series saturated down to the Lyman limit. The Voigt profile fit gives  \nhi~ of 18.4. The  MgI, OI, SiII, FeII and FeIII, which are the key lines forming  the basis for the solution in this case, do not give consistent solutions at column densities below 10$^{18.9}$ cm$^{-2}$.  At this column density, the detection of MgI and the saturation of OI at 1302 \AA constrain [X/H] to be close to a tenth solar, while the ratio of FeII to FeIII in this range sets the log U to be between -3 and -2.5. However, we are concerned that the COS data for this system do not show well-defined damping wings in the Ly$\alpha$ absorption as we would expect at this column density, and so we nonetheless adopt the lower value of \nhi~ at 18.5, consistent with the Voigt profile fit. At this column density and metallicity, the MgI is inconsistent with the lower limit on the OI absorption, although the rest of the low ions remain consistent.  Despite the MgI inconsistency (which may be complicated to model for reasons stated above), and due to the relatively narrow range of allowed ionization parameter values, we give this solution a quality flag of 4. 
}
\label{fig:cloudyi}
\end{figure}

\begin{figure}[ht]
\vspace{0.15in}

\begin{centering}
\includegraphics[width=0.95\linewidth]{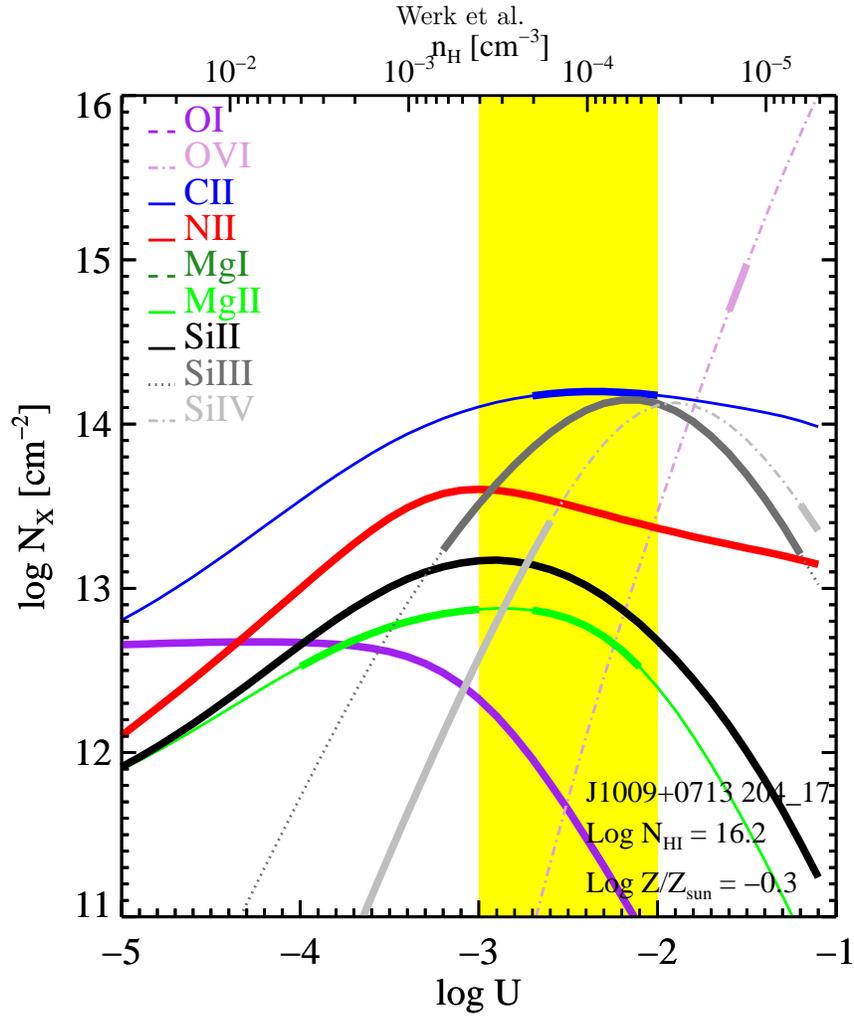}
\end{centering}
\caption{J1009+0713 204\_17:  The lower limit to \nhi~ from the AODM method is 15.25,  and the best Voigt profile fits to the Ly$\alpha$ and Ly$\beta$ absorption lines are consistent with this value as well. Due to the clear absence of damping wings, and the equivalent width of the strongest component being 650 m\AA, we place an additional upper limit on \nhi~ to be 18.2. The detection of CII at both 1036 and 1334 \AA~ 
leads us to adopt a \nhi~ of 16.2, since all solutions below this limit are super solar. The detection of MgII constrains [X/H] at this HI column to be close to solar as well, and the SiII and SiIII are consistent at these values. Given the large range of allowed \nhi~ (at higher values, the corresponding solution for log U increases slightly), the range of acceptable log U for the different ionization species lies between -3 and -2, and we therefore give this system a quality flag of 3. 
}
\label{fig:cloudyj}
\end{figure}

\begin{figure}[ht]
\vspace{0.15in}

\begin{centering}
\includegraphics[width=0.95\linewidth]{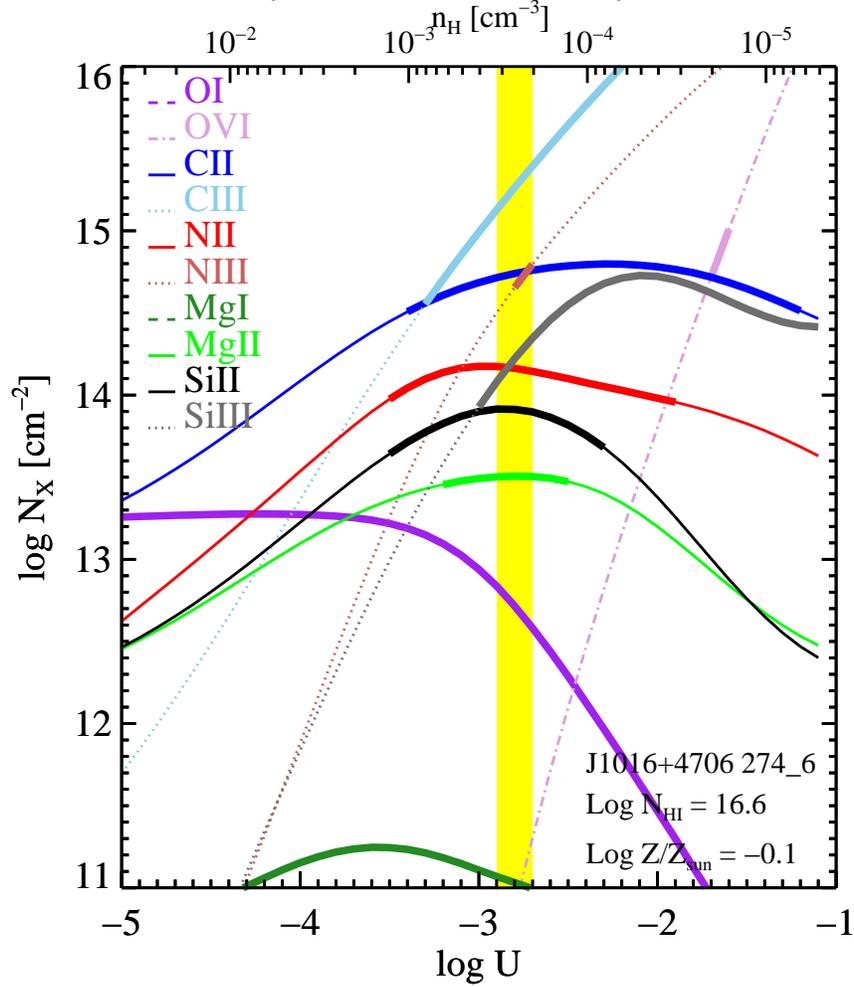}
\end{centering}
\caption{J1016+4706 274\_6:  The lower limit to \nhi~ from the AODM method is 16.6, yet the Voigt profile fits give a value of 17.9 based on two component fits to absorption lines in the Lyman series down to 930 \AA.  The absence of damping wings in the HI absorption lines allows us to place an additional upper limit of 18.5 on the value of \nhi. There are a number of detections of adjacent ionization states of Si, C, and N which allow us to independently tightly constrain log U. Regardless of the preferred HI, which we conservatively adopt to be 16.6 in this case, and the preferred [X/H], for which the data give a consistent solution at -0.1, the log U is between -2.9 and -2.7, and is thus very tightly constrained. We give this system our highest quality flag of 5. 
}
\label{fig:cloudyk}
\end{figure}

\begin{figure}[ht]
\vspace{0.15in}

\begin{centering}
\includegraphics[width=0.95\linewidth]{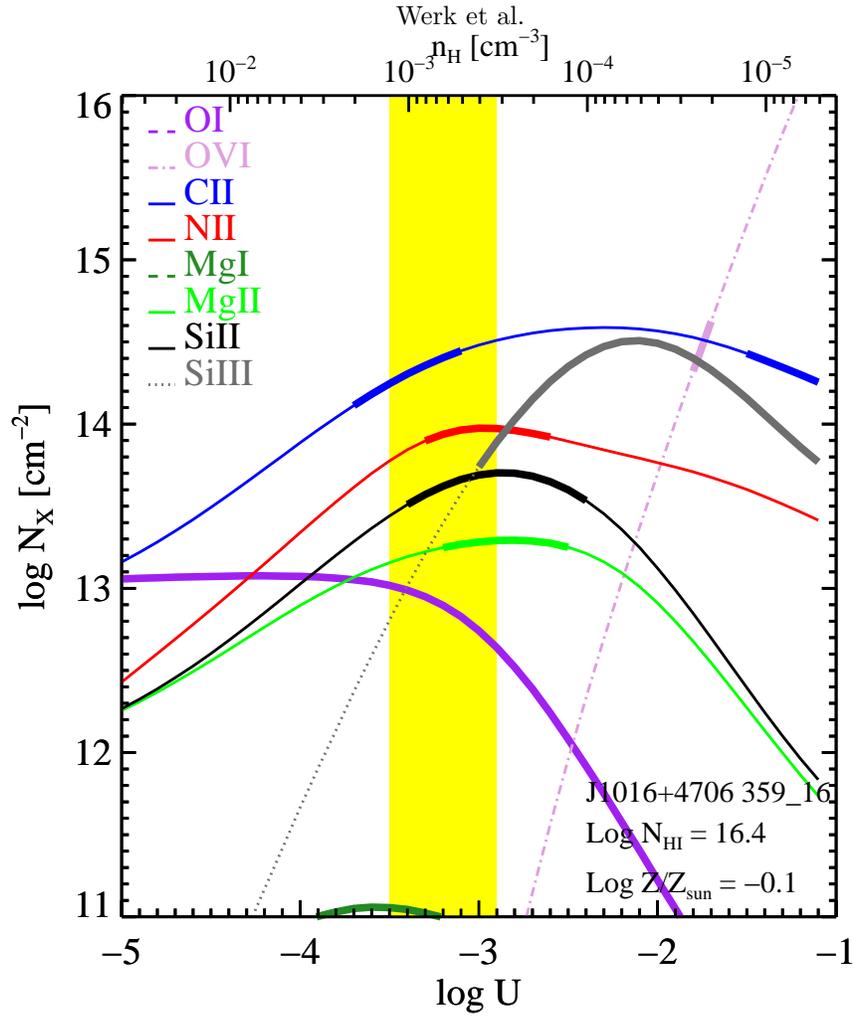}
\end{centering}
\caption{J1016+4706 359\_16: The COS data cover only Ly$\alpha$ and Ly$\beta$ in this system, both of which are saturated, giving a lower limit to \nhi~ of 15.4 in the AODM measurement. The single component Voigt profile fit to the data gives a best-fit value of 17.3, with a  Doppler $b$ value of 42 km/s. Solutions for [X/H] go super solar below a \nhi~ of 16.4, which we adopt here. The range of log U is well constrained by our low ion column density measurements, especially CII combined with SiII/SiIII,  and is found to lie between -3.5 and -2.9. We rate this system with a quality flag of 5 given the consistency of the result and the tightness of the allowed log U range, which is consistent across the full allowed range of ~\nhi. 
}
\label{fig:cloudyl}
\end{figure}

\begin{figure}[ht]
\vspace{0.15in}

\begin{centering}
\includegraphics[width=0.95\linewidth]{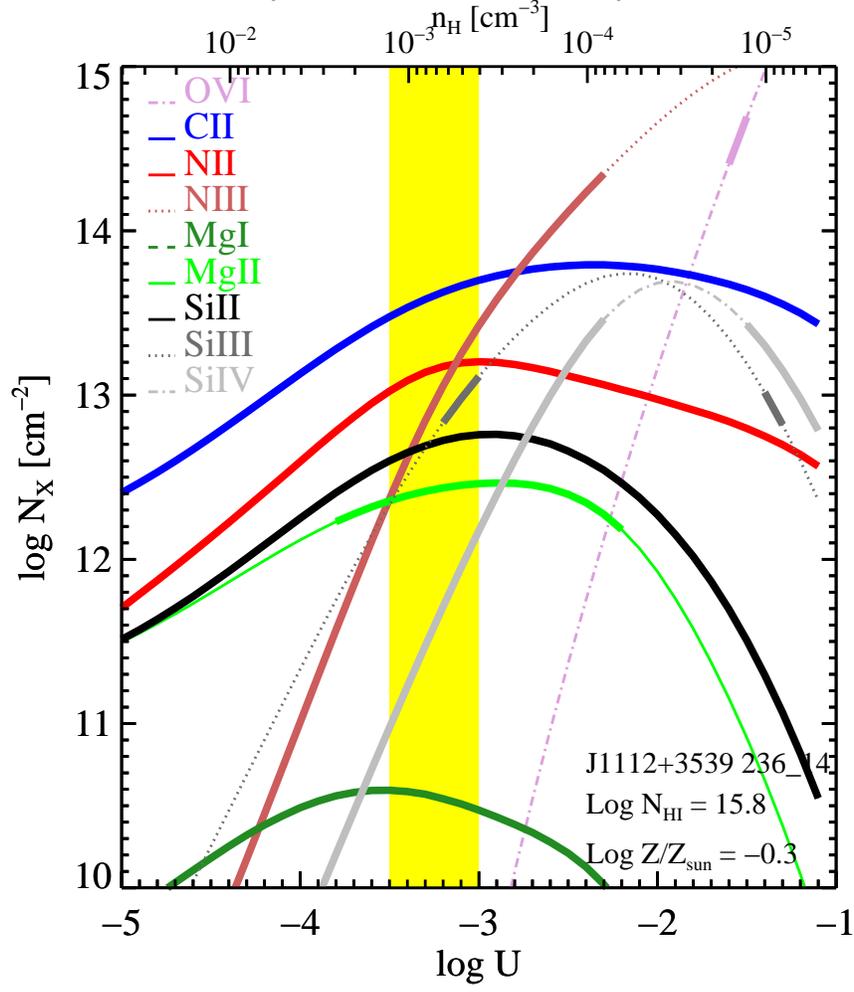}
\end{centering}
\caption{J1112+3539 236\_14: The COS data cover the Lyman series down to 937 \AA, which provides a lower limit on \nhi~ of 15.8.  The Voigt profile fits to the absorption is consistent with this value, and show no damping wings. Good detections of MgII and SiIII help pin down the solution for log U in the range to be between -3.5 and -3 over the full range of allowable HI column density. This solution is both consistent and tight, earning it our highest rating of 5. 
}
\label{fig:cloudym}
\end{figure}

\begin{figure}[ht]
\vspace{0.15in}

\begin{centering}
\includegraphics[width=0.95\linewidth]{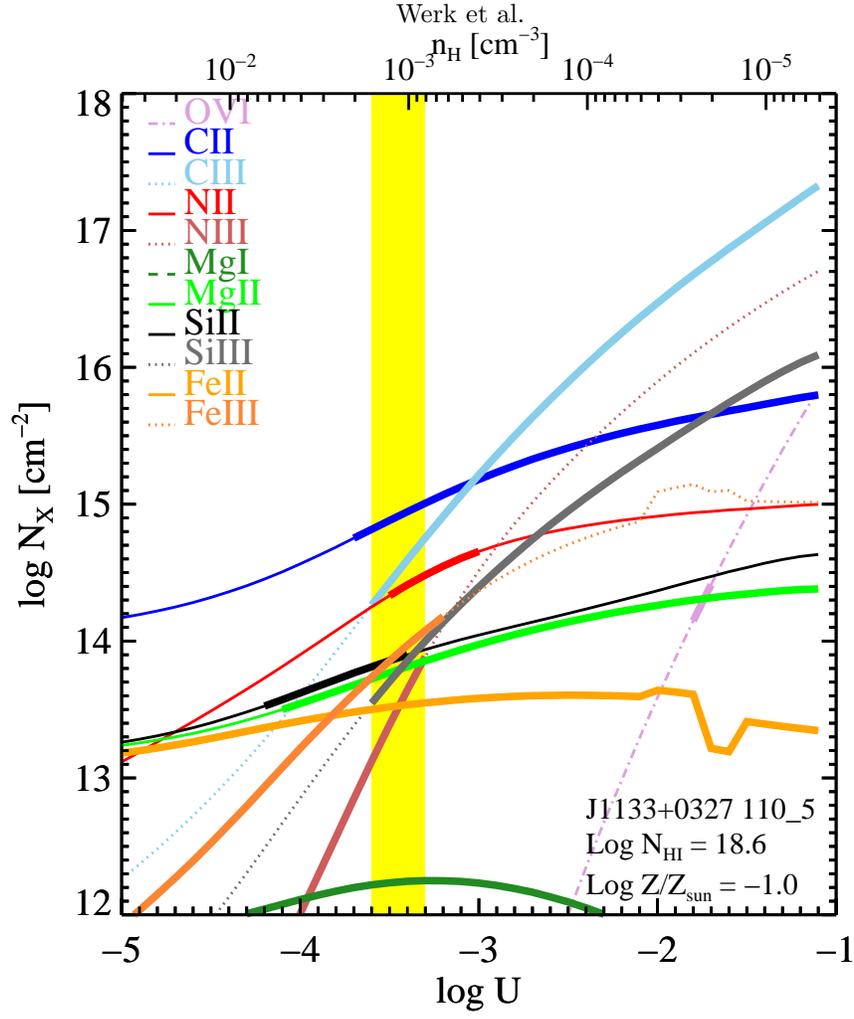}
\end{centering}
\caption{J1133+0327 110\_5: Here, the Ly$\alpha$ profile exhibits clear damping wings, which constrain the \nhi~ to be 18.6. The COS-data cover down to 919 \AA, where the HI absorption is still quite saturated. At this column density, the solution for the metallicity is fairly robust, with [X/H] of -1. We get good measurements of NII and SiII, which combined with lower limits on a range of other low ions , and upper limits on NIII and FeIII, provide a very consistent solution.  The models and data give a consistent solution for log U between -3.6 and -3.3, and we rate this solution with a 5. 
}
\label{fig:cloudyn}
\end{figure}

\begin{figure}[ht]
\vspace{0.15in}

\begin{centering}
\includegraphics[width=0.95\linewidth]{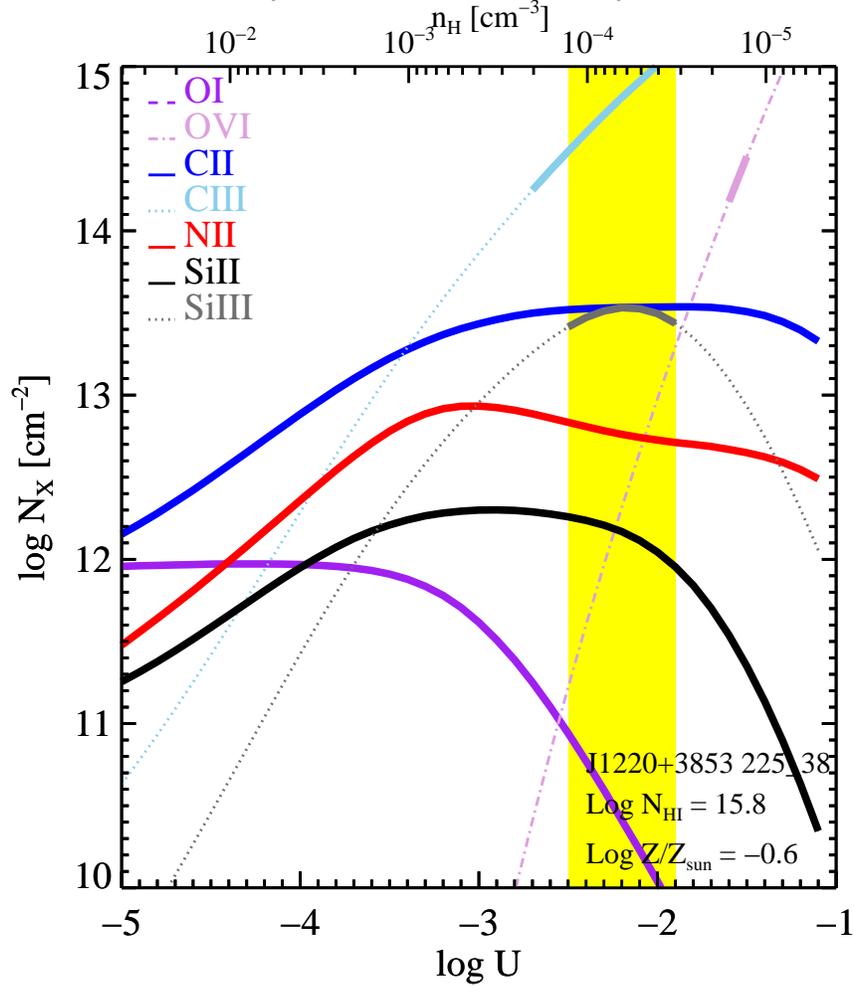}
\end{centering}
\caption{J1220+3853 225\_38:  Here, we have a tight constraint on the HI column density to be \nhi~ = 15.8, from the AODM measurement. The first line of the Lyman series that is not saturated in the data is  Lyman$\eta$, and the COS data cover down to 915 \AA. Good detections of SiIII and CIII, combined with non-detections of SiII and CII, constrain the metallicity and ionization parameter very well.  [X/H]  lies between -0.6 and solar, and log U for this set of models is well-constrained to lie between -2.5 and -2. 
}
\label{fig:cloudyo}
\end{figure}

\begin{figure}[ht]
\vspace{0.15in}

\begin{centering}
\includegraphics[width=0.95\linewidth]{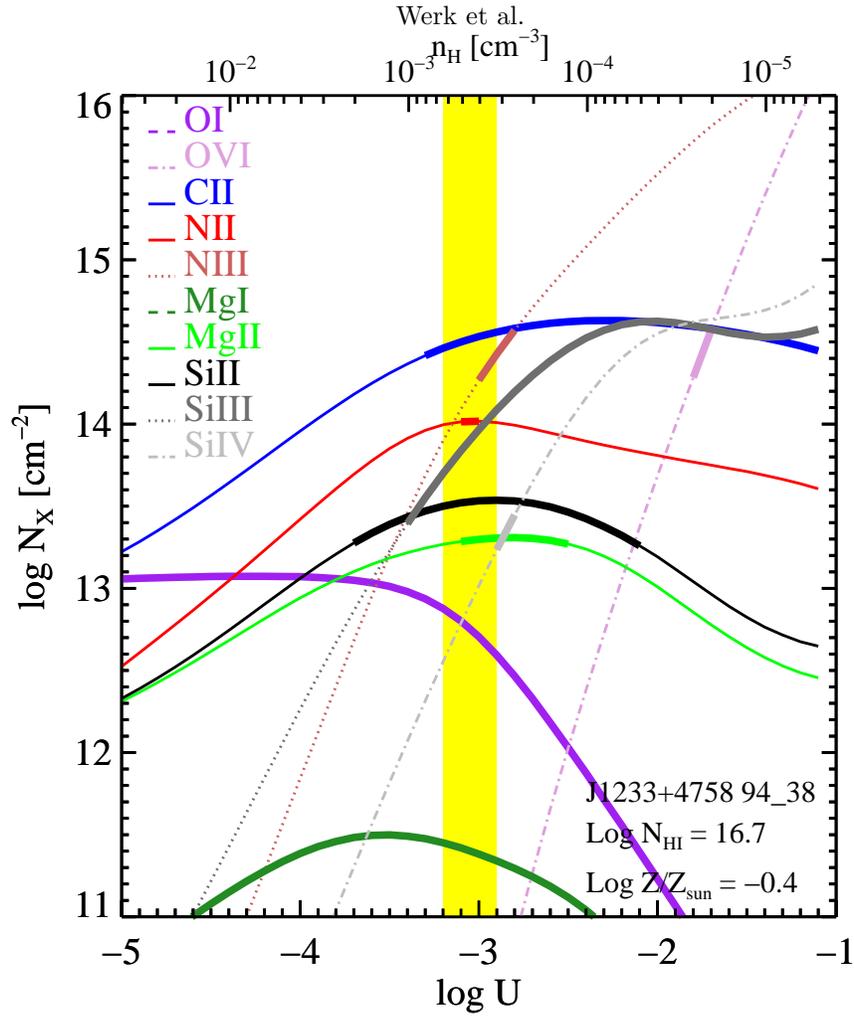}
\end{centering}
\caption{J1233+4758 94\_38:  HI is saturated down a nearly complete Lyman series, with the AODM lower limit to \nhi~ being 16.3, while the Voigt fitting prefers 18. We see no damping wings, and therefore we can place an additional upper limit on the \nhi~ to be 18.3.  The CII lower limit combined with the requirement that the gas not be super solar, permits models where \nhi~= 16.7 and above. At this HI column density, adopted here,  the low ion detections become self-consistent at a [X/H] of -0.4. The solution for log U is tightly constrained near -3, regardless of the HI column and metalliictity owing to the good detections of NII/NIII, and SiII, and SiIV. We rate this system a 5. 
}
\label{fig:cloudyp}
\end{figure}

\begin{figure}[ht]
\vspace{0.15in}

\begin{centering}
\includegraphics[width=0.95\linewidth]{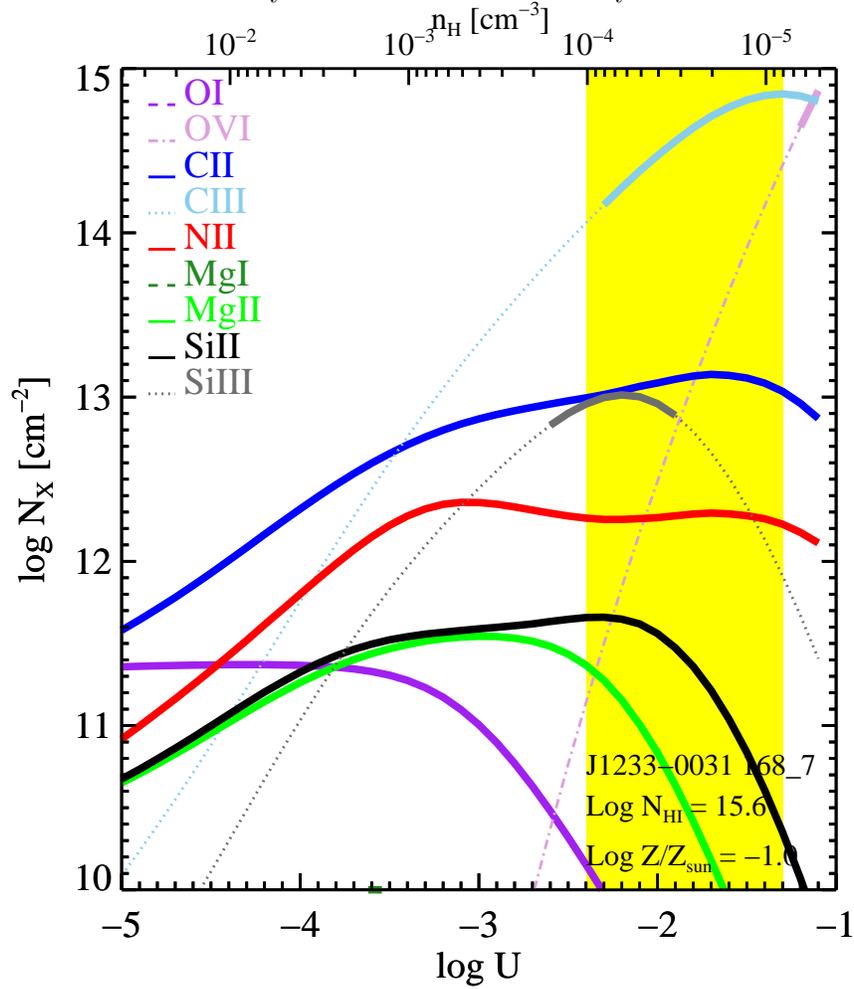}
\end{centering}
\caption{ J1233-0031 168\_7:  \nhi~ is securely measured to be 15.6 from the unsaturated COS data witch covers down to 915 \AA.  Non-detections of SiII, CII, NII, and detections of SiIII, CIII (saturated) work to constrain U decently over the range of allowed metallicity, -1.1 $<$ [X/H] $<$ 0.0. In this range of models, log U ranges between -2.4 and -1.3. We rate this system a 3. 
}
\label{fig:cloudyq}
\end{figure}

\begin{figure}[ht]
\vspace{0.15in}

\begin{centering}
\includegraphics[width=0.95\linewidth]{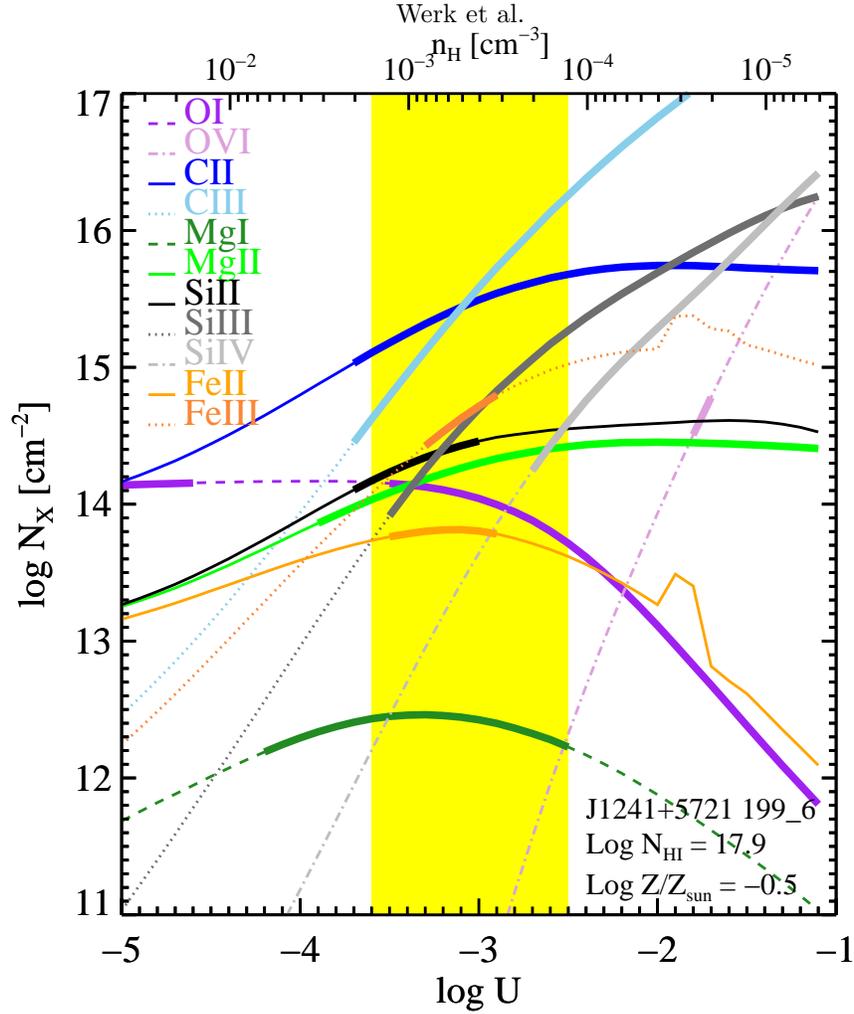}
\end{centering}
\caption{J1241+5721 199\_6: The HI column density is not well constrained by the COS data, with an AODM lower limit of 16.1 based on saturated Ly$\alpha$ and Ly$\beta$, while the Voigt profile fitting is best at \nhi~ of 18. There are not obvious damping wings seen in the HI absorption, and so we place an additional upper limit of 18.5 on the HI column density. Based on the detection of MgI, the saturation of FeII in our HIRES data, and a good detection of FeIII, we consider models with \nhi~ above 17.9 so that the gas metallicity is not super-solar. The solutions at this column density fit best with [X/H] of -0.5. In the range of allowed \nhi, with metallicities adjusted accordingly, the solutions for log U remain consistent between -3.5 and -2.5. We rate this system a 3. 
}
\label{fig:cloudyr}
\end{figure}

\begin{figure}[ht]
\vspace{0.15in}

\begin{centering}
\includegraphics[width=0.95\linewidth]{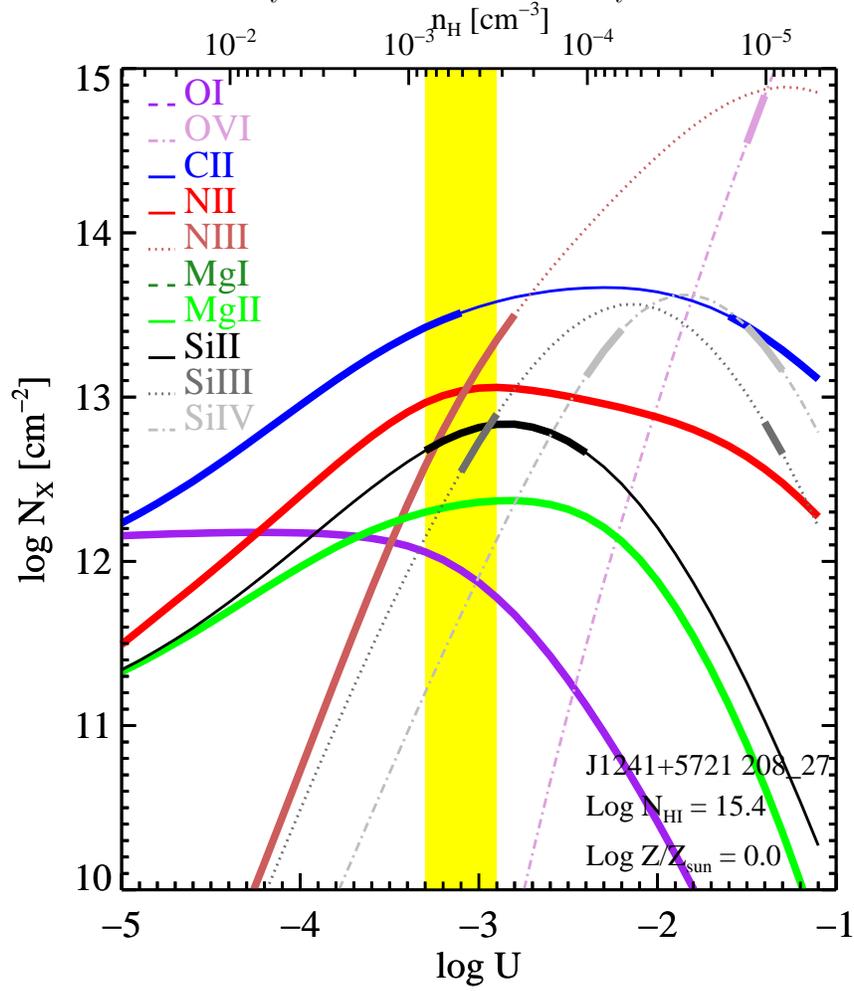}
\end{centering}
\caption{ J1241+5721 208\_27:  The HI column density is well-determined at \nhi~= 15.4. There are detections of SiII, SiIII, and SiIV in the COS spectral data, with the rest of the ions remaining undetected (no coverage of CIII). SiII/SiIII tightly constrain log U to be between -3.3 and -2.9, and the metallicity of the gas to be solar. This model, and all models at this HI column density underproduce SiIV by a large degree. This inconsistency is somewhat surprising given the good correspondence of the absorption of all the silicon ions over the same velocity range, with the same profile. Nonetheless, we are forced to accept a model 
in which some fraction of the SiIV is in a different phase from the SiII and SiIII. Although the range of log U is tight here, we give this a solution a rating of 4 because of its multiphase nature. 
}
\label{fig:cloudys}
\end{figure}

\begin{figure}[ht]
\vspace{0.15in}

\begin{centering}
\includegraphics[width=0.95\linewidth]{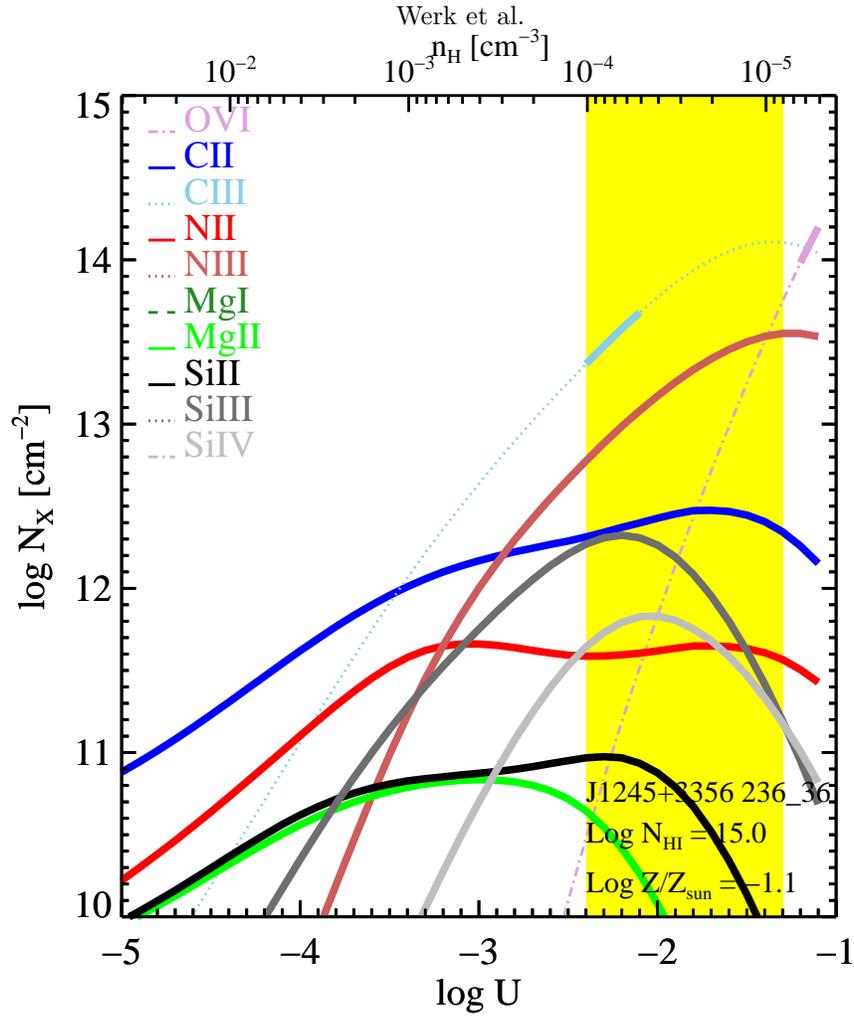}
\end{centering}
\caption{J1245+3356 236\_36: The HI is well-constrained in this system with \nhi~=14.7, well in the optically thin limit (we plot \nhi~= 15 here, which makes no difference in the results). At this column density, the non-detection of SiIII implies that the gas is less than a tenth solar, while the good detection of CIII over this range of metallicity requires -2.4 $<$ log U $<$ -1.3. We rate this system a 3. 
}
\label{fig:cloudyt}
\end{figure}

\begin{figure}[ht]
\vspace{0.15in}

\begin{centering}
\includegraphics[width=0.95\linewidth]{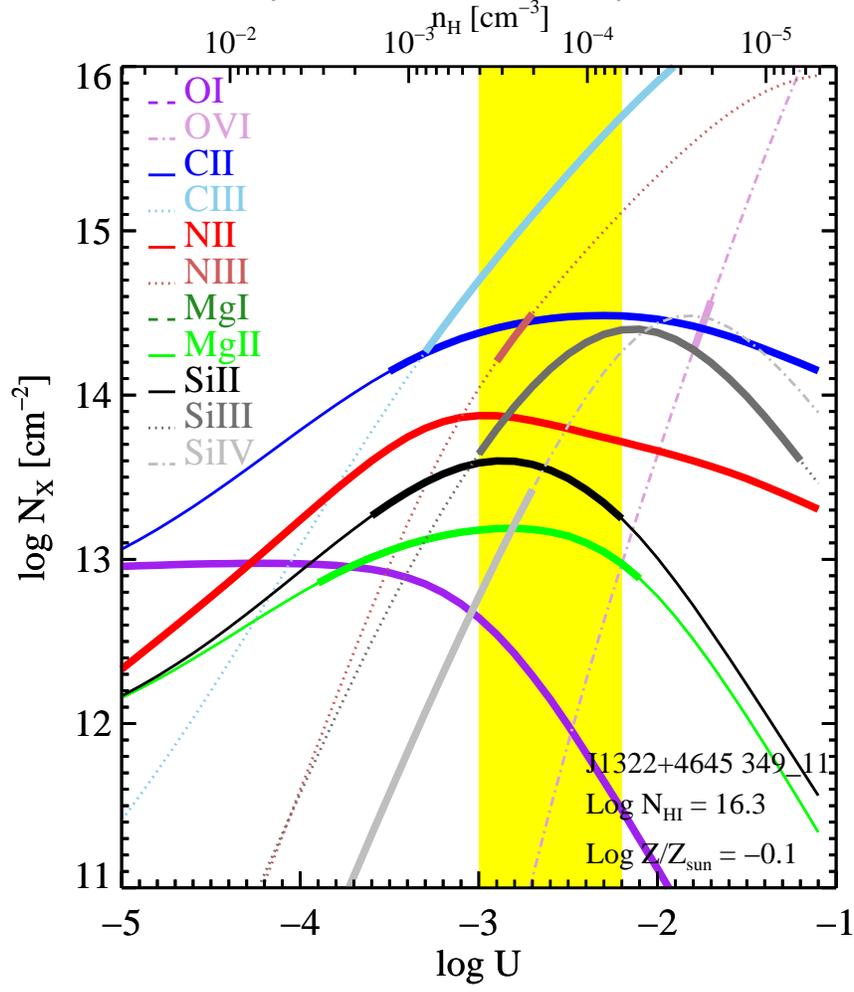}
\end{centering}
\caption{J1322+4645 349\_11:  The HI is not well-constrained with coverage down to 937 \AA as  all lines saturated. \nhi~$>$16.3, based on the AODM measurement, while the Voigt profile analysis prefers \nhi~= 18.0. The lack of damping wings in the Ly$\alpha$ profile provide an additional upper limit of 18.3. We model the gas at the AODM lower limit, since at this value of \nhi~ we get consistent solutions with the low-ions, noting that the solution for log U remains between -3 and -2.2 over the full range of HI and metallicity. We rate this solution a 4. 
}
\label{fig:cloudyu}
\end{figure}

\begin{figure}[ht]
\vspace{0.15in}

\begin{centering}
\includegraphics[width=0.95\linewidth]{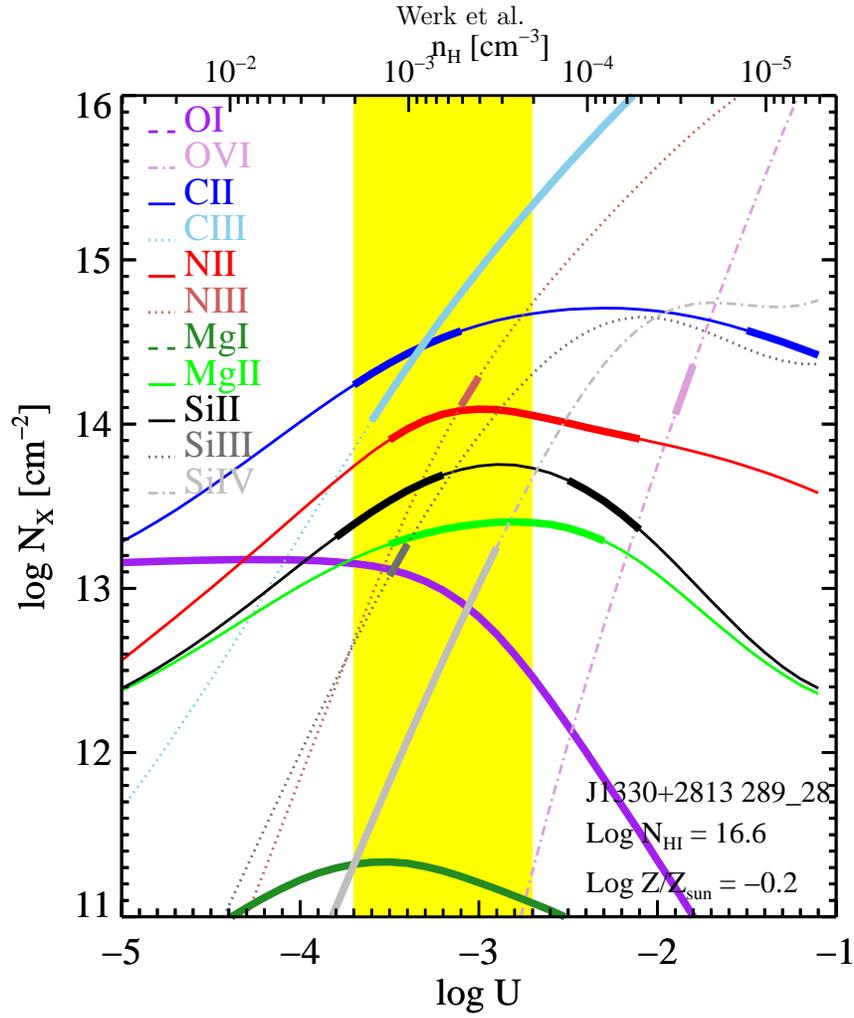}
\end{centering}
\caption{ J1330+2813 289\_28: While the AODM lower limit to the HI column density is 15.9, the Ly$\alpha$ profile shows damping wings, which allow us to constrain \nhi~=18.3. Nonetheless, the CLOUDY models can fit the data at \nhi~ of 16.6, so to be conservative we show the solution for log U at this adopted column density. The solution for log U is the same at higher column density, though the metallicity at the higher column is best fit at [X/H] of -1.6 (see also Figure 3 in Appendix A). The NII/NIII ratio gives a solution for log U that is slightly higher than the one solution for CII/CIII and SiII/SiIII. The upper limit on SiIV is also well-represented by this model. We rate this solution a 4. 
}
\label{fig:cloudyv}
\end{figure}

\begin{figure}[ht]
\vspace{0.15in}

\begin{centering}
\includegraphics[width=0.95\linewidth]{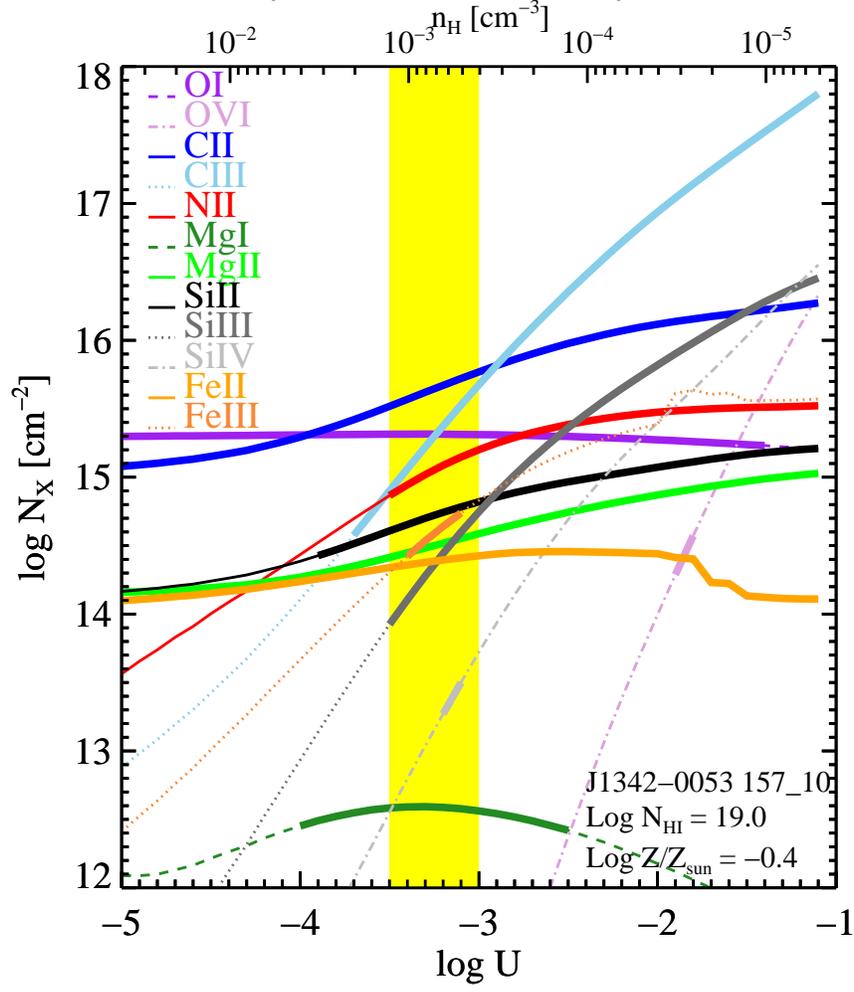}
\end{centering}
\caption{J1342-0053 157\_10: Based on the fits to the very obvious damping wings present in the Ly$\alpha$ profile, and Voigt profile fits down the Lyman series to Ly$\gamma$, the HI column density of this system is well constrained with \nhi~= 19.0. A good measurement of OI at 1039 \AA allows us to constrain the metallicity to be [X/H] = -0.4. At this metallicity, there is a remarkably consistent solution for log U between -3.5 and 3, based on FeIII, MgI, and SiIV, which is well-modeled here. This solution gets our highest rating of 5. 
}
\label{fig:cloudyw}
\end{figure}

\begin{figure}[ht]
\vspace{0.15in}

\begin{centering}
\includegraphics[width=0.95\linewidth]{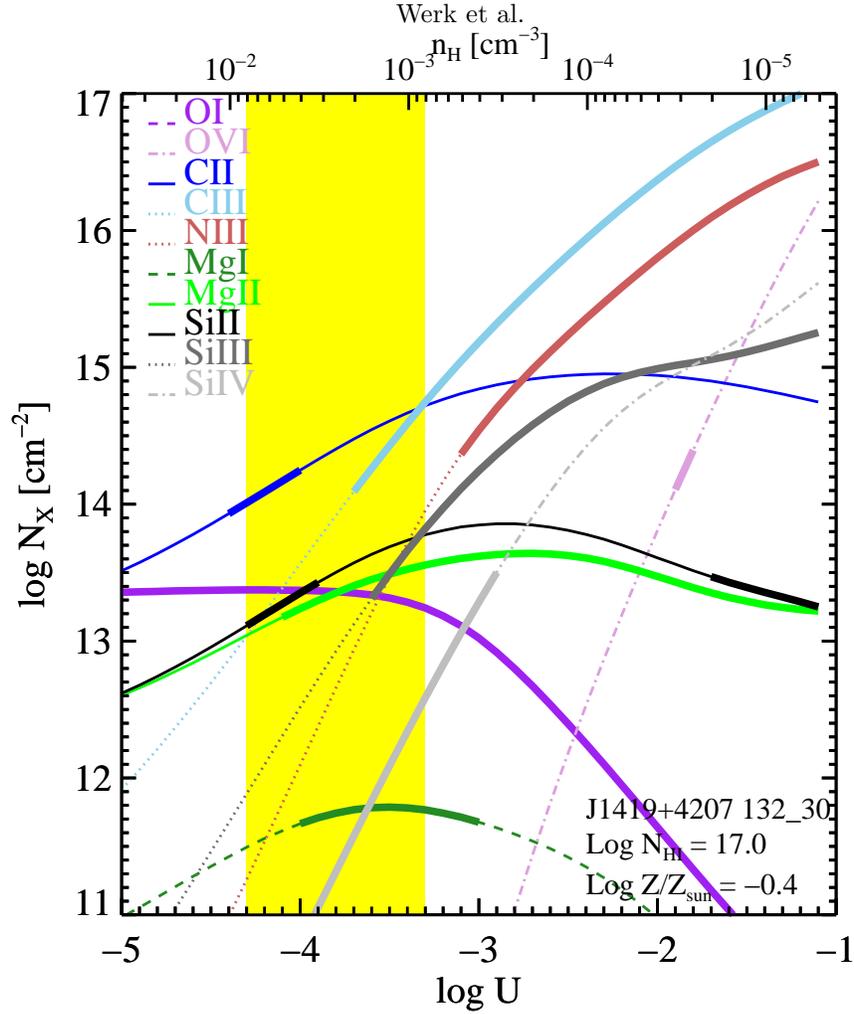}
\end{centering}
\caption{J1419+4207 132\_30:  The COS data cover HI down to Ly$\gamma$ which is quite saturated, providing a constraint on \nhi~ $>$ 15.4. The best fits to the data seem to show slight damping wings, and settle at a \nhi~ of 18.3. The constraint that the gas not be super solar requires that the HI column density be at least 17.0, based on several low ions. At this column density and above, there are no consistent solutions for the low ions in the singly ionized phase and doubly ionized phase.  In particular, the lower limit of NIII is quite discrepant from the rest of the data, which tend toward lower values of log U. We have adopted the lowest value of \nhi~ allowed by the data, and the metallicity that seems to best represent the solution for log U, [X/H] = -0.4, with -4.3 $<$ log U $<$ -3.3. We have rated this system a 2 owing to the potential multi-phase nature of the low ionization state gas.The conservative \nhi~ and the lower value of log U should nonetheless represent conservative limits on the amount of the  lowest ionization state gas that goes along with the neutral hydrogen. 
}
\label{fig:cloudyx}
\end{figure}

 \clearpage

\begin{figure}[ht]
\vspace{0.15in}

\begin{centering}
\includegraphics[width=0.95\linewidth]{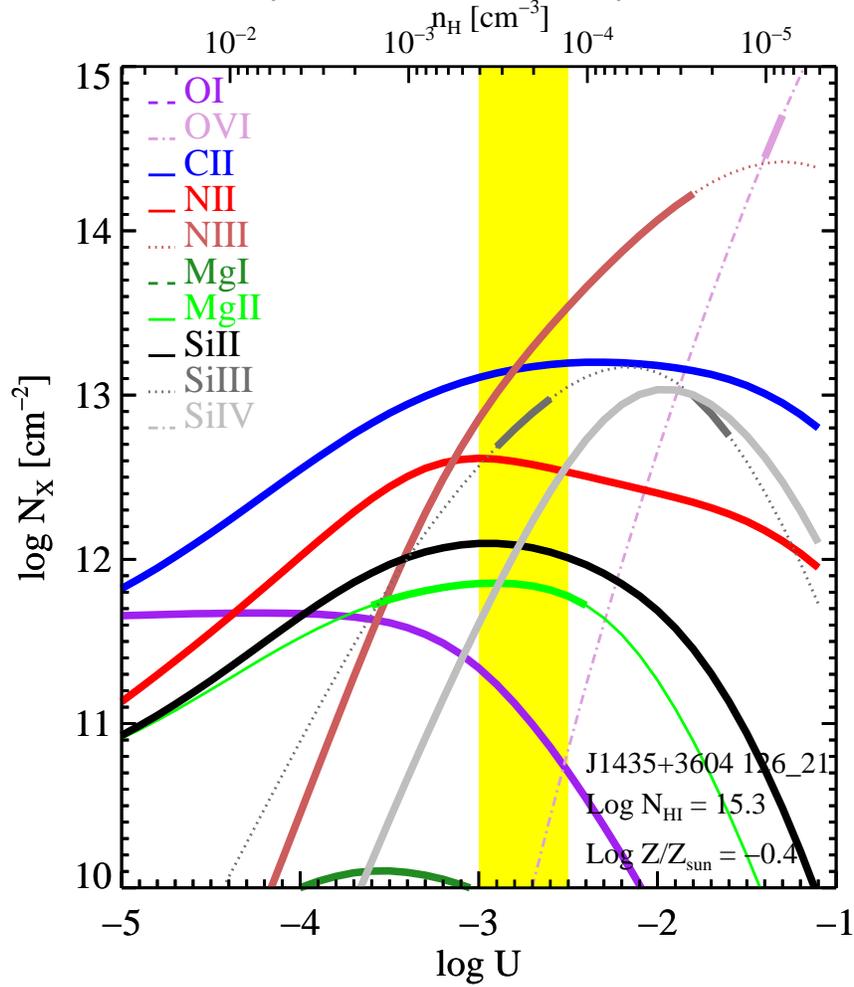}
\end{centering}
\caption{J1435+3604 126\_21: The \nhi~ is well-measured for this system to be 15.3, and there are good detections of both MgII and SiIII, which work well together to constrain both metallicity and ionization parameter. The model solution converges at -0.5 $<$ [X/H] $<$ 0.1 and -3 $<$ log U $< $ -2.5. The MgII detection is very weak, and the weaker of the doublet at 2852 \AA is not detected, though the upper limit is consistent with the measurement of the stronger line at 2796 \AA. Though the constraints on this system are good, we are a bit wary that they are based on very weak detections of single absorption lines for each of the ions, and so we nonetheless rate it a 4. 
}
\label{fig:cloudyy}
\end{figure}
\clearpage

\begin{figure}[ht]
\vspace{0.15in}

\begin{centering}
\includegraphics[width=0.95\linewidth]{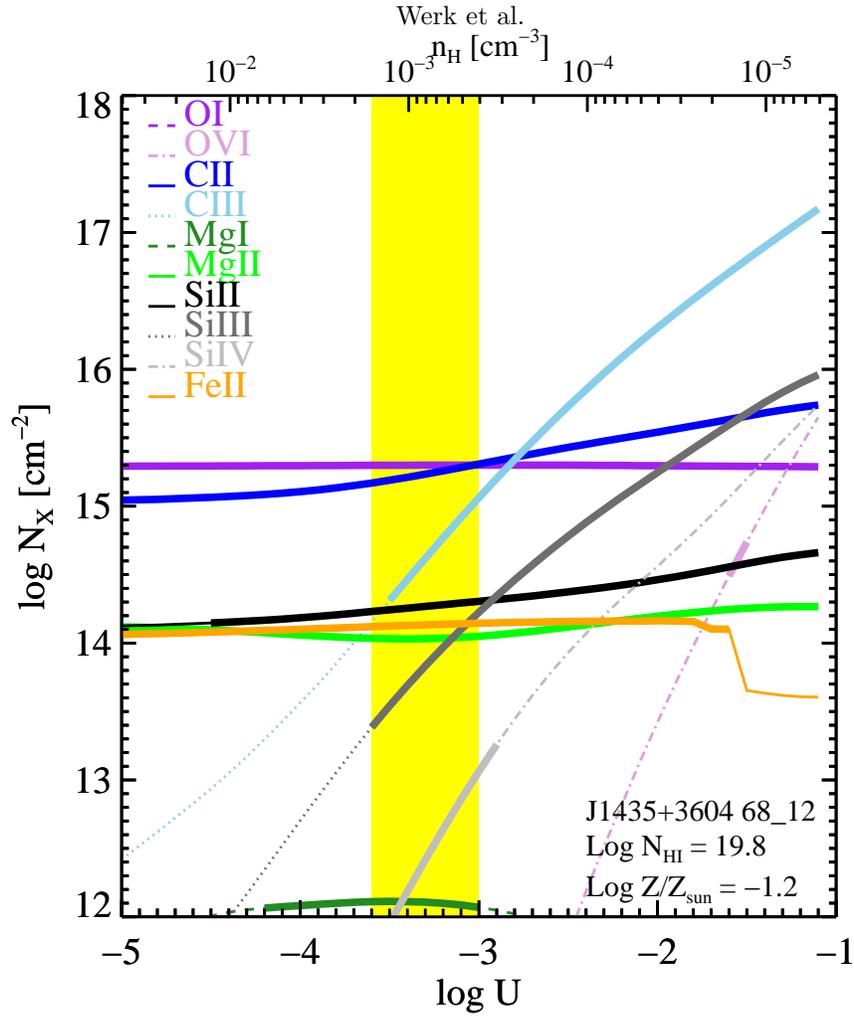}
\end{centering}
\caption{J1435+3604 68\_12: The Ly$\alpha$ and Ly$\beta$ for this system exhibit significant damping wings, constraining the \nhi~ of this system well at 19.8. Most of the low ion lines are saturated, with the exception of FeII and MgI, which provide a good solution for the metallicity at [X/H] = -1.2. At these two model parameters, the SiIII and CIII lower limits provide a lower bound to log U at -3.6, while the MgI bounds the upper value of log U to be -3. SiIV also fits well with this solution. We rate this system a 5. 
 }
\label{fig:cloudyz}
\end{figure}

\begin{figure}[ht]
\vspace{0.15in}

\begin{centering}
\includegraphics[width=0.95\linewidth]{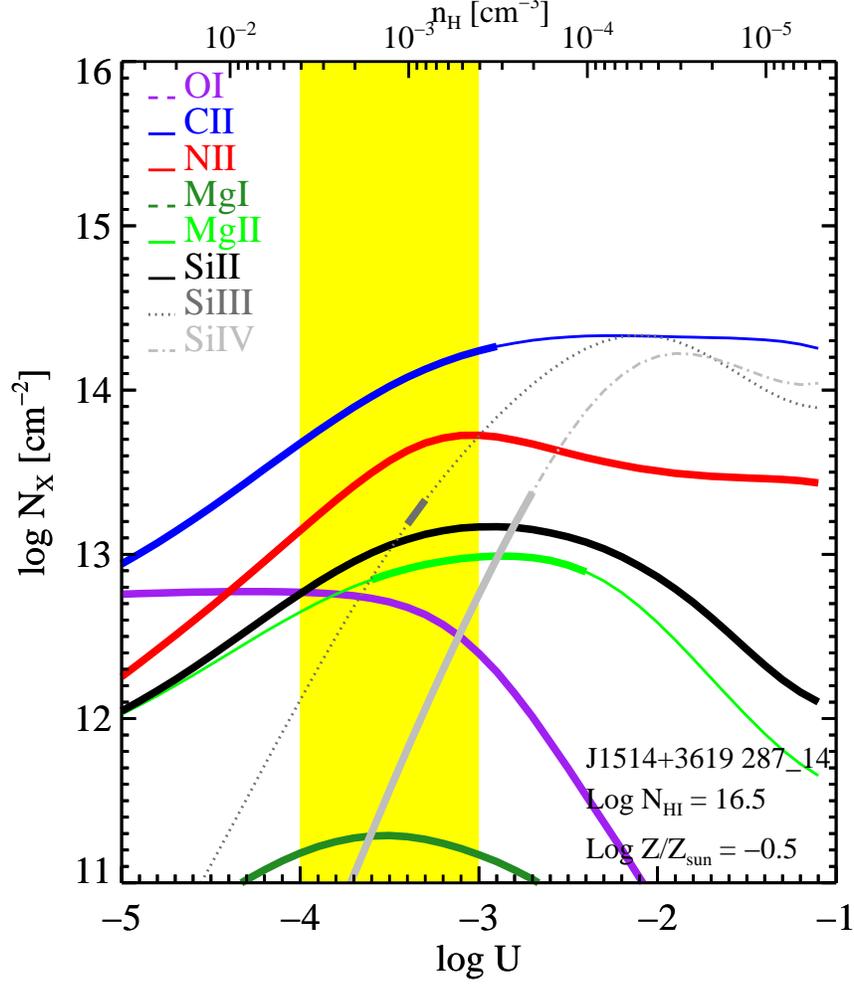}
\end{centering}
\caption{ J1514+3619 287\_14:  Because we cover only Ly$\alpha$ with the COS data for this system, which has several saturated components over a wide velocity range, \nhi~ is not well constrained. The AODM limit is 14.7, and the best fit to HI seems to prefer closer to 18.  The lack of damping wings is hardly reassuring, but sets the upper bound at 18.5 decisively. This system is also unfortunately hampered by blending in the silicon, non-detections of several key ions, and poor S/N over the carbon lines.  The SiIII detection is slightly dubious, as it is blended with a saturated CII 1036 \AA~ line at $z\sim0.411$, and so we do not emphasize the limits to the model based on this line.  The MgII robust detection in the HIRES data is very helpful for the solution.  The prior that the gas not be super solar requires a \nhi~ $>$ 16.3. For the MgII to be consistent with the CII upper limit, we are limited to log U $<$ -3 over the full allowable range of HI column density. We can use the SiIII to place an additional bound on log U $>$ -3.7. This solution for log U for this system is pretty well constrained despite several complications, and we rate it a 4. 
}
\label{fig:cloudyaa}
\end{figure}

\begin{figure}[ht]
\vspace{0.15in}

\begin{centering}
\includegraphics[width=0.95\linewidth]{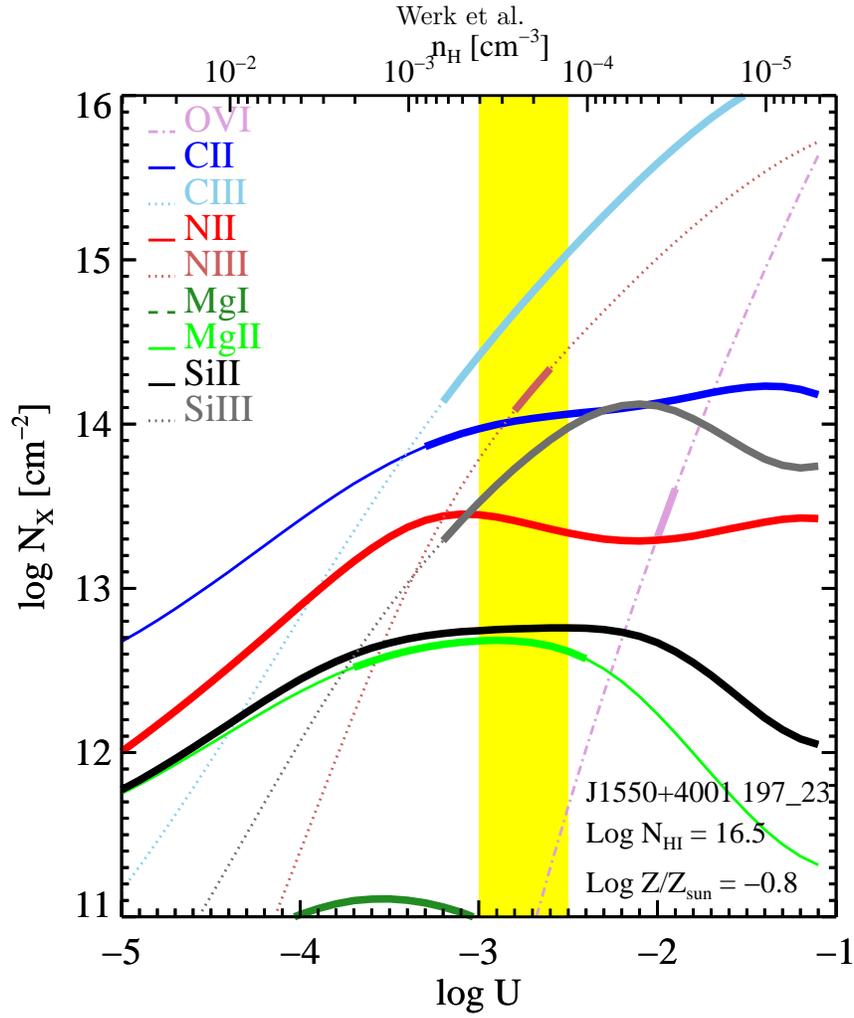}
\end{centering}
\caption{ J1550+4001 197\_23:  The COS data show the full Lyman series here which allows for a good measurement of \nhi~ = 16.5. A range of low ion detections allow us to tightly constrain both the metallicity and ionization parameter of this gas. At this column density, the MgII and NII set the metallicity and ionization to have the allowable range of -0.8 $<$ [X/H] $<$ -0.5 and -3.0 $<$ log U $<$ -2.5. Because the data are very consistent with the preferred model, we rate this system a 5. 
}
\label{fig:cloudyab}
\end{figure}

\begin{figure}[ht]
\vspace{0.15in}

\begin{centering}
\includegraphics[width=0.95\linewidth]{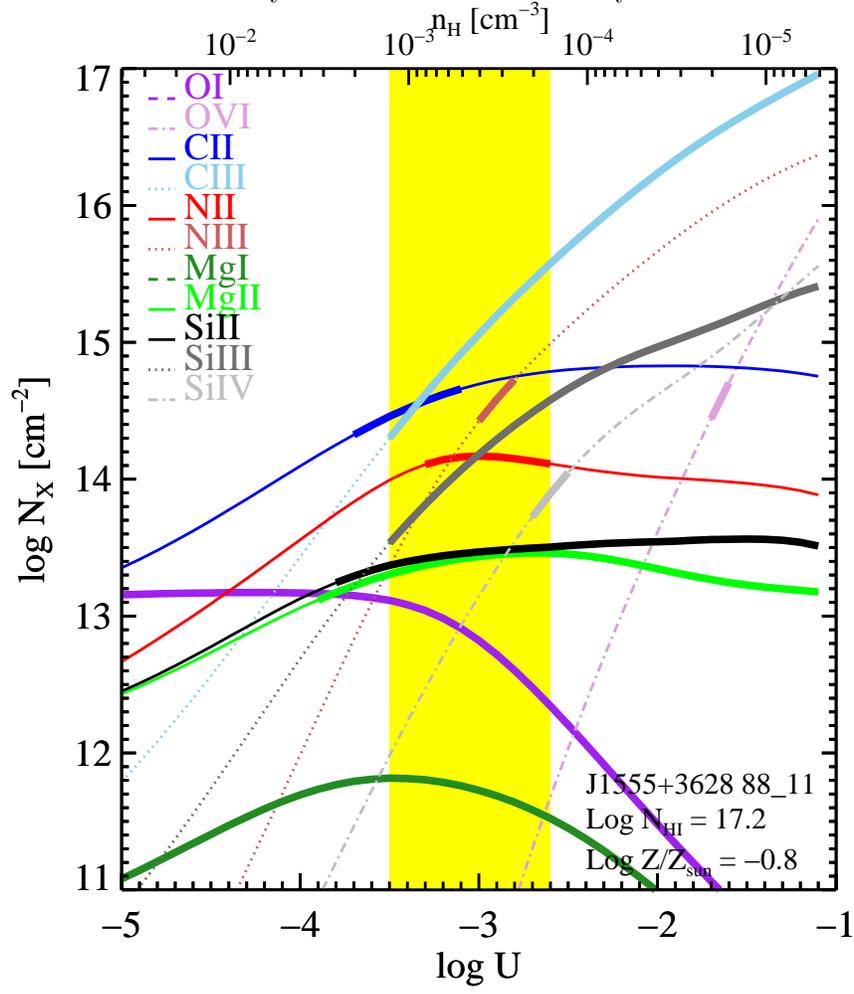}
\end{centering}
\caption{J1555+3628 88\_11: \nhi~ is not well-constrained by the data -- The AODM lower limit based on saturated Ly$\alpha$ - Ly$\gamma$ is 15.7, and the best fit single-component Voigt profile estimates an \nhi~ of 18.2. The most consistent solutions for the range of detected low ions occurs  at \nhi~$>$  17.2, with the requirement that the gas not be super solar limiting us to \nhi~ $>$ 16.6. We show the solution at the adopted HI column density of 17.2, since below this value the NII is inconsistent with MgII and SiII. In this case, the ratio of NII/NIII, CII/CIII, and SiII/SIII tightly constrain log U to lie between -3.5 and -2.6. We rate this solution a 4. 
}
\label{fig:cloudyac}
\end{figure}

\begin{figure}[ht]
\vspace{0.15in}

\begin{centering}
\includegraphics[width=0.95\linewidth]{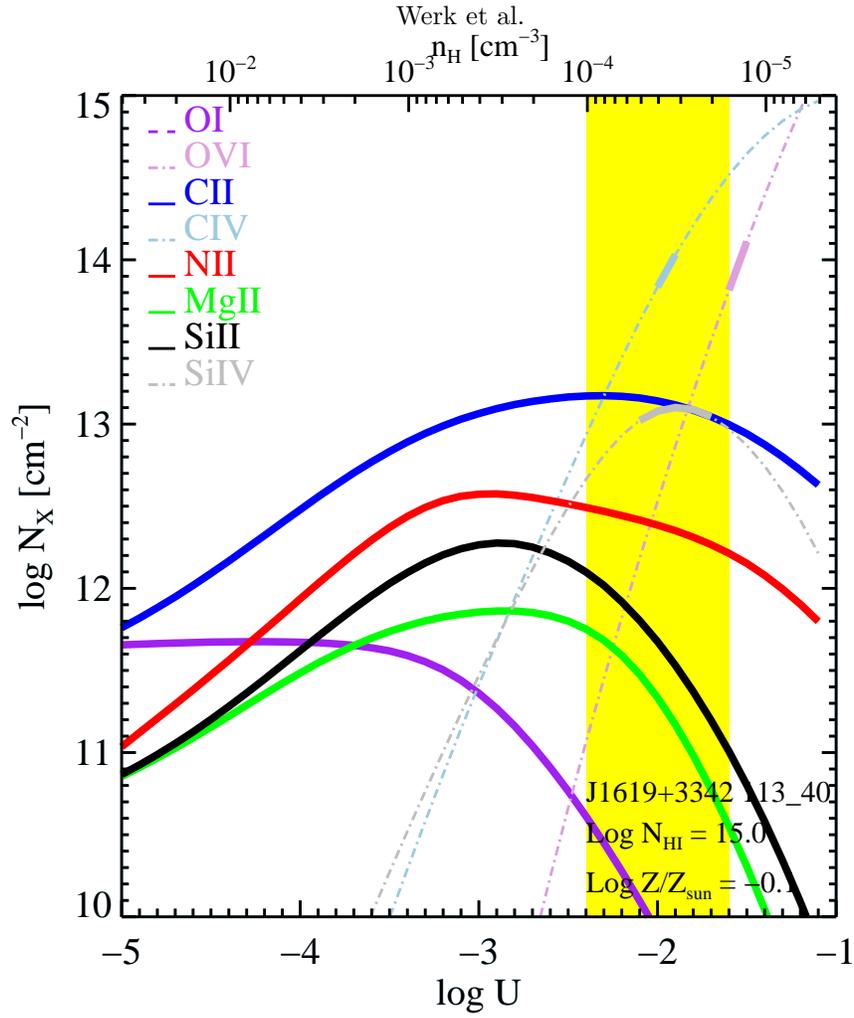}
\end{centering}
\caption{J1619+3342 113\_40: The HI absorption is slightly saturated in Ly$\beta$, with both AODM lower limit and best-fit Voigt profile giving a \nhi~ of 15.0. We adopt that value here. We can place an additional upper limit on the HI column density of 17.5 based on the equivalent width of H$\beta$. The only two detections are of CIV and SiIV, which together constrain log U to be between -2.4 and -1.6, and [X/H] to be between -0.7 and 0. We rate this solution a 4. 
}
\label{fig:cloudyad}
\end{figure}

\begin{figure}[ht]
\vspace{0.15in}

\begin{centering}
\includegraphics[width=0.95\linewidth]{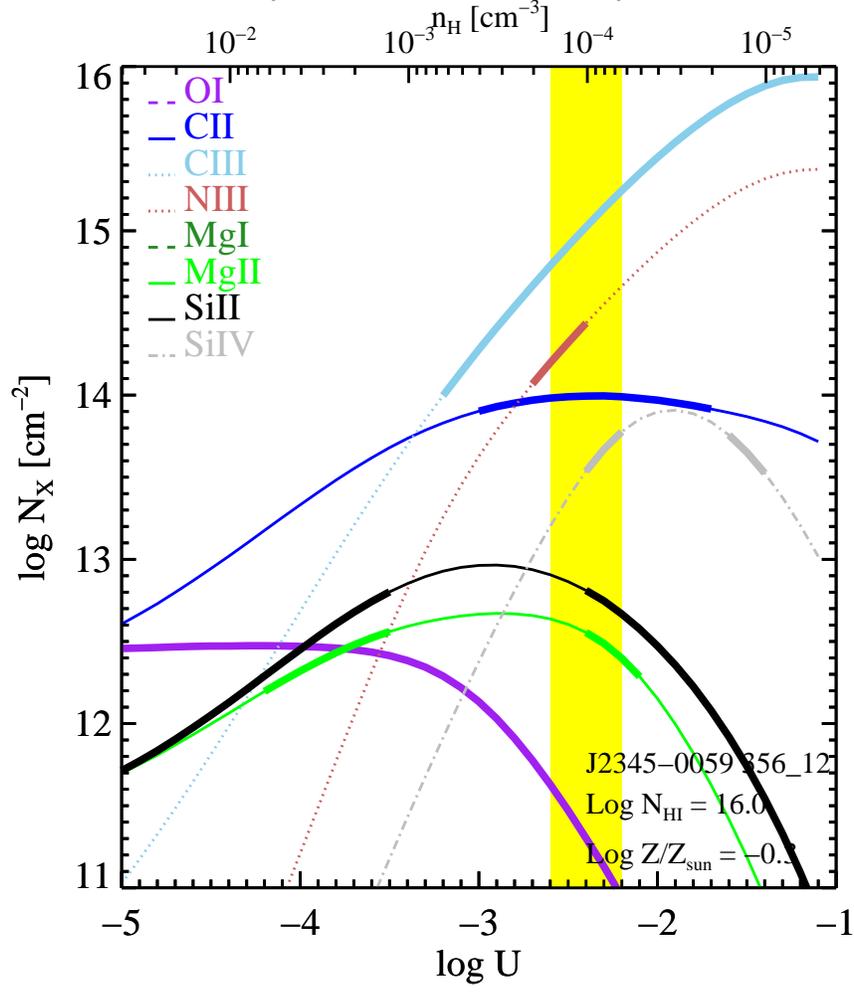}
\end{centering}
\caption{J2345-0059 356\_12: The COS data cover the Lyman series down to 930 \AA, which allows a measurement of \nhi~ at 16.0. CII/CIII,  and NIII along with a weak detection of MgII at 2796 \AA tightly constrain log U to be between -2.6 and -2.2. At this column density, the metallicity is -0.3 $<$ [X/H] $<$ 0.0. If we instead treat the weak detections of MgII and CII as upper limits, which may be justified by the data, the preferred metallicity drops a bit and higher values of log U are allowed. We adopt the lower range of log U here.  SiIV is consistent with the low ion data at these model parameters. We rate this solution a 4, due to the slightly doubtful nature of MgII and CII. 
}
\label{fig:cloudylast}
\end{figure}

 \clearpage

\end{document}